\documentclass[aps,reprint,superscriptaddress,balancelastpage,showpacs,prx]{revtex4-2}
\usepackage[svgnames]{xcolor}
\usepackage{centernot}
\usepackage{graphicx}
\usepackage{amsmath}
\usepackage{times}
\usepackage{amssymb}
\usepackage{mathrsfs}
\usepackage{chemarr}
\usepackage{color}
\usepackage{bbold}
\usepackage{url}
\usepackage{version}
\usepackage{csquotes}
\definecolor{linkcolor}{rgb}{0,0,0.6}
\usepackage[pdftex,colorlinks=true,
pdfstartview=FitV,
linkcolor= linkcolor,
citecolor= linkcolor,
urlcolor= linkcolor,
hyperindex=true,
hyperfigures=false]
{hyperref}
\usepackage{bookmark}
\DeclareMathSymbol{\shortminus}{\mathbin}{AMSa}{"39}

\newcommand{\cF}{\mathcal{F}}

\newcommand{\cC}{\mathcal{C}}

\newcommand{\cU}{\mathcal{U}}

\newcommand{\cA}{\mathcal{A}}

\newcommand{\cS}{\mathcal{S}}
\newcommand{\cO}{\mathcal{O}}

\newcommand{\cN}{\mathcal{N}}

\newcommand{\dd}{\text{d}}

\newcommand{\OO}{\text{O}}

\newcommand{\bu}{\textbf{u}}

\newcommand{\bX}{\textbf{X}}
\newcommand{\XX}{\text{X}}

\newcommand{\br}{\textbf{r}}

\newcommand{\bF}{\textbf{F}}

\newcommand{\red}[1]{{\color{red}#1}}

\newcommand{\bluelink}[1]{{\color{linkcolor}#1}}

\newcommand{\eref}[1]{(\ref{#1})}
\newcommand{\Fref}[1]{Fig.~\ref{#1}}

\DeclareMathOperator{\argsh}{argsh}
\DeclareMathOperator{\sech}{sech}
\DeclareMathOperator{\erfc}{erfc}
\DeclareMathOperator{\TExp}{TExp}

\definecolor{greenpyplot}{RGB}{0,128,0}
\definecolor{greenswap}{RGB}{55,92,50}
\definecolor{greenchaos}{RGB}{103,191,92}
\definecolor{greenleaf}{RGB}{118, 188, 124}
\definecolor{bordeau}{RGB}{167, 1, 69}
\definecolor{airforceblue}{RGB}{39, 80, 176}
\definecolor{purpledisord}{RGB}{128, 0, 128}
\definecolor{orangeord}{RGB}{255, 165, 0}
\definecolor{c_rho_2}{RGB}{106, 174, 214}
\definecolor{c_m_2}{RGB}{250, 105, 73}
\definecolor{c_species_1}{RGB}{106, 174, 214}
\definecolor{c_species_2}{RGB}{250, 105, 73}
\definecolor{shift_up}{RGB}{110, 5, 33}
\definecolor{shift_down}{RGB}{199, 62, 98}
\definecolor{green_dens}{RGB}{11, 84, 6}

\graphicspath{{./}{./Images/}}

\usepackage{tikz}
\usetikzlibrary{calc}
\usetikzlibrary{arrows, decorations.markings}
\usetikzlibrary{positioning}
\usetikzlibrary{decorations.pathreplacing}
\usetikzlibrary{arrows.meta}
\usetikzlibrary{shapes}
\usetikzlibrary{decorations.text}
\usetikzlibrary{math}

\begin{document}

\title{The transition to collective motion in nonreciprocal active matter: coarse graining agent-based models into fluctuating hydrodynamics}

\author{David Martin}
\affiliation{University of Chicago, Kadanoff Center for Theoretical Physics and Enrico Fermi Institute, 933 E 56th St, Chicago, IL 60637}

\author{Daniel Seara}
\affiliation{University of Chicago, James Franck Institute, 929 E 57th Street, Chicago, IL 60637}

\author{Yael Avni}
\affiliation{University of Chicago, James Franck Institute, 929 E 57th Street, Chicago, IL 60637}

\author{Michel Fruchart}
\affiliation{University of Chicago, James Franck Institute, 929 E 57th Street, Chicago, IL 60637}
\affiliation{ESPCI, Laboratoire Gulliver, 10 rue Vauquelin,
75231 Paris cedex 05}

\author{Vincenzo Vitelli}
\affiliation{University of Chicago, James Franck Institute, 929 E 57th Street, Chicago, IL 60637}
\affiliation{University of Chicago, Kadanoff Center for Theoretical Physics, 933 E 56th St, Chicago, IL 60637}

\date{\today}

\begin{abstract}
  Two hallmarks of non-equilibrium systems, from active colloids to animal herds, are agents motility and nonreciprocal interactions.
  Their interplay creates feedback loops that lead to complex spatiotemporal dynamics crucial to understand and control the non-linear response of active systems.
  Here, we introduce a minimal model that captures these two features at the microscopic scale, while admitting an exact hydrodynamic theory valid also in the fully-nonlinear regime. 
Using statistical mechanics techniques we exactly coarse-grain our non-reciprocal microscopic model into a fluctuating hydrodynamics and use dynamical systems insights to analyze the resulting equations.
  In the absence of motility, we find two transitions to oscillatory phases occurring via distinct mechanisms: a Hopf bifurcation and a Saddle-Node on Invariant Circle (SNIC) bifurcation.
  In the presence of motility, this rigorous approach, complemented by numerical simulations, allows us to quantitatively assess the hitherto neglected impact of inter-species nonreciprocity on a paradigmatic transition in active matter: the emergence of collective motion.
  When nonreciprocity is weak, we show that flocking is accelerated and bands tend to synchronize with a spatial overlap controlled by nonlinearities.
  When nonreciprocity is strong, flocking is superseded by a Chase \& Rest dynamical phase where each species alternates between a chasing state, when they propagate, and a resting state, when they stand still. Phenomenological models with linear non-reciprocal couplings fail to predict the Chase \& Rest phase which illustrates the usefulness of our exact coarse-graining procedure.
  Finally, we demonstrate how fluctuations in finite systems can be harnessed to characterize microscopic non-reciprocity from macroscopic time-correlation functions, even in phases where nonreciprocal interactions do not affect the thermodynamic steady-state.
\end{abstract}

\pacs{}

\maketitle

Flocking is an emergent nonequilibrium phenomenon in which interactions between individual agents produce collective motion at large scale. It has been observed both in natural systems such as flocks of starlings \cite{cavagna2010scale,ballerini2008interaction} or human crowds \cite{Bain2019} and in artificial systems ranging from self-propelled colloids \cite{Bricard2013} to driven filaments \cite{Schaller2010}. Models of flocking typically boil down to three ingredients: agents are self-propelled, they tend to align with each other, and they are subject to a random noise. Collective motion occurs when the alignment tendency beats the noise.
In addition to these three ingredients -- self-propulsion, alignment, and noise -- systems where flocking occurs often exhibit non-reciprocal interactions between the constituents~\cite{fruchart2021non,you2020nonreciprocity,Saha2020,ivlev2015statistical,dinelli2022self,yllanes2017many,Brauns2023,Kreienkamp2022,Meredith2020,Zhang2023,Osat2022,Loos2020,Alston2023,Mandal2022}. These non-reciprocal interactions may have behavioral origins such as the presence of multiple competing populations~\cite{you2020nonreciprocity,fruchart2021non,ivlev2015statistical,dinelli2022self,yllanes2017many,Brauns2023,Saha2020,haluts2024active} or hierarchical structures within a single populations \cite{Nagy2010,yllanes2017many}, as well as more mechanistic origins such as hydrodynamic or chemical interactions~\cite{Liebchen2021,Baek2018,Saha2019,Poncet2022}, cones of visions~\cite{Loos2023,Lavergne2019,saavedra2024swirling}, or even just motility \cite{dadhichi2020nonmutual,Chepizhko2021,kreienkamp2024non}. 
While the transition to collective motion is perhaps the most iconic manifestation of active matter, little is known on how it is affected by non-reciprocal interactions between flocking agents~\cite{dadhichi2020nonmutual,Chepizhko2021,seara2023non,yllanes2017many,bagarti2019milling,chen2017fore,Chatterjee2023,Ferretti2022,
Dinelli2022,Zhang2021,Duan2023,Maity2023,Kreienkamp2022,chatterjee2023flocking,kursten2023flocking}.

\begin{figure*}
\hspace*{-0.7cm}\includegraphics{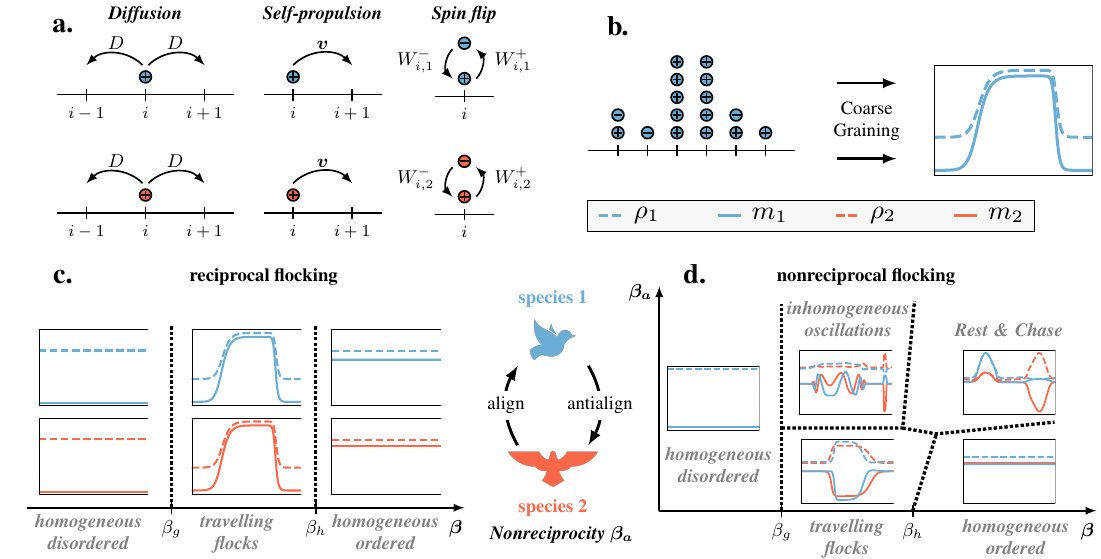}
  \caption{
  \textbf{Microscopic model of non-reciprocal flocking.} Two minimal ingredients are needed for flocking: i) microscopic agents have to be self-propelled and ii) they have to align their direction of motion along the ones of their fellows. When the alignment strength $\beta$ is increased, two non-interacting species exhibit a discontinuous transition to collective motion featuring bands at onset. However, two species generically interacts in a nonreciprocal way which will impact the phase diagram of this transition to collective motion.
  {\bf a.} Schematic summary of the different microscopic moves \protect\hyperlink{rule:diffusion}{RI}, \protect\hyperlink{rule:hopping}{RII}, \protect\hyperlink{rule:flip1}{RIII}, \protect\hyperlink{rule:flip2}{RIV} for species 1 (blue) and species 2 (red). {\bf b.} Coarse-graining the on-lattice dynamics of a single species yields two smooth fields: the density $\rho$ (dashed line) and magnetization $m$ (plain line). {\bf c.} Schematic representation of the phase diagram upon varying $\beta$ without nonreciprocity between the species ($\beta_a = 0$).
  {\bf d.} Schematic representation of the complete nonreciprocal phase diagram in the $\beta-\beta_a$ plane.
  }
  \label{fig:schematic_moves}
\end{figure*}

In this work, we introduce a minimal microscopic model of non-reciprocal flocking with two populations with competing goals: species 2 aligns with species 1 while species 1 antialigns with species 2 (Fig.~\ref{fig:schematic_moves}). 
Crucially, this microscopic model can be exactly coarse-grained into fluctuating hydrodynamic equations (the limit of vanishing lattice spacing)
which takes the form
\begin{equation}
\label{eq:summary_eq}
    \partial_t \boldsymbol{\psi} = \bF(\boldsymbol{\psi},\nabla\boldsymbol{\psi},\dots) + \sqrt{a}\:\boldsymbol{\mathcal{M}}(\boldsymbol{\psi},\nabla)\cdot\boldsymbol{\xi} + \mathcal{O}(a)\;,
\end{equation}
where the components of the vector $\boldsymbol{\psi}$ are the densities and polarizations of the species, $a$ is the lattice size and $\bF$ gives the deterministic evolution in the thermodynamic limit $a\to 0$. 
The fluctuations in finite systems are described, at first order in $a$, by the second term on the right hand side of \eref{eq:summary_eq} where $\boldsymbol{\mathcal{M}}$ is a matrix and $\boldsymbol{\xi}$ is a vector of Gaussian white noises such that $\langle\xi_i(\br,t)\xi_j(\br^{\prime},t^{\prime})\rangle=\delta_{i,j}\delta(\br-\br^{\prime})\delta(t-t^{\prime})$. Studying the effect of this noise, while technically challenging, is particularly important in
experimental contexts: animal herds and colloidal swarms are rarely in the thermodynamic limit where particle number fluctuations can be completely ignored.
Here, we explicitly derive the expression of $\bF$ and $\boldsymbol{\mathcal{M}}$ directly from the microscopic dynamics of the agents.

Equation~\eref{eq:summary_eq} allows us to rigorously analyze non-reciprocal correlation functions as well as the effect of density fluctuations, which were neglected in previous analytical investigations of non-reciprocal flocking~\cite{fruchart2021non}, despite their crucial role at the onset of flocking.
Even in reciprocal flocking with a single population, the onset of collective motion is not homogeneous (i.e. the density is not uniform). 
Instead, the emergence of flocking is accompanied by the formation of dense polar bands, which are non-linear wave solutions (akin to solitons) moving in a dilute disordered gas~\cite{solon_flocking_2015,toner1995long,toner2005hydrodynamics,bertin2006boltzmann,mishra2010fluctuations,solon2013revisiting,bialek2012statistical,vicsek1995novel,gregoire2004onset,VicsekFirstOrder,mahault2019quantitative,narayan2007long,deseigne2010collective,schaller2010polar,bricard2013emergence,iwasawa2021algebraic,ballerini2008interaction,cavagna2010scale}, as illustrated in Fig.~\ref{fig:schematic_moves}c. 
In a nutshell, the homogeneous disordered gas is subject to an instability leading to the formation of a pattern comprising many proto-bands, that eventually coarsen into a single travelling band. 
Based on our exact hydrodynamic theory \eref{eq:summary_eq}, we analyze how the phase diagram is modified by the presence of non-reciprocity, with a focus on the onset of flocking (Fig.~\ref{fig:schematic_moves}d).

In the absence of self-propulsion, a uniform oscillating phase emerges between the homogeneous static phases. 
It is heralded by two different transitions: a Hopf bifurcation from the disordered phase and a Saddle-Node on Invariant Circle (SNIC) bifurcation from the ordered phase. 
In the presence of self-propulsion, a weak non-reciprocity increases the range of parameters where bands are formed.
The bands in the two populations interact in a non-trivial way, leading to complex spatiotemporal dynamics.
When non-reciprocity is strong, flocking is superseded by a Chase \& Rest dynamical phase where each species alternates between a chasing state, when they propagate, and a resting state, when they stand still.
Even away from the onset of flocking, we observe that the phases with uniform oscillations present in the absence of self-propulsion are destabilized and replaced by more complex spatiotemporal phases when self-propulsion is turned on.

\begin{figure*}
\centering
\includegraphics{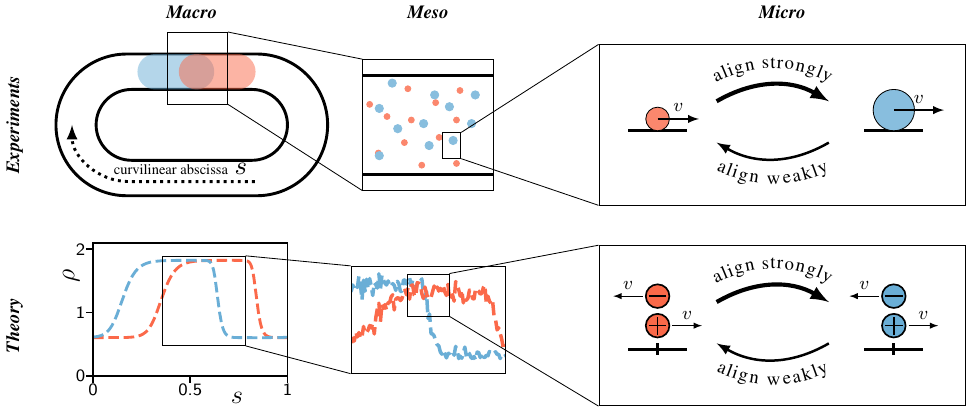}
\caption{{\bf Non-reciprocal flocking with two populations.}
{\bf Top:} Schematic details of an experimental realization where two populations of Quincke rollers with different sizes interact nonreciprocally in a quasi-1D gemoetry.
A smaller roller (red) strongly align its self-propulsion with the one of a bigger roller (blue) while the latter only align weakly with the former. {\bf Bottom:} Corresponding modelling by our equations \eref{eq:hydro_NR_and_R} describing the macroscopic hydrodynamics emerging from an assembly of self-propelled spins interacting nonreciprocally.}
\label{fig:schematic_experiment}
\end{figure*}

Our results are particularly relevant to highlight the collective phases emerging in mixtures of active particles, where we generically expect nonreciprocal interactions between two entities belonging to different populations.
For example, two phoretic Janus colloids with different surface coatings experience nonreciprocal interactions leading to the formation of dimers where one colloid is chasing the other \cite{Saha2019}. 
Likewise, two Quincke rollers with two different radii $a_2$ and $a_1$ such that $a_2>a_1$ will exhibit a nonreciprocal alignment between their polarizations: the magnetic torque exerted by $2$ on $1$ will be higher than the torque exerted by $1$ on $2$ \cite{Maity2023}. 
As schematically detailed in \Fref{fig:schematic_experiment}, one can then confine two such populations of active entities interacting nonreciprocally --Janus or Quincke colloids-- in order to study the collective properties of the mixture.
A first experiment of this nature has been realized very recently in a circular cell \cite{Maity2023}. 
Subjecting this system to quasi-1D confinement using a racetrack, ring, or annulus geometry would provide an ideal playground to test various predictions of our model and explore the virtually uncharted physics generated by the interplay of non-reciprocity, particle motility and number fluctuations.

From a methodological perspective, our work illustrates the usefulness of exact coarse-graining procedures~\cite{De1985,De1986,erignoux2016hydrodynamic,kourbane-houssene_exact_2018}. Continuum models of non-equilibrium systems written by hand have a tendency to display non-generic behavior if care is not given to include all possible terms (a daunting task when symmetry constraints are less stringent, as in non-reciprocal active matter) possibly leading to unphysical predictions~\cite{FrohoffHulsmann2023}. This is indeed the case here: simple models with linear non-reciprocal couplings fail to predict an entire phase (the Chase \& Rest phase) present in the exact hydrodynamic equations.

The paper is organized as follows.
In Sec.~\ref{sec:NRASM}, we introduce the Nonreciprocal Active Spin Model (NRASM), a minimal microscopic model amenable to exact coarse-graining describing two populations of self-propelled active spins with antagonistic alignment interactions. We then coarse-grain this model to obtain the fluctuating hydrodynamic equations that we use in the remainder of the paper.
In Sec.~\ref{sec:phase_diagram}, we sketch the phase diagram of this model using a combination of linear stability analysis and direct numerical simulations of the resulting partial differential equations.
We distinguish the passive case (no self-propulsion), for which a uniform oscillating phase exists, from the active case, where we instead observe a new ``Chase \& Rest'' phase.
In Sec.~\ref{sec:fluctuations}, we compare the phenomenology unveiled in Sec.~\ref{sec:phase_diagram} with microscopic agent-based simulations and show that our fluctuating hydrodynamics provides a quantitative description of the NRASM.
In addition, we detail how the fluctuations in finite systems can be harnessed as a probe to characterize microscopic non-reciprocity from macroscopic time-correlations. 
In Sec.~\ref{sec:bands}, we study the impact of non-reciprocity on flocking bands close to the emergence of collective motion.
In particular, we quantify how nonreciprocity induces spatial synchronization of flocks and affects the flocking speed.
Finally, in Sec.~\ref{sec:flying_XY}, we study the genericity of our results in a flying Kuramoto model where particles evolve off-lattice and are endowed with a continuous degree of freedom.
We conclude by suggesting how our results could provide control over flocks through nonlinear nonreciprocal interactions along with potential experimental platforms.

\section{The Nonreciprocal Active Spin Model (NRASM)}
\label{sec:NRASM}

\subsection{Microscopic model}
\label{sec:micro}

We consider a lattice gas model with two species of particles (labeled by $\alpha=1,2$) that can move on a discrete one-dimensional (1D) lattice of $N$ sites with lattice spacing $a$, see \Fref{fig:schematic_moves}. There is no exclusion or other spatial interaction between the particles. Each particle carries a classical spin taking values $\pm 1$, and in the remainder of this work we will often refer to it simply as \enquote{spin} instead of \enquote{particle}.

Particles of both species diffuse isotropically to their neighbouring site. In addition, they actively jump to the site on their right if their spin is $+$. 
Finally, the spin carried by a particle can flip depending on the values of the spins of the other particles on the same site. The two species have a propensity to align with their own kin. In addition, a non-reciprocal coupling is present: species $1$ anti-aligns with species $2$ while species $2$ aligns with species $1$.
The densities $\rho^{\alpha}_i$ and magnetizations $m^\alpha_i$ of species $\alpha$ at site $i$ are given by
\begin{align}
  \nonumber
  \rho^{\alpha}_i &= \eta^{+,\alpha}_i + \eta^{-,\alpha}_i & m^{\alpha}_i &= \eta^{+,\alpha}_i - \eta^{-,\alpha}_i \end{align}
where $\eta^{\pm,\alpha}_i$ are the number of $\pm$ spins belonging respectively to species $\alpha$ at site $i$.
The microscopic stochastic dynamics is ruled by the following four moves:
\begin{itemize}
  \item[\hypertarget{rule:diffusion}{{\bf \bluelink{RI}}}] Spins jump to a neighboring site at rate $D/a^2$.
  \item[\hypertarget{rule:hopping}{{\bf \bluelink{RII}}}] $+$ spins of both species actively hop to the site on their right at rate $v/a$.
  \item[\hypertarget{rule:flip1}{{\bf \bluelink{RIII}}}] A spin $s^1_i$ of species $1$ and located at site $i$ flips at rate $W_{i,1}^{s^1_i}=\gamma\exp[-\beta m^1_i s^1_i + (\beta_a-\beta_0) m^2_i s^1_i]$.
  \item[\hypertarget{rule:flip2}{{\bf \bluelink{RIV}}}] A spin $s^2_i$ of species $2$ and located at site $i$ flips at the rate $W_{i,2}^{s^2_i}=\gamma\exp[-\beta m^2_i s^2_i - (\beta_a+\beta_0) m^1_i s^2_i]$.
\end{itemize}
In \Fref{fig:schematic_moves}a, we summarize schematically these microscopic moves.
Note that in \hyperlink{rule:flip1}{RIII} and \hyperlink{rule:flip2}{RIV}, $\beta$ ($\beta_0$) controls the intra(inter)-species aligning strength while $\beta_a$ sets the inter-species nonreciprocal strength.
According to \hyperlink{rule:hopping}{RII}, we note that only $+$ spins are active while $-$ spins are merely subject to diffusion. This is not a restrictive feature of the NRASM, i.e. we could have chosen a different convention for the speeds, as discussed in detail at the end of Sec.~\ref{eq:cg_hydro_fluct}.

\subsection{Coarse-grained fluctuating hydrodynamics}
\label{eq:cg_hydro_fluct}

We now derive the coarse-grained hydrodynamics corresponding to the
microscopic evolution \hyperlink{rule:diffusion}{RI}, \hyperlink{rule:hopping}{RII}, \hyperlink{rule:flip1}{RIII} and  \hyperlink{rule:flip2}{RIV} by using lattice gas methods \cite{kourbane-houssene_exact_2018,erignoux2016hydrodynamic}.
In a nutshell, this method relies on the scaling we adopted for the transition rates and in particular on the fact that in the limit $a\to 0$ the most probable move is a diffusive one.
When diffusion dominates, the probability density becomes a product of independent Poisson distributions at each lattice site and the dynamics of the system boils down to the time-evolution of the corresponding Poisson parameters \cite{erignoux2016hydrodynamic}.
Using path integral techniques, we derive in Appendix \ref{app:solidification_coarse_graining} the probability $P(\{\eta\})$ to observe a given trajectory $\{\eta\}$ of the spins.
We then change variables from the discrete $\eta_i^{\pm,\alpha}$ to the averaged densities $\rho_{\alpha}(x=ia)=\langle \rho^{\alpha}_{i}\rangle$ and magnetizations $m_{\alpha}(x=ia)=\langle m^{\alpha}_{i}\rangle$. Finally, as shown in Appendix \ref{app:solidification_coarse_graining}, we use a saddle-point of the action in the limit of small lattice size $a\to 0$ to obtain the fluctuating hydrodynamics
\begin{subequations}
\label{eq:hydro_NR_and_R}
\begin{align}
\partial_t \rho_1 =& D\partial_{xx}\rho_1- \cA(\rho_1,m_1) + \sqrt{a}\;\partial_x \eta^1_1\\
  \partial_t m_1 =& D\partial_{xx}m_1 -\cA(m_1,\rho_1) - l_{\beta_0}(\rho_1,\rho_2,m_1,m_2) \notag\\
  &+ \sqrt{a}\;\partial_x \eta^1_2 + \sqrt{a\:\bar{l}_{\beta_0}(\rho_1,\rho_2,m_2)}\; \eta^1_3 \\
\partial_t \rho_2 =& D\partial_{xx}\rho_2 - \cA(\rho_2,m_2) + \sqrt{a}\;\partial_x \eta^2_1 \\
  \partial_t m_2 =& D\partial_{xx}m_2 - \cA(m_2,\rho_2) - l_{\shortminus\beta_0}(\rho_2,\rho_1,m_2,\shortminus m_1)\notag\\ 
  &+ \sqrt{a}\;\partial_x \eta^2_2 
  + \sqrt{a\:\bar{l}_{\shortminus\beta_0}(\rho_2,\rho_1,\shortminus m_1)}\; \eta^2_3\;,
\end{align}
\end{subequations}
where $\cA(\rho_{\alpha},m_{\alpha})=\frac{v}{2}\partial_x(\rho_{\alpha}+m_{\alpha})$ while the functions $l_{\beta_0}$ and $\bar{l}_{\beta_0}$ stem from the microscopic flipping dynamics of the spins and are given by
\begin{align}
  \label{eq:landau_NR_and_R}
  l_{\beta_0}(\rho_1,\rho_2,m_1,m_2)=&\tilde{\gamma}(\rho_1,\rho_2) (\cosh(Z) m_1 - \sinh(Z)\rho_1) \\
  \label{eq:landau_NR_and_R_bar_l}
  \bar{l}_{\beta_0}(\rho_1,\rho_2,m_2)=&\tilde{\gamma}(\rho_1,\rho_2)e^{-m_2\sinh(\beta_0-\beta_a)}\;,
\end{align}
with $\tilde{\gamma}$ and $Z$ being field-dependents and read
\begin{align}
\label{eq:gamma_eff_reciprocal_plus_nonreciprocal}
    \tilde{\gamma}(\rho_1,\rho_2) =& 2\gamma e^{-\beta + (\cosh(\beta)-1)\rho_1 + (\cosh(\beta_0-\beta_a)-1)\rho_2} \\
    Z=&\sinh(\beta)m_1+\sinh(\beta_0-\beta_a)m_2\; .
\end{align}
Finally, note that the $\eta^{\alpha}_{n}$ ($\alpha\in\{1,2\}$ and $n\in\{1,2,3 \}$) in \eqref{eq:hydro_NR_and_R} are Gaussian white noises such that $\langle\eta^{\alpha}_{n}\eta^{\alpha^{\prime}}_{n'}\rangle=\delta_{\alpha,\alpha^{\prime}}M^{\alpha}_{n,n^{\prime}}$, with the species-dependent matrix $M^{\alpha}$ being given by
\begin{align}
\label{eq:matrix_noise_beta0}
M^{\alpha}=\begin{pmatrix}
2D\rho_{\alpha} & 2Dm_{\alpha} & 0 \\
2Dm_{\alpha} & 2D\rho_{\alpha} & 0 \\
0 & 0 & S^{m,\alpha}_{f}
\end{pmatrix}\;,
\end{align}
where the species-dependent amplitude of the non-conserved noise $S^{m,\alpha}_{f}$ reads
\begin{align}
S^{m,\alpha}_{f}=\frac{\rho_{\alpha}\cosh(m_{\alpha}\sinh(\beta))-m_{\alpha}\sinh(m_{\alpha}\sinh(\beta))}{2}\;.
\end{align}
In the limit $a\to 0$, the fluctuating hydrodynamics \eref{eq:hydro_NR_and_R} becomes deterministic as the strength of the noisy terms vanish. 
In this regime, we can thus neglect the noises $\eta^{\alpha}_{n}$ in \eref{eq:hydro_NR_and_R} to obtain the following thermodynamic coarse-grained evolution
\begin{subequations}
\label{eq:hydro_NR_and_R_non_fluct}
\begin{align}
	\partial_t \rho_1 =& D \partial_{xx}\rho_1 -\cA(\rho_1,m_1)\\
  \partial_t m_1 =& D\partial_{xx}m_1 -\cA(m_1,\rho_1) - l_{\beta_0}(\rho_1,\rho_2,m_1,m_2) \\
\partial_t \rho_2 =& D\partial_{xx}\rho_2-\cA(\rho_2,m_2) \\
  \partial_t m_2 =& D\partial_{xx}m_2 -\cA(m_2,\rho_2) - l_{\shortminus\beta_0}(\rho_2,\rho_1,m_2,\shortminus m_1) \;.
\end{align}
\end{subequations}
For the sake of simplicity, we will often consider the $\beta_0=0$ case for which the functions $l_{\beta_0}$ and $\bar{l}_{\beta_0}$ respectively become $l_0$ and $\bar{l}_0$.
Note that Eqs.~\eref{eq:hydro_NR_and_R_non_fluct} are deterministic: the stochastic nature of the microscopic moves of the spins seems to have vanished.
This is because the number of sites in a mesoscopic volume becomes very large in the limit $a\to 0$: the average mesoscopic occupancy becomes a well-defined non-fluctuating quantity through the central limit theorem.
However, for finite system sizes we have to take into account the stochastic deviations of the occupancies from their thermodynamic limits through \eref{eq:hydro_NR_and_R}.
These fluctuations might ultimately impact the phenomenology observed in \eref{eq:hydro_NR_and_R_non_fluct}, however small they are \cite{martin_fluctuation-induced_2021}.
Therefore, we detail the analysis of the fluctuating hydrodynamics \eref{eq:hydro_NR_and_R} along with the results of finite size microscopic simulations of \hyperlink{rule:diffusion}{RI}-\hyperlink{rule:flip2}{RIV} in section~\ref{sec:fluctuations}.
While the averaged evolution of the agent-based simulations exhibits a phenomenology quantitatively similar to the one observed in \eref{eq:hydro_NR_and_R_non_fluct}, the fluctuating theory \eref{eq:hydro_NR_and_R} is needed to quantitatively capture the correlation functions.

We now pause to discuss in detail the dynamical rule \href{rule:hopping}{{\bf \bluelink{RII}}} along with its generality.
At first sight, choosing that only $+$ spins self-propel on the lattice might appear restrictive and arbitrary. 
However, it turns out that choosing \href{rule:hopping}{{\bf \bluelink{RII}}} is equivalent to choosing a specific Galilean reference frame.
Therefore, all our results obtained with \href{rule:hopping}{{\bf \bluelink{RII}}} remain valid for other microscopic self-propulsion rules up to a change of reference frame.
Let us illustrate this point by considering a different hopping rule where $+$ spins self-propel in the right direction with speed $v_{+}$ while $-$ spins self-propel in the left direction with speed $v_{-}$.
In this case, the hydrodynamics \eref{eq:hydro_NR_and_R_non_fluct} remains valid up to the change $\cA \to \tilde{\cA}$ where $\tilde{\cA}$ is given by 
\begin{align}
    \tilde{\cA}(\rho,m)=\tfrac{v_{-}-v_{+}}{2}\partial_x\rho -\tfrac{v_{+}+v_{-}}{2}\partial_x m
\end{align}
We now apply the Galilean change of reference frame $t^{\prime}=t$ and $x^{\prime}=x-v_g t$.
This amounts to performing the replacement $\partial_t \to \partial_{t^{\prime}} -v_g\partial_{x^{\prime}}$ and $\partial_x\to \partial_{x^{\prime}}$, thereby only changing advective terms in \eref{eq:hydro_NR_and_R_non_fluct}.
Performing this Galilean boost thus transforms $\tilde{\cA}$ into $\tilde{\cA}_g$ which is given by
\begin{align}
    \tilde{\cA}_g(\rho,m)=\tfrac{v_{-}-v_{+}+2v_g}{2}\partial_x\rho -\tfrac{v_{+}+v_{-}}{2}\partial_x m
\end{align}
We can then choose $v_g=-v_{-}$ to obtain  
\begin{align}
    \tilde{\cA}_g(\rho,m)=\tfrac{v}{2}\partial_x(\rho+m)=\cA(\rho,m)\;,
\end{align}
where $v=v_{+}+v_{-}$. 
Starting from a generic hopping rule where $+$ and $-$ spins hop at different speeds $v_{+}$ and $v_{-}$, we thus showed that a Galilean change of reference brings us back to hydrodynamics \eref{eq:hydro_NR_and_R_non_fluct} obtained for the  hopping rule \href{rule:hopping}{{\bf \bluelink{RII}}}.
Therefore all the properties and phenomenology of \eref{eq:hydro_NR_and_R_non_fluct} is insensitive to the specific choice of the underlying hopping rule.

\subsection{Relations with reciprocal active spin models}

When $\beta_a = 0$, the two species are non-interacting and their coarse-grained evolution is given by the same hydrodynamics which reads
\begin{subequations}
\label{eq:hydro_non_interacting}
\begin{align}
\partial_t \rho &= D\partial_{xx}\rho -\cA(\rho,m)\;,  \\ 
    \partial_t m &= D\partial_{xx}m -\cA(m,\rho) -g_r(\rho,m)
\end{align}
\end{subequations}
where the function $g_r$ is given by
\begin{align}
\nonumber
  g_r(\rho,m)=&\gamma_{r}( \cosh\left[\sinh(\beta)m\right]m -\sinh\left[\sinh(\beta)m\right]\rho)\;,
\end{align}
in which $\gamma_{r} = 2\gamma e^{-\beta + (\cosh(\beta)-1)\rho}$.
Equation \eqref{eq:hydro_non_interacting} corresponds to the hydrodynamics of the Active Spin Model (ASM) derived in \cite{kourbane-houssene_exact_2018} with the exception of the advective terms.
In \cite{kourbane-houssene_exact_2018}, the advective terms in the evolution of $\rho$ and $m$ are respectively $v\partial_x m$ and $v\partial_x \rho$.
This minor discrepancy stems from the microscopic rule \hyperlink{rule:hopping}{RII} which states that only $+$ spins are self-propelled and able to hop to the site on their right.
By contrast, in the active spin model of \cite{kourbane-houssene_exact_2018}, $-$ spins are also self-propelled and able to hop on the site on their left.
This discrepancy leads to different advective terms in the hydrodynamics but leaves the phenomenology as well as the phase diagram of the model unchanged.
In particular, as explained in the previous section, the advective terms of \eref{eq:hydro_non_interacting} can be mapped to the ones of \cite{kourbane-houssene_exact_2018} by performing the change of reference frame $t'=t$ and $x'=x-vt/2$.
Thus, when $\beta_a=0$, the NRASM is phenomelogically equivalent to the ASM.
In this case, upon varying $\beta$, the hydrodynamics \eref{eq:hydro_non_interacting} displays the classical flocking phase diagram: the homogeneous disordered phase $\rho(x)=\rho_0$, $m(x)=m_0=0$ loses stability when $\beta>\beta_{g}\equiv\argsh(\rho_0^{-1})$.
A fully ordered solution then appears: it is given by $\rho_0$ and $m_0$ being the solutions of
\begin{align}
  \label{eq:uniform_reciprocal}
  m_0=\rho_0\tanh(m_0\sinh(\beta))\; .
\end{align}
This ordered solution is linearly unstable for $\beta_g\leq\beta\leq\beta_l$ and the dynamics instead leads to travelling bands \cite{solon_flocking_2015,kourbane-houssene_exact_2018,scandolo2023active}.
At higher aligning strength, for $\beta\geq\beta_l$, the uniform ordered profile \eref{eq:uniform_reciprocal} finally becomes linearly stable.
The details of these results as well as the quantitative phase diagram with its binodals and spinodals in the $\beta^{-1}-\rho$ plane were worked out in \cite{kourbane-houssene_exact_2018} for the ASM.
In this paper, we want to assess the impact of the inter-species nonreciprocity $\beta_a$ on this flocking phenomenology.

\section{Phase diagram of the dynamical systems}
\label{sec:phase_diagram}
In this part, we set $\gamma=D=1$ and we focus our analysis on the case where both species have the same averaged density $\rho_0^1=\rho_0^2$.
We then perform an exploration of the phase diagram of \eref{eq:hydro_NR_and_R_non_fluct} in the $\beta$-$\beta_a$ plane and distinguish two different cases: the non-motile one where $v=0$ and the self-propelled one where $v\neq 0$, whose analysis is reported below.

\subsection{Non-reciprocity but no self-propulsion (\texorpdfstring{$v=0$}{v=0})}
\label{subsec:nr_passive_analysis}
In this section, we analyze the hydrodynamic equations \eref{eq:hydro_NR_and_R_non_fluct} without self-propulsion ($v=0$). 
In the absence of non-reciprocity ($\beta_a = 0$), the NRASM exhibits both a disordered phase and a long-ranged ordered phase~\cite{De1985,De1986}. The latter is present even in 1D, under the lower critical dimension of Ising. This is because the diffusive dynamics, which is absent in the Ising model, drives the system out of equilibrium (the system is effectively connected to two baths at different temperatures~\cite{De1985,De1986}). Intuitively, long-range order can survive since the diffusive moves~\hyperlink{rule:diffusion}{RI} are the most probable (i.e., fastest) in the hydrodynamic limit $a \to 0$: they tend to homogenize the system. As we shall see, this feature remains present when non-reciprocity is turned on, leading to an homogeneous oscillating phase.

To see that, we first look for homogeneous solutions of \eref{eq:hydro_NR_and_R_non_fluct} by removing all spatial derivatives from the equation.
As a consequence, the densities are constant, and we write $\rho_0^\alpha$ their values.
In addition, the uniform magnetizations must satisfy 
\begin{subequations}
\label{eq:system_zero_d}
\begin{align}  
  \dot{m}_1=&-l_{\beta_0}(\rho_0^1,\rho_0^2,m_1,m_2) \\ \dot{m}_2=&-l_{\shortminus\beta_0}(\rho_0^2,\rho_0^1,m_2,-m_1)
\end{align}
\end{subequations}
in which $l_{\beta}$ corresponds to Eq.~\eqref{eq:landau_NR_and_R}. 
We now assess the stability of homogeneous but possibly time-dependent solutions of \eref{eq:hydro_NR_and_R_non_fluct} when $v=0$.
To this aim, we perform a linear stability analysis, as detailed in Appendix \ref{app:floquet}.
The result of this analysis is that the growth rates of perturbations in the spatially extended system are $\sigma_1(q) = \sigma_1 - D q^2$, $\sigma_2(q) = \sigma_2 - D q^2$, $\sigma_3(q) = \sigma_4(q) = - D q^2$ in which $\sigma_1$ and $\sigma_2$ are the growth rates of perturbations in the zero-dimensional dynamical system Eq.~\eqref{eq:system_zero_d}, and $q$ is the wavevector.
Therefore, when $v=0$, any (possibly time-dependent) stable solution observed in \eref{eq:system_zero_d} is also a stable solution in \eref{eq:hydro_NR_and_R_non_fluct}.
We can therefore study the dynamical system \eref{eq:system_zero_d} to characterize the phase diagram in the absence of self-propulsion.

\begin{figure*}
\includegraphics{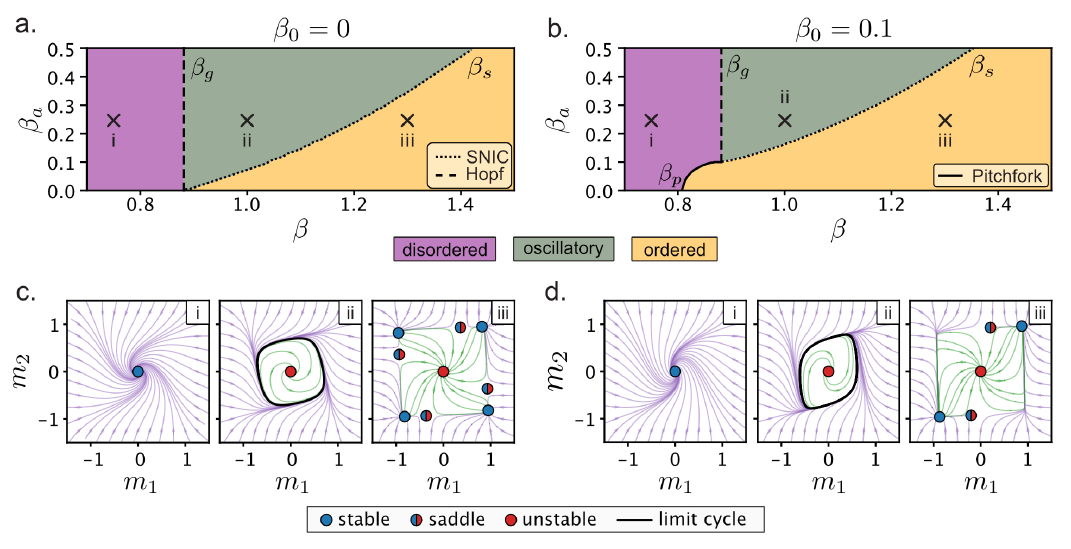}
\caption{\textbf{Phase diagram of the non-motile homogeneous system \eref{eq:system_zero_d}: a dynamical systems perspective.} When nonreciprocity is nonzero, a new dynamical phase (green) corresponding to a stable limit cycle emerges.  
  {\bf a.-b.} Phase diagram of dynamical system \eref{eq:system_zero_d} with either purely non-reciprocal couplings ($\beta_0 = 0$) or with non-zero reciprocal couplings ($\beta_0=0.1$). Each cross refers to a specific coordinate $(\beta,\beta_a)$ within the three phases for which the dynamical flow is plotted below.
  The $\beta_g$ and $\beta_s$ lines respectively corresponds to a Hopf (dashed line) and a SNIC (dotted line) bifurcation. Finally, a Pitchfork bifurcation (solid line) also exists when $\beta_0 = 0.1$ along the line labeled $\beta_p$: it corresponds to the direct transition from the disordered state to the ordered state.
  As discussed in the main text, when $\beta_0=0.1$, a secondary bifurcation also occurs in the ordered domain when $\beta > \beta_g$ and $\beta_a < \beta_s$. Since it only leads to the appearance of the two saddle solutions with magnetizations of alternating signs, it does not affect the stability of the system and we choose not to represent it here.
  {\bf c.-d.} When $\beta<\beta_d$, the system flows toward the stable disordered solution ({\bf i}).
  Whenever $\beta_g\leq\beta\leq\beta_s$, the dynamical flow instead exhibits a stable limit cycle ({\bf ii}).
  Finally, when $\beta>\beta_s$ the system flows toward one of the four fixed points ({\bf iii}).
  Note that the dynamics breaks the 4-fold symmetry when $\beta_0 = 0.1$, most strikingly by the annihilation of two pairs of stable and saddle nodes.
  }
  \label{fig:flow_passive}
\end{figure*}

We analyze the vector field defined by the right-hand side of Eq.~\eqref{eq:system_zero_d} and represented it in \Fref{fig:flow_passive}b for different values of the parameters.
It shows the existence of three different phases upon varying $\beta$ and $\beta_a$, as shown in \Fref{fig:flow_passive}a: a disordered phase, in which the system has no magnetization ($(m_1,m_2)=(0,0)$; purple region), an ordered phase in which non-trivial fixed points $(m_1,m_2) \neq (0,0)$ are stable (orange region), and an oscillatory phase in which the only stable solution is a limit cycle (green region).

The fixed points are solutions of the equations
\begin{subequations}
\label{eq:homogeneous_1}
\begin{align}  
	m_1 &= \rho_0^1\tanh\left[\sinh(\beta)m_1+\sinh(\beta_0-\beta_a)m_2\right] \\
	m_2 &= \rho_0^2\tanh\left[\sinh(\beta)m_2+\sinh(\beta_0+\beta_a)m_1\right]\;.
\end{align}
\end{subequations}
Nonzero solutions of the coupled equations \eref{eq:homogeneous_1} corresponding to non-trivial fixed points exist only in the domain $\beta_{s}(\beta_a) \leq \beta $, where the function $\beta_{s}(\beta_a)$ (continuous line in \Fref{fig:flow_passive}a-b) is determined implicitly. 
Note that the trivial solution $(m_1,m_2)=(0,0)$ corresponding to the disordered phase, always exists. It is stable if
\begin{equation}
\label{eq:disord_stab}
    \beta \leq \min(\beta_g,\beta_s)\equiv \beta_d\;,
\end{equation}
where $\beta_g = \argsh(1/\rho^1_0)$ ( represented by dashed lines in \Fref{fig:flow_passive}a-b).
Furthermore, in this range it is the only stable solution. 

Let us now specify our analysis by differentiating the cases $\beta_0=0$ and $\beta \neq 0$.
When $\beta_0=0$, \eref{eq:homogeneous_1} is invariant under a rotation of the magnetizations $(m_1,m_2)\rightarrow(-m_2,m_1)$ and exchange of the densities.
This invariance allows us to focus on solutions of \eref{eq:homogeneous_1} that are located in the upper right quadrant of the $(m_1,m_2)$ plane.
Applying rotations of $\pi/2$ on this upper right quadrant will then yield the complete set of solutions in the full $(m_1,m_2)$ plane.
As shown in \Fref{fig:flow_passive}c, we find 8 non-trivial fixed points out of which 4 are always stable while the remaining ones are saddle points, with both a stable and an unstable direction. 
As discussed above, these solutions exist only in the domain $\beta_{s}(\beta_a) \leq \beta $.

When $\beta_0\neq 0$, the four-fold symmetry of \eref{eq:homogeneous_1} is broken into a two-fold symmetry $(m_1,m_2)\to-(m_1,m_2)$.
Therefore, as shown in \Fref{fig:flow_passive}d, we find $4$ fixed points out of which only $2$ are now stable.

Independently of $\beta_0$, the critical value $\beta_g$ (dashed line in \Fref{fig:flow_passive}a) marks the transition from the disordered phase to the oscillatory phase via a supercritical Hopf bifurcation, where a stable limit cycle grows continuously from the trivial fixed point (thick black line in \Fref{fig:flow_passive}b, middle). This is signaled by the complex eigenvalues of Eq.~\eref{eq:system_zero_d} gaining a positive real part with a non-zero imaginary part.

The crossing of the line $\beta = \beta_s(\beta_a)$ can correspond to two different transition depending on the value of $\beta$.
When $\beta \geq \beta_g$, it corresponds to the transition from the oscillatory phase to the ordered phase, where $(m_1, m_2) \neq (0, 0)$. 
The nature of this transition is distinct from the transition from the disordered phase to the oscillatory phase discussed above. 
While the latter occurs via a Hopf bifurcation, the former occurs via a so-called Saddle Node on Invariant Circle (SNIC) bifurcation~\cite{Strogatz2015, Izhikevich2007}. 
At the SNIC bifurcation, the limit cycle is broken by the emergence of four (two) pairs of fixed points when $\beta_0=0$ ($\beta_0 \neq 0$), as represented in \Fref{fig:flow_passive}. 
We expect such SNIC bifurcation to be generically present in non-reciprocal systems~\cite{fruchart2021non,Liu2023,Guislain2023,Marques2013} and give details about its features in Appendix \ref{subsec:SNIC_bifurcation}.
Finally, when $\beta < \beta_g$, the crossing of $\beta_s$ corresponds to a ferromagnetic transition from the disordered to the ordered solutions (see \Fref{fig:flow_passive}b).
In the ordered phase, when $\beta > \beta_g$ and $\beta_a < \beta_s$, we also found an additional secondary bifurcation when $\beta_0 > 0$. In this bifurcation, the disordered solution transitions from saddle to fully instable while two other pair of saddle solutions appear.
This secondary bifurcation originates from the crossing of a domain where the only two solutions have magnetizations of the same sign to a domain where the two additional solutions with magnetizations of alternating signs emerge. 
However, since it is not relevant for the stability of the system and as such for the phase diagram, we only detail this secondary bifurcation in Appendix \ref{subsec:simple_model_mf_nrasm} (see \Fref{fig:o2phaseDiagram}), where we study a minimal model exhibiting both a Hopf and SNIC bifurcation.

\begin{figure*}
\includegraphics{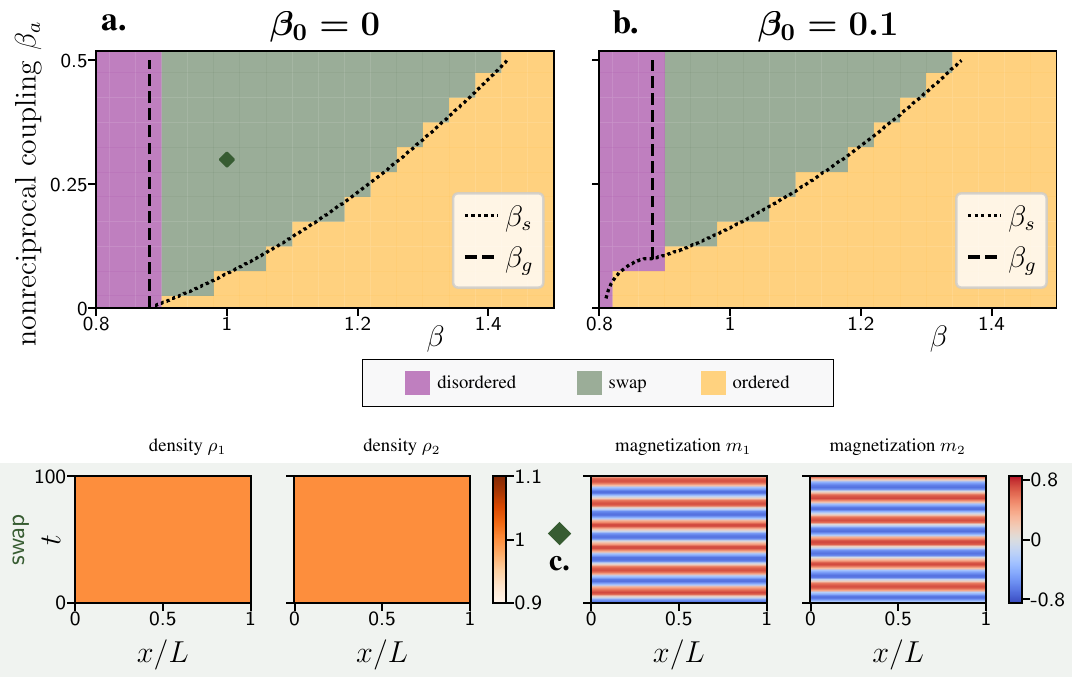}
  \caption{\textbf{Phase diagram of the non-reciprocal hydrodynamics \eref{eq:hydro_NR_and_R_non_fluct} without motility (v=0).} In this case, nonreciprocity induces the emergence of a new dynamical phase (swap) where the magnetizations of both species stay homogeneous while oscillating in time.
  {\bf a.} Numerical phase diagram of \eref{eq:hydro_NR_and_R_non_fluct} in the purely nonreciprocal case ($\beta_0=0$).
  {\bf b.} Numerical phase diagram of \eref{eq:hydro_NR_and_R_non_fluct} with reciprocal coupling ($\beta_0=0.1$).
  In both cases, the dashed line $\beta_g\equiv\argsh(\rho_0^{-1})$ and dotted line $\beta_s$ respectively indicates the stability of the disordered profile and the existence of a nonzero homogeneous solution.
  Parameters: $D=\gamma=1$, $v=0$, $\rho_0^1=\rho_0^2=1$, $dx=0.2$, $dt=0.002$.
  {\bf c.} Kymograph of the swap phase corresponding to the colored diamond in the upper phase diagram.
  Parameters: $D=\gamma=1$, $\rho_0^1=\rho_0^2=1$, $dx=0.05$, $dt=0.0002$, $L=100$.
  }
  \label{fig:phase_diagram_v_0}
\end{figure*}

We now verify our predictions from dynamical systems analysis by numerically integrating \eref{eq:hydro_NR_and_R_non_fluct} for different values in the $\beta$-$\beta_a$ plane (see Appendix \ref{app:numerics} for details about the numerical methods).
As shown in \Fref{fig:phase_diagram_v_0}, we indeed observe phase diagrams similar to the ones of \eref{eq:system_zero_d} which were represented on \Fref{fig:flow_passive}a-b.
When $\beta \leq \beta_g$ and $\beta \leq \beta_s$, the system remains homogeneously disordered (purple region) while when $\beta \geq \beta_s$ it remains in the ordered state (orange region).
Finally, when $\beta_g\leq\beta\leq\beta_s$, we observe a swap phase in which the magnetizations $m_1$ and $m_2$ stay homogeneous while following a periodic limit cycle where they alternate between two extreme values (see \Fref{fig:phase_diagram_v_0}c. and movie 1).

\subsection{Nonreciprocity {\it and} self-propulsion (\texorpdfstring{$v\neq 0$}{nonzero v})}
In this section, we analyze the hydrodynamic equations \eqref{eq:hydro_NR_and_R_non_fluct} in the motile case for both $\beta_0=0$ and $\beta_0\neq 0$.
When $v\neq0$, the linear stability of the hydrodynamics \eqref{eq:hydro_NR_and_R_non_fluct} is not directly related to the linear stability of the dynamical system \eref{eq:system_zero_d}.
Furthermore, the equations of motion are no longer invariant upon rotating the magnetizations and exchanging the densities.
We can, however, numerically assess the stability of the three phases present in the passive case: the disordered phase, the ordered phase and the homogeneously oscillating phase (swap phase).
\par
\begin{figure*}
\includegraphics{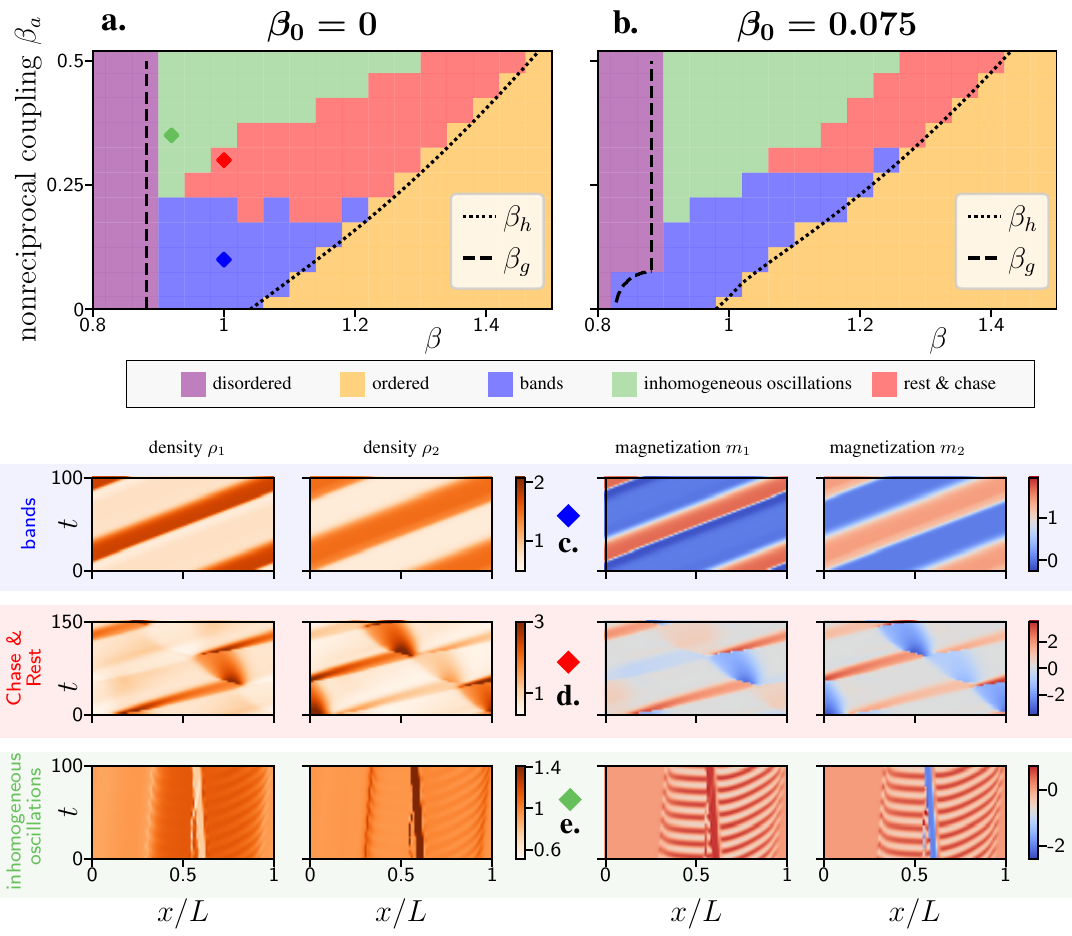}
  \caption{\textbf{Phase diagram of the non-reciprocal flocking hydrodynamics \eref{eq:hydro_NR_and_R_non_fluct} with motility (v=1).} 
  In the presence of motility, the swap phase of \Fref{fig:phase_diagram_v_0} becomes unstable: we instead observe a phase of flocking bands, a phase of inhomogeneous oscillations and a new Chase \& Rest phase where clusters made of spins from different species endlessly alternate between a rest state and a chasing state.
  {\bf a.} Numerical phase diagram of \eref{eq:hydro_NR_and_R_non_fluct} in the purely nonreciprocal case ($\beta_0=0$).
  {\bf b.} Numerical phase diagram of \eref{eq:hydro_NR_and_R_non_fluct} with reciprocal coupling ($\beta_0=0.075$).
  In {\bf a.} and {\bf b.}, the dashed line $\beta_g$ and dotted line $\beta_h$ respectively indicates the stability of the disordered and ordered profiles.
  Parameters: $D=\gamma=1$, $v=1$, $\rho_0^1=\rho_0^2=1$, $dx=0.05$, $dt=0.0002$.
  {\bf c-e.} Kymographs of the three phases corresponding to the colored diamonds in the upper phase diagrams. From top to bottom: band phase, Chase \& Rest phase and inhomogeneous oscillations.
  Parameters: $D=\gamma=1$, $\rho_0^1=\rho_0^2=1$. For inhomogeneous oscillations, $dx=0.2$, $dt=0.01$, $L=1200$. For all other phases $dx=0.05$, $dt=0.0002$, $L=100$.
  }
  \label{fig:phase_diagram}
\end{figure*}

 We start with the time-independent phases which correspond to the four (two) stable solutions of Eq.~\eref{eq:system_zero_d} when $\beta_0=0$ ($\beta_0\neq 0$) found in the Sec~\ref{subsec:nr_passive_analysis} and represented in \Fref{fig:flow_passive}iii.
In contrast to the non-motile case, the lack of rotational symmetry implies that these fixed points can have a priori different stabilities.
However, no significant differences emerge from our numerical analysis: the fixed points become stable at the same parameter values on the $\beta-\beta_a$ plane.
We therefore focus on the homogeneous stable fixed point in the upper right quadrant for which we both have $m_1>0$ and $m_2>0$.
Linearizing \eref{eq:hydro_NR_and_R_non_fluct} around this solution allows us to obtain numerically the growth rate of the Fourier modes $q$ in a range $[0,q_m]$ (in practice, we took $q_m=2$).
Whenever one of these modes shifts from stable to unstable, we report the corresponding set of parameters $(\beta,\beta_a)$ as a stability boundary.
Applying this method, we find that the ordered profile is always unstable for $\beta \leq\beta_h$ where the critical value $\beta_h$ depends on $\beta_a$ (see dotted line in \Fref{fig:phase_diagram}).
Extending our analysis to the stability of the disordered solution, we report it to be unchanged with respect to the passive case: $(m_1,m_2)=(0,0)$ remains stable when $\beta\leq \beta_d$, where $\beta_d$ is given by Eq~\eref{eq:disord_stab}.
Our stability analysis thus divides the $\beta$-$\beta_a$ plane into three domain: for $\beta\leq\beta_d$ and $\beta\geq\beta_h$ the disordered and ordered solution are respectively stable while for $\beta_d\leq\beta\leq\beta_h$ none of the uniform constant profiles are stable.
The region of most interest to us is the latter.
Thanks to our analysis of the non-motile case, we know that in this region there exists an homogeneously oscillating solution where $m_1$ and $m_2$ follow a periodic limit cycle described by the dynamical system in Eq. \eref{eq:system_zero_d}.
However, when $v\neq 0$, we show in Appendix \ref{app:floquet} that this oscillating solution is unstable to perturbations at small wavelength $q$ using a numerical Floquet analysis.
In addition, numerical simulations show that self-propulsion destroys the swap phase in the whole region $\beta_d\leq\beta\leq\beta_h$.

\Fref{fig:phase_diagram}a-b show phase diagrams obtained by direct numerical simulations of Eqs.~\eqref{eq:hydro_NR_and_R_non_fluct} for $\beta_0=0$ and $\beta_0=0.075$. 
We first discuss the $\beta_0=0$ case corresponding to Fig~\ref{fig:phase_diagram}a. 
We observe that on the reciprocal line ($\beta_a = 0$) there is no direct transition between the disordered and ordered phases. 
Instead, a phase with polar bands is present (blue points in \Fref{fig:phase_diagram}a-b), as in similar models of flocking~\cite{solon_pattern_2015,caussin_emergent_2014,solon_flocking_2015,kourbane-houssene_exact_2018,erignoux2016hydrodynamic}.
This phase survives a small amount of non-reciprocity $(\beta_a \neq 0)$; in this case the bands of each species are coupled and, as detailed later in Sec~\ref{sec:bands}, have a tendency to keep a fixed distance from each other. 
In addition, we observe that for the values of parameters accessible in our simulations, the swap phase with uniform oscillations observed when $v=0$ is replaced by non-uniform time-dependent phases: inhomogeneous oscillations, polar bands, or a phase  having a complex temporal structure that we dub the \enquote{Chase \& Rest} phase.

\par
As shown in \Fref{fig:phase_diagram}a where it is indicated by a blue region, we numerically find that the band phase extends, in the $\beta-\beta_a$ plane, to a finite domain close to the $\beta$ axis.
At higher values of $\beta_a$, above the flocking bands, we find two new additional phases that we dubbed inhomogeneous oscillations and Chase \& Rest which respectively correspond to the green domain on the right hand side of $\beta_d$ and to the red domain on the left hand side of $\beta_h$.
In the inhomogeneously oscillating phase, we observe a rich variety of behaviors all involving some spatiotemporal oscillations both in the densities and in the magnetizations.
For example, as highlighted in \Fref{fig:phase_diagram}e, we can observe the coexistence of flocking bands with oscillating regions (see movie 4).
But we were also able to observe a fully spatially-oscillating phase whose spectrum contains multiple wavelengths and whose amplitude varies with time (see movie 5).
\par
By contrast, in the Rest \& Chase phase (see \Fref{fig:phase_diagram}d), we distinguish a clear pattern repeating itself.
In this phase, we observe two different types of spins' clusters.
There are clusters of arrested spins where the magnetization is negative and clusters of moving spins where the magnetization is positive.
As highlighted in \Fref{fig:phase_diagram}d, whenever a cluster of moving spins hits an arrested cluster it becomes arrested itself while the previously arrested cluster starts to run away (see also movie 3).
We studied the coarsening of these spins' clusters for several values of $(\beta,\beta_a)$ and found that they ultimately coalesce to a single couple of two chasing clusters (see movie 3).
Therefore, in steady-state, there remains only one cluster of species 1 and one cluster of species 2 that are alternating between a moving state and a stand still state.
We further assessed the robustness of this coarsening upon increasing the system size: increasing $L$ did not alter the observed phenomenology (see movie 6).
Because the switch between a moving and a stand still state involves strong spatial derivatives, it is increasingly difficult to study numerically the Chase \& Rest phase at large system sizes.
Remarkably, the mechanism at the origin of the Chase \& Rest dynamics crucially relies on the nonreciprocal nonlinearities emerging from the exact coarse-graining as we did not observe it in the phenomenological hydrodynamics \eref{eq:toner_tu_coupled_1}-\eref{eq:toner_tu_coupled_1_NL} that we will study in the next parts.

Before concluding this section, we now briefly discuss the case $\beta_0 \neq 0$ with $\beta_0 < \beta_a$. 
The phase diagram corresponding to this situation is reported in \Fref{fig:phase_diagram}b. 
We observe that a nonzero $\beta_0$ does not change the qualitative features of the phenomenology, which is still constituted by the five different phases: the inhomogeneous oscillations, the Rest \& Chase, the bands, the ordered phase and the disordered phase.
However, as shown in \Fref{fig:phase_diagram}b, the location of these different phases in the $(\beta-\beta_a)$ plane changes quantitatively. 
The transition from bands to Rest \& Chase or to inhomogeneous oscillations now occurs at higher values of $\beta_a$.
We observe the same phenomenon for the transition from the ordered phase to the bands or to the Rest \& chase.
Finally, we also remark that the destabilization of the disordered phase occurs at a smaller value of $\beta$ compared with the $\beta_0=0$ case. 
\red{Note that when $\beta_0>\beta_a$, a case not reported in \Fref{fig:phase_diagram}, reciprocal interactions overcome their nonreciprocal counterparts: we lose both the inhomogeneously oscillating phase and the rest \& chase phase in favor of the band phase. }

\section{Testing fluctuating hydrodynamics using microscopic simulations}
\label{sec:fluctuations}
In this section, we perform microscopic simulations of \hyperlink{rule:diffusion}{RI}-\hyperlink{rule:flip2}{RIV} to verify our findings from section \ref{sec:phase_diagram} as the latter were obtained by direct analysis of the deterministic coarse-grained evolution \eref{eq:hydro_NR_and_R_non_fluct}, which neglects the fluctuations arising in finite systems.
The details of our numerical implementation are reported in App.~\ref{app:microscopic_simu}.
For conciseness, we focus on the reciprocal case when $\beta_0=0$.

\subsection{Phase diagram}

\begin{figure*}
\includegraphics{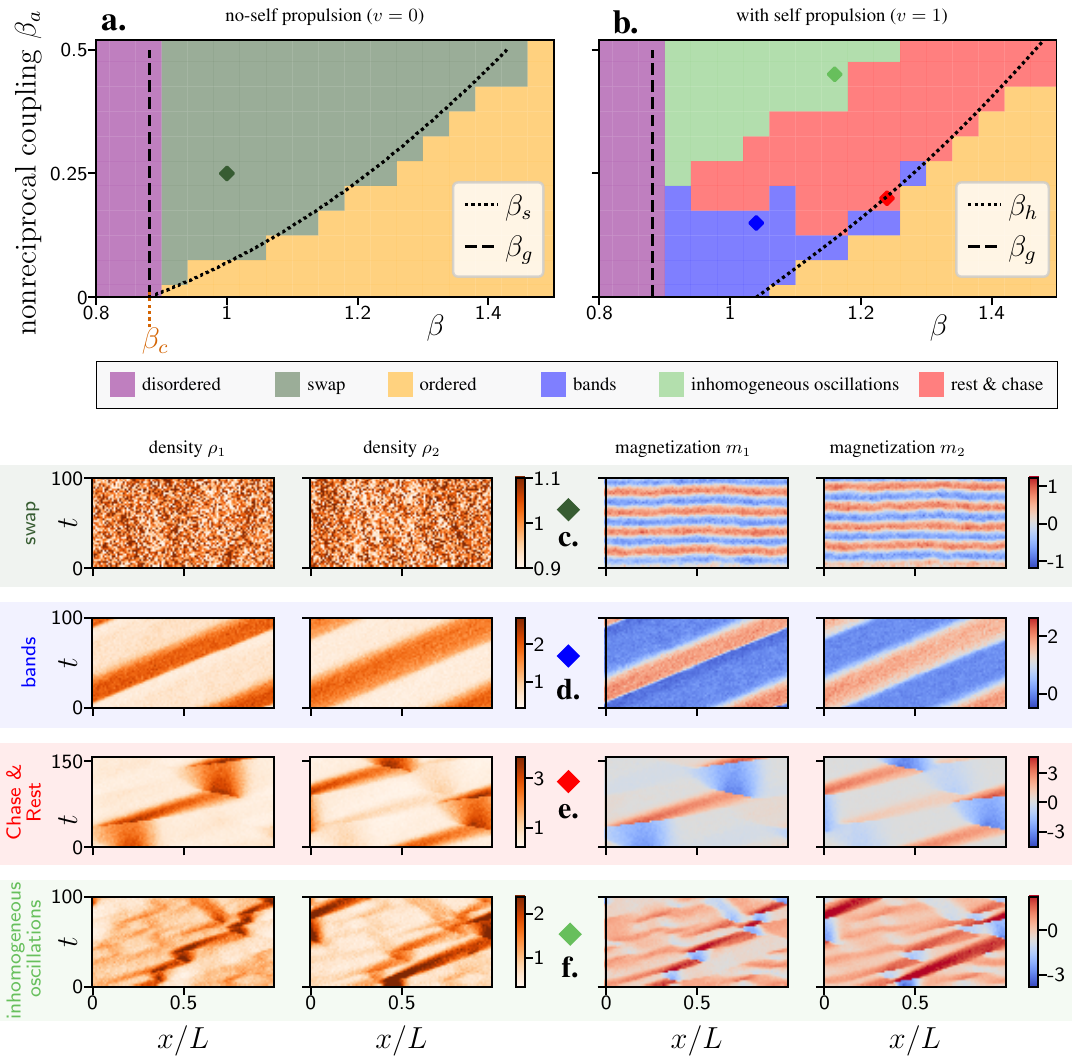}
\caption{\textbf{Phase diagrams observed in purely non-reciprocal ($\beta_0=0$) microscopic agent-based simulations.} 
They confirm the quantitative validity of the phase diagrams shown in \Fref{fig:phase_diagram_v_0} and \Fref{fig:phase_diagram} which were obtained by numerical simulations of the coarse-grained hydrodynamics \eref{eq:hydro_NR_and_R_non_fluct}.
\textbf{a.-b.} Phase diagrams obtained from microscopic simulations of the active lattice gas dynamics \protect\hyperlink{rule:diffusion}{RI} to \protect\hyperlink{rule:flip1}{RIV} with either $v=0$ or $v=1$. 
{\bf c.-f.} Kymographs of the four phases corresponding to the colored diamonds in the upper phase diagrams.
From top to bottom: swap phase, band phase, Chase \& Rest phase and inhomogeneous oscillations.
Parameters: $\beta_0=0$, $D=\gamma=1$, $\rho_0^1=\rho_0^2=1$, $L=100$, $a=0.003$, $dt=4\times 10^{-6}$, $T_f=400$ and $v=1$ except for the swap phase for which $v=0$. 
For the swap phase, $\beta=1$ and $\beta_a=0.25$.
For the bands $\beta=1$, $\beta_a=0.05$. 
For the Chase \& Rest phase $\beta=1.24$, $\beta_a=0.2$.
For inhomogeneous oscillations, $\beta=1.16$, $\beta_a=0.45$.
}
\label{fig:kymo_micro}
\end{figure*}

As shown in \Fref{fig:kymo_micro}, the agent-based simulations exhibit a phase diagram quantitatively similar to the one we reported on \Fref{fig:phase_diagram} for \eref{eq:hydro_NR_and_R_non_fluct}.
In particular, the microscopic spins indeed self-organize into the different macroscopic phases described in section \ref{sec:phase_diagram} from the swap and band phases to the Chase \& rest and inhomogeneously oscillating dynamics.

In addition, the microscopic simulations contain more information than the corresponding coarse-grained hydrodynamics \eref{eq:hydro_NR_and_R_non_fluct}.
Indeed, the latter only gives the deterministic evolution of the spins' occupancies in the thermodynamic limit while the former retain the stochastic fluctuations arising from finite system size, thereby allowing the determination of finer signatures such as correlation functions.

\subsection{Correlation functions}

\begin{figure*}
\begin{center}
\hspace*{-0.2cm}
\includegraphics{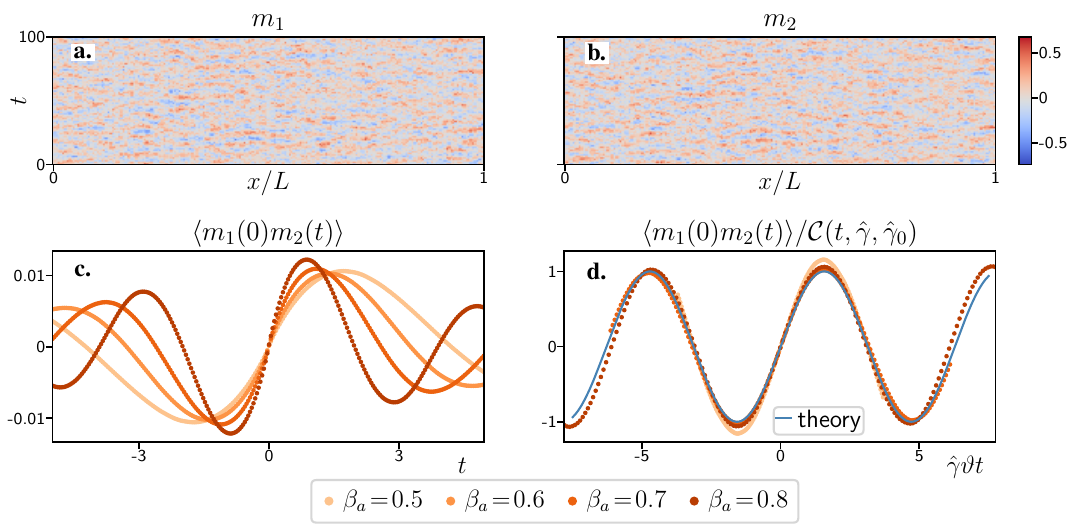}
\end{center}
\caption{{\bf Inter-species magnetization time-correlation in the disordered phase.} Interestingly, $\langle m_1(0)m_2(t)\rangle$ betrays the presence of microscopic nonreciprocal interactions even though the averaged density and magnetization fields are not affected by $\beta_a$. {\bf a.-b.} Kymographs of the magnetization fields $m_1$ and $m_2$ observed in microscopic agent-based simulations when $\beta_a=0.5$. {\bf c.} Raw time-correlation between $m_1$ and $m_2$.
{\bf d.} 
Time-correlation between $m_1$ and $m_2$ rescaled by $\mathcal{C}(t,\hat{\gamma},\hat{\gamma}_0)$ as a function of $\hat{\gamma}\vartheta t$ for different values of $\beta_a$.
All curves obtained from microscopic agent-based simulations (orange-shaded dots) universally collapse on our theoretical prediction \eref{eq:odd_time_correlation} (blue line) without any fitting parameters.
{\bf Parameters:} $\beta_0=0$, $\beta=0.84$, $\gamma=1$, $D=0.2$, $v=0$, $a=1/300$.}
\label{fig:time_corr}
\end{figure*}

To illustrate the importance of these correlations, we now show how they can be used to differentiate nonreciprocal from reciprocal interactions when both cases exhibit a similar average thermodynamic behavior, as is the case, \textit{e.g.}, in the disordered phase of the NRASM.
Figure \ref{fig:time_corr}a-b shows the evolution of the magnetizations $m_1(t,x)$ and $m_2(t,x)$ in the disordered phase: they are noisy and mostly featureless.
Yet, we observe some structure in the fluctuations, which is captured by the correlation functions $\langle m_\alpha(t,x) m_{\alpha^{\prime}}(0,x) \rangle$. 
In order to derive these correlation functions, we use our fluctuating hydrodynamic \eref{eq:hydro_NR_and_R} which captures the stochastic terms beyond the thermodynamic limit at first order in the lattice spacing $a$.
Linearizing \eref{eq:hydro_NR_and_R} around the solution $\rho_{\alpha}=\rho_0$, $m_{\alpha}=0$, we derive in Appendix~\ref{app:fluctuating_hydro} the time-correlations of the different magnetization fields.
In particular, we show that the inter-species time-correlation of the magnetizations is given by
\begin{align}
\label{eq:odd_time_correlation}
\langle m_1(0)m_2(t)\rangle=\mathcal{C}(t,\hat{\gamma},\hat{\gamma}_0)\sin\big(\hat{\gamma}\vartheta t\big)\;,
\end{align}
where $\langle\cdot\rangle$ designate averaging over the $\eta^{\alpha}_{n}$ noises, $\vartheta=\sinh(\beta_a)$, $\mathcal{C}(t,\hat{\gamma},\hat{\gamma}_0)$ is an even-time function reported in \eref{eq:mathcal_c_final}, 
$\hat{\gamma}=\tilde{\gamma}(\rho_0,\rho_0)$ and $\hat{\gamma}_0=\tilde{\gamma}(\rho_0,0)$ with $\tilde{\gamma}$ given by \eref{eq:gamma_eff_reciprocal_plus_nonreciprocal}.
As a consequence, the cross-correlation $\langle m_1(0)m_2(t)\rangle$ is an odd function of time (alternatively, the matrix $\langle m_\alpha(0)m_\beta(t)\rangle$ has an antisymmetric component), which heralds the nonequilibrium nature of the system~\cite{Onsager1931,Onsager1931b,Casimir1945,Seifert2012,oddreview,Roberts2021,Ohga2023}.
Figure \ref{fig:time_corr}c-d show the correlation function $\langle m_1(0)m_2(t)\rangle$ obtained from agent-based simulations (panel c), and compares these with our analytical predictions (blue curve in panel d, compared to the rescaled data of panel c plotted as yellow to red points).
The comparison shows that Eq~\eref{eq:odd_time_correlation} indeed provides a quantitative description of the time-correlation measured in the agent-based simulations. 
In addition, we note that \eref{eq:odd_time_correlation} also allows us to assess a universal thermodynamic bound recently introduced in Refs~\cite{Ohga2023,Vu2023}.
The details of this test can be found at the end of appendix \ref{app:fluctuating_hydro}.

To sum up, while the average thermodynamic fields do not bear the mark of nonreciprocity in the disordered phase, their time-correlations do, through the emergence of a time-antisymmetric component.

\section{Bands at the onset of flocking}
\label{sec:bands}

We now study more precisely the influence of non-reciprocity on the emergence of collective motion.
For simplicity, we focus on the $\beta_0=0$ case in this section.

This emergence is almost always accompanied by the formation of travelling bands of polar liquid that move without changing shape~\cite{gregoire2004onset,martin_fluctuation-induced_2021}.
As illustrated in \Fref{fig:classical_flocking_band}, the inside of these flocking bands consists of a dense ordered liquid while outside the system adopts a dilute disordered gas phase.
We first review how to describe these bands in a reciprocal system using the Toner-Tu equations.

\subsection{Phenomenology Lost: the reciprocal Toner-Tu model and its failings}

\subsubsection{Review of polar bands in the reciprocal Toner-Tu model}

\begin{figure}
\hspace*{-0.2cm}\includegraphics{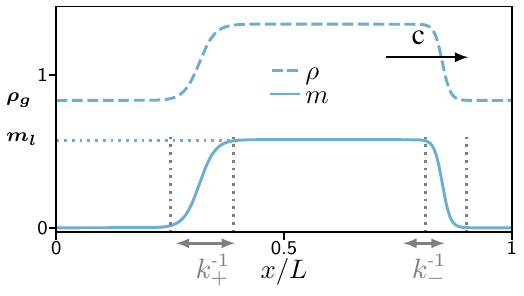}
  \vspace{-0.4cm}
  \caption{
  \textbf{Flocking band at onset of collective motion.}
  Travelling wave solution of \eref{eq:hydro_toner_tu} propagating to the right. Such profiles are observed at the onset of the transition when both the homogeneous disordered and ordered solutions are unstable.
  The magnetization (plain line) and the density (dashed line) are reported as functions of the position.
  The band's parameters $\rho_g$, $m_l$ and $k_{\pm}$ defined in \eref{eq:wave_param_single} are reported to highlight their physical meaning. 
  Parameters: $L=200$, $dx=0.5$, $dt=0.001$, $D=v=a_4=\phi_g=\lambda=\xi=1$, $\rho_0=1.1$.}
  \label{fig:classical_flocking_band}
\end{figure}

In this section, we review the so-called Toner-Tu equations, one of the canonical hydrodynamic theories highlighting this phase separation in bands at onset of collective motion.A simple version of the Toner-Tu equations has proven useful to study and derive the properties of flocking bands as well as the characteristics of the transition to collective motion \cite{solon_flocking_2015,caussin_emergent_2014}.
In this simple version, the retained coarse-grained fields in 1D are again the density $\rho$ and magnetization $m$.
Their time-evolution reads
\begin{align}
  \label{eq:hydro_toner_tu}
  \partial_t \rho &= -v_0\partial_x m \\
  \nonumber
   \partial_t m &= \mathcal{D}(m,\rho) + \mathcal{L}(m,\rho) \;,
\end{align}
where the differential operator $\mathcal{D}$ and Landau terms $\mathcal{L}$ are given by
\begin{align}
  \label{eq:def_operators_simplified_TT}
  \mathcal{D}(m,\rho) =& D \partial_{xx} m - \xi m\partial_x m -\lambda \partial_x \rho \;,\\
  \mathcal{L}(\rho,m) =& (\rho-\varphi_g)m -a_4 m^3\;,
\end{align}
with $v_0$, $\xi$, $D$, $\lambda$, $\varphi_g$ and $a_4$ being fixed parameters of the model.
The hydrodynamic Eqs. \eref{eq:hydro_toner_tu} exhibit flocking bands whenever $\rho$ is close to but higher than $\varphi_g$: we report in \Fref{fig:classical_flocking_band} a typical profile of the fields $m$ and $\rho$ in this band phase.
In Ref. \cite{solon_flocking_2015,caussin_emergent_2014}, the analytical form of these travelling bands was derived together with their speed using a
 travelling wave ansatz of the form $\rho(x-ct)$ and $m(x-ct)$ in Eq. \eref{eq:hydro_toner_tu}, with $c$ being the speed of propagation.
This substitution allows to map solutions of the PDE \eqref{eq:hydro_toner_tu} to solutions of a simpler dynamical system~\cite{Scott1999,vanSaarloos2003}.
In the case of a single flocking species whose fields evolve according to \eref{eq:hydro_toner_tu}, this dynamical system reads \cite{caussin_emergent_2014,solon_pattern_2015}
\begin{align}
 \label{eq:dyn_sys_single}
 D\ddot{m} &= -f(m)\dot{m} - \frac{dH}{dm}\;,
\end{align}
where the friction $f$ and potential $H$ are given by
\begin{align}
 \label{eq:friction_single_species}
 f(m)&=c - \frac{\lambda v_0}{c}-\xi m \\
 \label{eq:pot_single_species}
 H(m)&=-\left(\varphi_g - \rho_g \right)\frac{m^2}{2} + \frac{v_0}{3c}m^3 - a_4 \frac{m^4}{4}\;.
\end{align}
In \eref{eq:friction_single_species}-\eref{eq:pot_single_species}, $c$ and $\rho_g$ are respectively the band's speed and the density of the disordered gas outside the band (see \Fref{fig:classical_flocking_band}).
In \eref{eq:dyn_sys_single}, we also indicated derivatives with respect to $\zeta=x-ct$ by a dot to highlight the correspondance with Newtonian dynamics.
A solution of \eref{eq:dyn_sys_single} corresponds to a travelling band of the original flocking model \eref{eq:hydro_toner_tu} with a given speed $c$ and a given gas density $\rho_g$.
Two different solutions $m_{\pm}$ corresponding to the upward and downward fronts of the band have been reported in \cite{solon_pattern_2015}.
They read
\begin{align}
  \label{eq:wave_form_single}
  m_{\pm}(\zeta)&=\frac{m_l}{2}\left[1+\tanh(k_{\pm}\zeta)\right]\;,
\end{align}
where $m_l$ is the magnetization inside the band, $c$ is the wave speed, $\rho_g$ is the density outside the band and $k_{\pm}$ are the inverse of the fronts' widths.
They can be expressed in terms of the parameters of the flocking model only \cite{solon_pattern_2015} according to
\begin{align}
  \label{eq:wave_param_single}
  m_l&=\frac{2\sqrt{v_0}}{\sqrt{3a_4(3a_4\lambda+\xi)}}\;, \quad  c=\sqrt{\frac{v_0(3a_4\lambda+\xi)}{3a_4}}\;, \\
  \nonumber
  \rho_g &=\varphi_g-\frac{2v_0}{9a_4\lambda + 3\xi}\;,
\ k_{\pm} = -\sqrt{v_0}\frac{\xi\mp\sqrt{(8a_4D+\xi^2)}}{4D\sqrt{3a_4(3a_4\lambda+\xi)}}\;.
\end{align}
We will now proceed to explore how the flocking wave characterized by \eref{eq:wave_form_single}-\eref{eq:wave_param_single} is modified in the presence of nonreciprocity.

\subsubsection{A Non-reciprocal Toner-Tu model?}
\label{eq:subsec_simplified_toner_tu}
\begin{figure*}
\includegraphics{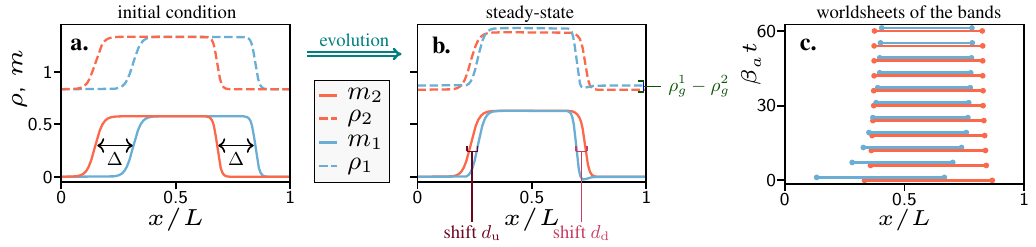}
  \vspace{-0.5cm}
  \caption{\textbf{Effect of linear nonreciprocity on flocking bands in Toner-Tu model \eref{eq:toner_tu_coupled_1}.}
  The linear nonreciprocal coupling in \eref{eq:toner_tu_coupled_1} induces a spatial synchronization of flocking bands which is characterized by two shifts $d_{\rm{u}}$ and $d_{\rm{d}}$.
  Species $1$ is represented in blue while species $2$ is represented in red.
  Magnetization fields are represented in plain lines while density fields are represented in dashed lines.
  {\bf a.} Initial profile corresponding to simulations of \eref{eq:toner_tu_coupled_1} with $\beta_a =0$.
  Note that the spatial delay $\Delta$ between the two bands will not affect the steady-state observed.
  {\bf b.} Stationary profile observed in simulations of \eref{eq:toner_tu_coupled_1} after switching on nonreciprocity ($\beta_a=0.02$).
  We define $d_{\rm{u}}$ and $d_{\rm{d}}$ as the spatial shifts between the upward and downward fronts of the two species.
  Note that the travelling bands end up overlapping and that the gas densities $\rho^1_g$ and $\rho^2_g$ have been modified due to nonreciprocity.
  {\bf c.} Time-evolution of species 1's band length (blue) and species 2's band length (red) in the reference frame of species 1. Starting from an initial condition where the bands are spatially shifted, they reach a configuration with fixed overlaps $d_{\rm{u}}$ and $d_{\rm{d}}$ in the steady state.
  {\bf Parameters:} $\beta_0=0$, $D=v=\phi_g=\lambda=\xi=a_4=1$, $\rho^1_0=\rho^2_0=1.1$, $dx=0.5$, $L=200$, $dt=0.002$.
  }
  \label{fig:definition_linear}
\end{figure*}

We now study two different species of flockers with purely nonreciprocal alignment between their respective magnetization fields $m_1$ and $m_2$.
A simple way of modelling this situation is to consider two replicas of the system in Eqs. \eref{eq:hydro_toner_tu}, which is valid for a single species, and to add a minimal {\it linear} nonreciprocal coupling as follows
\begin{subequations}
\label{eq:toner_tu_coupled_1}
\begin{align}
\partial_t \rho_1 &= -v_0\partial_x m_1\\
    \partial_t m_1 &= \mathcal{D}(m_1,\rho_1) + \mathcal{L}(m_1,\rho_1) - \beta_a m_2\\
    \partial_t \rho_2 &= -v_0\partial_x m_2 \\
\partial_t m_2 &= \mathcal{D}(m_2,\rho_2) + \mathcal{L}(m_2,\rho_2) + \beta_a m_1
\end{align}
\end{subequations}
In \eref{eq:toner_tu_coupled_1}, similarly to the NRASM, we remark that species $1$ aligns with species $2$ while species $2$ anti-aligns with species $1$, giving rise to a nonreciprocal coupling whose strength is controlled by $\beta_a$.
Inserting the travelling wave solutions $\rho_1(x-c_1 t)$, $\rho_2(x-c_2 t)$, $m_1(x-c_1 t)$ and $m_2(x-c_2 t)$ in \eref{eq:toner_tu_coupled_1} leads to two nonreciprocally coupled replicas of \eref{eq:dyn_sys_single} where,
\begin{subequations}
\label{eq:dyn_sys_NR}
\begin{align}
D\ddot{m}_1 &= -f_1(m_1)\dot{m}_1 - \frac{dH_1}{dm_1} - \beta_a m_2 \\ D\ddot{m}_2 &= -f_2(m_2)\dot{m}_2 - \frac{dH_2}{dm_2} + \beta_a m_1
\end{align}
\end{subequations}
and for which $f_1$, $f_2$, $H_1$ and $H_2$ have the same expression as in \eref{eq:friction_single_species}-\eref{eq:pot_single_species} but with species-dependent parameters for the travelling bands:
\begin{subequations}
\label{eq:pot}
\begin{align}
 \label{eq:pot_NR}
 f_1(m_1)&=c_1 - \frac{\lambda v_0}{c_1}-\xi m_1\;, \\
 H_1(m_1)&=-\left(\varphi_g - \rho^1_g \right)\frac{m_1^2}{2} + \frac{v_0}{3c_1}m_1^3 - a_4 \frac{m_1^4}{4} \\
 f_2(m_2)&=c_2 - \frac{\lambda v_0}{c_2}-\xi m_2\;, \\
 \label{eq:pot_NR_2}
 H_2(m_2)&=-\left(\varphi_g - \rho^2_g \right)\frac{m_2^2}{2} + \frac{v_0}{3c_2}m_2^3 - a_4 \frac{m_2^4}{4}\;.
\end{align}
\end{subequations}
In \eref{eq:pot}, $c_1$, $c_2$, $\rho^1_g$ and $\rho^2_g$ are the bands' speeds and the gas densities for species $1$ and $2$ respectively.
We will show in Section \ref{sec:simplified_nonlinear_toner_tu} that the travelling wave solutions of the NRASM originating from \eref{eq:hydro_NR_and_R_non_fluct} are actually also described, at linear order in $\beta_a$ and the $m_i$'s, by \eref{eq:dyn_sys_NR}.
We now assess how the single-species travelling solution \eref{eq:wave_form_single}-\eref{eq:wave_param_single} is affected by nonreciprocity.
To this end, we look for solutions of \eref{eq:dyn_sys_NR} perturbatively close to \eref{eq:wave_form_single}-\eref{eq:wave_param_single} in the form
\begin{subequations}
\label{eq:wave_form_NR}
\begin{align}
\displaystyle m^{\pm}_{1}(\zeta)&=\frac{m_1^l}{2}\left[1+\tanh(k^{\pm}_{1}\zeta)\right]\\
  \displaystyle m^{\pm}_{2}(\zeta)&=\frac{m_2^l}{2}\left[1+\tanh(k^{\pm}_{2}\zeta)\right]
\end{align}
\end{subequations}
with the bands' parameters being perturbed from their noninteracting values as
\begin{subequations}
\label{eq:wave_param_perturb}
\begin{align}
  m^l_1 &= m^l + \beta_a\delta m^l_1\;, & c_1&=c+\beta_a\delta c_1 \\ \rho^1_g &=\rho_g+\beta_a\delta\rho_g^1\;,
  & k^1_{\pm} &= k_{\pm} + \beta_a\delta k_{\pm}^1\;, \\
  m^l_2 &= m^l + \beta_a\delta m^l_2\;, & c_2&=c+\beta_a\delta c_2 \\
\rho^2_g &=\rho_g+\beta_a\delta\rho_g^2\;,
  & k^2_{\pm} &= k_{\pm} + \beta_a \delta k_{\pm}^2\;.
\end{align}
\end{subequations}
Inserting \eref{eq:wave_form_NR} into \eref{eq:dyn_sys_NR} and linearizing up to order $\beta_a$ yields a system of equations which allows us to find the values of the perturbations $\delta m^l_{\alpha}$, $\delta \rho^{\alpha}_g$, $\delta c_\alpha$ and $\delta k_{\pm}^{\alpha}$ for the two species ($\alpha\in\{1,2\}$).
The details of this computation are outlined in Appendix \ref{app:NR_bands} --  here we only report the result as
\begin{subequations}
\label{eq:results_first_order}
\begin{align}
  \delta m^l_1 &=\delta m^l_2 = 0\;, & \delta c_1 &=\delta c_2 = 0\;, \\
\delta \rho^1_g &=-\delta \rho^2_g = 1\;, & \delta k^1_{\pm} &= \delta k^2_{\pm} = 0\;.
\end{align}
\end{subequations}
In particular, we remark that only the $\delta\rho_g^{\alpha}$'s are nonzero.
Strikingly, the solution \eref{eq:wave_form_NR} that we have just found is spatially synchronized: the bands of the two species lie on top of each other.
Let us now provide a qualitative discussion of how two flocking bands spatially synchronize.
Consider two travelling bands initially separated by a distance $\Delta$. 
Using collective coordinates, we show in appendix \ref{app:synchronization} that $\Delta(t)$ decrease exponentially with time, thereby leading to spatial synchronization, according to
\begin{equation}
  \label{eq:synchronization}
  \Delta(t)=\Delta\exp\left(-\frac{2\beta_a m_l}{\hat{\alpha}_0} t\right)\;,
\end{equation}
where $\hat{\alpha}_0$ is given in \eref{eq:collective_coordinate_alpha_0}.
Let us now verify our predictions in simulations.
We consider periodic boundary conditions and we numerically integrate \eref{eq:toner_tu_coupled_1} for a set of parameters that lies in the band regime.
We start from an initial profile exhibiting two travelling bands of equal length $L^b$ at the same averaged density $\rho_0$ for both species.
These two travelling bands are initially described by equations \eref{eq:wave_form_single}-\eref{eq:wave_param_single} and we take them to be spatially separated by a distance $\Delta$ at time $t=0$ (see \Fref{fig:definition_linear}a for an illustration).
Independently of $\Delta$, we find that in the steady-state the travelling bands of the two species spatially synchronize (see \Fref{fig:definition_linear}c).
Furthermore, we remark that species 2's band lies inside the one of species 1: its final length $L_2^b$ is smaller than species 1's stationary band length $L_1^b$.
To characterize this overlap, we introduce $d_{\rm{u}}$ and $d_{\rm{d}}$ as respectively the spatial shifts between the two upward and downward fronts of the bands (see \Fref{fig:definition_linear}b for an illustration).
\par
\begin{figure*}
  \includegraphics{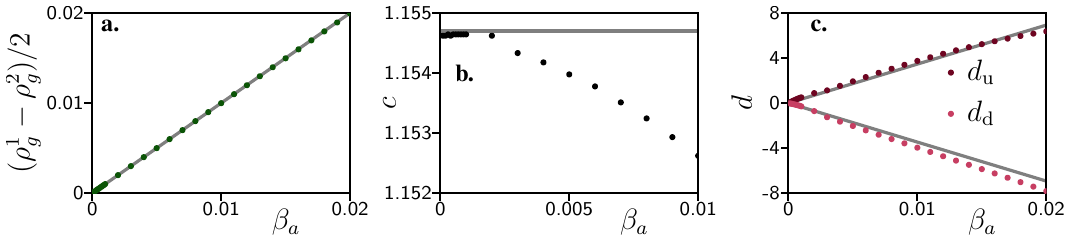}
  \vspace{-0.8cm}
  \caption{\textbf{Evolution of flocking bands' characteristics with linear nonreciprocity in Toner-Tu model \eref{eq:toner_tu_coupled_1}.}
  Three bands' characteristics are plotted as a function of the nonreciprocal strength $\beta_a$ for model \eref{eq:toner_tu_coupled_1}: the difference between gas densities, the speed and the spatial shifts.
  Dots are obtained through numerical integration of \eref{eq:toner_tu_coupled_1}.
  {\bf a.} Difference between the gas densities of species $1$ and $2$. The plain line corresponds to prediction \eref{eq:pred_densities}.
  {\bf b.} Speed of the stationary state made of two overlapping bands as described in \Fref{fig:definition_linear}.
  Our prediction \eref{eq:results_first_order} of a vanishing slope at $\beta_a=0$ (plain line) is confirmed in the numerics.
  Beyond this regime, we observe a decrease of the speed compared to its non-interacting single specie value.
  {\bf c.} Steady-state spatial shifts $d_{\rm{u}}$ and $d_{\rm{d}}$ as defined in \Fref{fig:definition_linear}b.
  The plain lines correspond to our prediction \eref{eq:shift_predictions}.
  {\bf Parameters:} $\beta_0=0$, $D=v=\phi_g=\lambda=\xi=a_4=1$, $\rho^1_0=\rho^2_0=1.1$, $dx=0.5$, $L=200$, $dt=0.002$.
  }
  \label{fig:linear_bands}
\end{figure*}

In \Fref{fig:linear_bands}a, we report the difference between the gas densities of both species and find an excellent agreement with our result \eref{eq:results_first_order} which entails that
\begin{align}
  \label{eq:pred_densities}
  \frac{\rho^1_g-\rho^2_g}{2}=\beta_a + \cO(\beta_a)\;.
\end{align}
Now that we have verified how the gas densities are affected by $\beta_a$, we can quantify the behavior of $d_{\rm{u}}$ and $d_{\rm{d}}$ with $\beta_a$ using mass conservation.
Let us determine how the final band lengths $L_1^b$ and $L_2^b$ vary with nonreciprocity.
Equating mass at time $t=0$ with mass in the steady-state for species 1, we obtain
\begin{align}
    \label{eq:mass_conservation_1}
    (\rho_g+m_l) L^b + \rho_g(L-L^b) = \rho^1_g(L-L_1^b) + (\rho^1_g + m_l)L_1^b\;.
\end{align}
We are looking for a perturbation of $L_1^b$ around $L^b$ so we insert $L_1^b = L^b + \beta_a \delta L_1^b$ into \eref{eq:mass_conservation_1}.
Further using our result \eref{eq:results_first_order}, which states that $\rho^1_g=\rho_g+\beta_a$, we get
\begin{align}
    \label{eq:delta_L_1}
    \delta L_1^b = -\frac{L}{m_l} \;.
\end{align}
Repeating the argument for species $2$, we can show that
\begin{align}
    \label{eq:delta_L_2}
    \delta L_2^b = \frac{L}{m_l} \;.
\end{align}
Combining \eref{eq:delta_L_1}-\eref{eq:delta_L_2} and further assuming that the change of bands' lengths is evenly distributed, we obtain
\begin{align}
  \label{eq:shift_predictions}
  d_{\rm{u}}=-d_{\rm{d}}=\frac{L}{m_l}\beta_a + \cO(\beta_a)\;.
\end{align}
In \Fref{fig:linear_bands}c, we compare our prediction \eref{eq:shift_predictions} with simulations of the non-reciprocal Toner-Tu model \eqref{eq:toner_tu_coupled_1}, and find a quantitative agreement.

\subsection{Bands in the exact hydrodynamic equations}
\label{sec:bands_NRASM}
\begin{figure*}
  \includegraphics{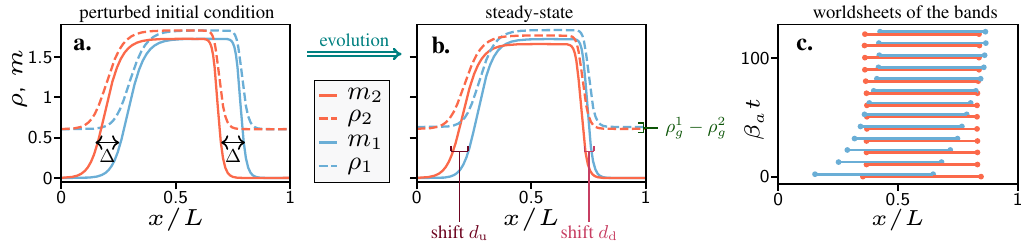}
  \vspace{-0.5cm}
  \caption{
  \textbf{Effect of nonreciprocity on flocking bands in the exact hydrodynamics \eref{eq:hydro_NR_and_R_non_fluct}.}
  Nonreciprocity in \eref{eq:hydro_NR_and_R_non_fluct} induces a spatial synchronization of flocking bands which is characterized by two shifts $d_{\rm{u}}$ and $d_{\rm{d}}$. 
  Species $1$ is colored in blue while species $2$ is colored in red.
  Magnetization fields are represented in plain lines while density fields are represented in dashed lines.
  {\bf a.} Initial profile used in the simulations.
  This profile is obtained by integrating numerically \eref{eq:hydro_NR_and_R_non_fluct} with $\beta_a=0$.
  Note that the spatial delay $\Delta$ between the two bands does not affect the steady-state observed.
  {\bf b.} Stationary profile observed in the simulations of \eref{eq:hydro_NR_and_R_non_fluct} after switching on nonreciprocity ($\beta_a = 0.02$).
  We define $d_{\rm{u}}$ and $d_{\rm{d}}$ as the spatial shifts between the upward and downward fronts of the two species.
  Note that the travelling bands end up overlapping despite their initial delay $\Delta$ and that the gas densities $\rho^1_g$ and $\rho^2_g$ have been modified due to nonreciprocity.
  {\bf c.} Time-evolution of species 1's band length (blue) and species 2's band length (red) in the reference frame of species 1. Starting from an initial condition where the bands are spatially shifted, they reach a configuration with fixed overlaps $d_{\rm{u}}$ and $d_{\rm{d}}$ in the steady state.
  {\bf Parameters:} $\beta_0=0$, $D=\gamma=1$, $v=2$, $\rho_0^1=\rho_0^2=1.2$, $L=100$, $dx=0.5$, $dt=0.002$.}
  \label{fig:definition_NRAIM}
\end{figure*}

We now compute the evolution of travelling wave solutions for the NRASM.
Using techniques similar to the ones detailed in \cite{solon_pattern_2015,caussin_emergent_2014}, we insert $m_1(x-(c_1+v/2) t)$, $m_2(x-(c_2+v/2) t)$, $\rho_1(x-(c_1+v/2) t)$ and $\rho_2(x-(c_2+v/2) t)$ into \eref{eq:hydro_NR_and_R_non_fluct}.
We show in appendix \ref{app:dynamical_NRAIM} that in the weak alignment, high density limit, the travelling profiles $m_1$ and $m_2$ are given, to first order in $\beta_a$, by equations similar to \eref{eq:dyn_sys_NR} which reads
\begin{align}
  \nonumber
  D\left(1+\frac{v^2}{4c_1^2}\right)\ddot{m}_1 &= -\left(\bar{f}_1(m_1)-\bar{\beta}_a \Gamma_1^{01} m_2\right)\dot{m}_1 - \frac{d\bar{H}_1(m_1)}{dm_1} \\
  \label{eq:wave_nraim_1}
  & - \bar{\beta}_a \alpha_1^{01} m_2 + \bar{\beta}_a \alpha_1^{11} m_2 m_1 - \bar{\beta}_a \alpha_1^{21} m_1^2 m_2 \;, \\
  \nonumber
  D\left(1+\frac{v^2}{4c_2^2}\right)\ddot{m}_2 &= -\left(\bar{f}_2(m_2)+\bar{\beta}_a \Gamma^{01}_{2} m_1\right)\dot{m}_2 - \frac{d\bar{H}_2(m_2)}{dm_2}
  \\
  \label{eq:wave_nraim_2}
  & + \bar{\beta}_a \alpha_2^{01} m_1 - \bar{\beta}_a \alpha_2^{11} m_2 m_1 + \bar{\beta}_a \alpha_2^{21} m_2^2 m_1 \;.
\end{align}
where $\bar{f}_1$, $\bar{f}_2$ are linear functions, $\bar{H}_1$, $\bar{H}_2$ are fourth-degree polynomials as in \eref{eq:pot} and $\bar{\beta}_a=\rho_0\beta_a$ with $\rho_0$ the initial averaged density.
The exact expressions of the $\Gamma^{i,j}_{l}$'s, the $\alpha^{i,j}_{l}$'s, $\bar{f}_1$, $\bar{f}_2$, $\bar{H}_1$ and $\bar{H}_2$ in \eref{eq:wave_nraim_1}-\eref{eq:wave_nraim_2} are reported in appendix \ref{app:dynamical_NRAIM}.
We can now compare the coupled equations \eref{eq:wave_nraim_1}-\eref{eq:wave_nraim_2} describing travelling waves in the NRASM with their equivalent \eref{eq:dyn_sys_NR} for a phenomenological nonreciprocal Toner-Tu model.
We note in \eref{eq:wave_nraim_1}-\eref{eq:wave_nraim_2} the presence of nonlinear nonreciprocal terms (the ones proportional to $\alpha_{l}^{11}$, $\alpha_{l}^{21}$ and $\gamma_{l}^{01}$) that were neglected in \eref{eq:dyn_sys_NR}.
We now assess numerically the importance of these nonlinear nonreciprocal terms for the phenomenology exhibited by the flocking bands of both species.
We repeat the simulation setup described in Part.~\ref{eq:subsec_simplified_toner_tu}.
We consider periodic boundary conditions and numerically integrate the full hydrodynamic Eq. \eref{eq:hydro_NR_and_R_non_fluct} for a set of paremeters that lies in the band regime (see \Fref{fig:phase_diagram}a).
We start from an initial condition exhibiting two travelling bands of equal length $L^b$: one for species $1$ and one for species $2$.
These initial band profiles correspond to travelling solutions of \eref{eq:hydro_NR_and_R_non_fluct} with nonreciprocity switched off (\textit{ie} $\beta_a=0$).
We take these bands to be spatially separated by a distance $\Delta$ at time $t=0$.
In \Fref{fig:definition_NRAIM}a-b, we report a typical initial condition together with its typical stationary solution.
\par
\begin{figure*}
  \includegraphics{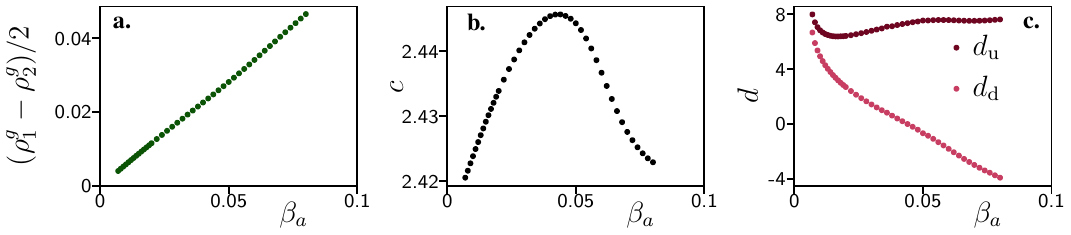}
  \vspace{-0.8cm}
  \caption{\textbf{Evolution of flocking bands' characteristics with nonreciprocity in the exact hydrodynamics \eref{eq:hydro_NR_and_R_non_fluct}.}
  Three bands’ characteristics are plotted as a function of the nonreciprocal strength $\beta_a$ for hydrodynamics \eref{eq:hydro_NR_and_R_non_fluct}: the difference between gas densities, the speed and the spatial shifts.
  Dots are obtained through numerical integration of \eref{eq:hydro_NR_and_R_non_fluct}.
  {\bf a.} difference between the gas densities of species $1$ and $2$.
  {\bf b.} speed of the stationary state made of two overlapping bands as described in \Fref{fig:definition_NRAIM} upon varying $\beta_a$. Obtained by numerical integration of \eref{eq:hydro_NR_and_R_non_fluct}.
  {\bf c.} steady-state spatial shifts $d_{\rm{u}}$ and $d_{\rm{d}}$ as defined in \Fref{fig:definition_NRAIM}b upon varying $\beta_a$.
  We remark that in the regime of small $\beta_a$ both shifts goes to a constant value $d^{\star}$.
  {\bf Parameters:} $\beta_0=0$, $D=\gamma=1$, $v=2$, $\rho_0^1=\rho_0^2=1.2$, $\beta=0.9$, $L=100$, $dx=0.5$, $dt=0.002$.}
  \label{fig:NR_AI_bands}
\end{figure*}

Independently of $\Delta$, we find that in the steady-state the travelling bands of the two species acquire an overlap with fixed spatial delays $d_{\rm{u}}$ and $d_{\rm{d}}$ between the upward and downward fronts respectively.
In \Fref{fig:definition_NRAIM}c, we report such a convergence toward $d_{\rm{u}}$ and $d_{\rm{d}}$ in steady state for a given initial condition $\Delta$.
Note that this convergence is not always monotonic: a collective coordinate description of the band would have to account for this non-trivial dynamics.
As we expect from our analysis in Part.~\ref{eq:subsec_simplified_toner_tu}, we remark that the stationary gas densities of both species have been modified compared to the noninteracting case $\beta_a=0$.
In \Fref{fig:NR_AI_bands}a, we report the difference between the gas densities of the two species as a function of $\beta_a$.
We find a behavior similar to the phenomenological model studied in Part.~\ref{eq:subsec_simplified_toner_tu}: there is a linear increase with a slope close to $1$.
In figure \Fref{fig:NR_AI_bands}c, we report the spatial delays $d_{\rm{u}}$ and $d_{\rm{d}}$ as a function of $\beta_a$.
Surprisingly, these delays do not vanish in the limit $\beta_a \to 0$ but rather reach the same nonzero value $d^{\star}$ which defines a unique spatial overlap between the two bands.
This stands in contrast with our phenomenological model \eref{eq:toner_tu_coupled_1} for which the delays $d_{\rm{u}}$ and $d_{\rm{d}}$ were vanishing linearly in the regime of small $\beta_a$.
Finally, in figure \Fref{fig:NR_AI_bands}b, we report the speed of the overlapping stationary bands upon varying $\beta_a$.
We remark that it is a non-monotonic function: the speed first increases with nonreciprocity before decreasing back again.
This is a feature which was not captured with the phenomenological Toner-Tu model of Part.~\ref{eq:subsec_simplified_toner_tu} for which we instead observed a steady decrease of the speed upon varying $\beta_a$.
\par
In the NRASM, we thus have two features that were not accounted for in the phenomenological Toner-Tu model \ref{eq:subsec_simplified_toner_tu}: the nonmonoticity of the speed upon varying nonreciprocity and the fixed stationary spatial delay $d^{\star}$ between the bands in the regime of weak nonreciprocity.
Let us now refine the phenomenological Toner-Tu model in a minimal way in order to account for these two new features.
To achieve this goal, we will have to consider nonlinear nonreciprocal couplings that we previously neglected in Part.~\ref{eq:subsec_simplified_toner_tu}.

\subsection{Phenomenology Regained: a nonlinear nonreciprocal Toner-Tu model}
\label{sec:simplified_nonlinear_toner_tu}
Informed by the exact hydrodynamic equations, we now develop a refined phenomenological model by including an additional nonlinear nonreciprocal coupling term between the two species.
Namely, we consider
\begin{align}
  \label{eq:toner_tu_coupled_1_NL}
\partial_t \rho_1 &= -v_0\partial_x m_1\;, \\
  \partial_t m_1 &= \mathcal{D}(m_1,\rho_1) + \mathcal{L}(m_1,\rho_1) - \beta_a m_2 + \beta_a\alpha m_2m_1 \notag \\
\partial_t \rho_2 &= -v_0\partial_x m_2 \notag \\
  \partial_t m_2 &= \mathcal{D}(m_2,\rho_2) + \mathcal{L}(m_2,\rho_2) + \beta_a m_1 - \beta_a\alpha m_2m_1 \notag
\end{align}
The above hydrodynamics \eref{eq:toner_tu_coupled_1_NL} leads to travelling wave solutions evolving according to
\begin{subequations}
\label{eq:dyn_sys_NR-NL}
\begin{align}
D\ddot{m}_1 &= -f_1(m_1)\dot{m}_1 - \frac{dH_1}{dm_1} - \beta_a m_2 + \beta_a \alpha m_2m_1 \\
 D\ddot{m}_2 &= -f_2(m_2)\dot{m}_2 - \frac{dH_2}{dm_2} + \beta_a m_1- \beta_a \alpha m_2m_1
\end{align}
\end{subequations}
where $f_1$, $f_2$, $H_1$ and $H_2$ are still given by \eref{eq:pot}.
We remark that the new nonlinear nonreciprocal terms $\pm\beta_a\alpha m_1m_2$ in \eref{eq:dyn_sys_NR-NL} are also present in the travelling wave evolution for the NRASM \eref{eq:wave_nraim_1}-\eref{eq:wave_nraim_2}.
Using the method detailed in Part.~\ref{eq:subsec_simplified_toner_tu} and Appendix \ref{app:NR_bands}, we compute how the noninteracting single species solution \eref{eq:wave_form_single} is affected by nonreciprocity at first order in $\beta_a$ for \eref{eq:dyn_sys_NR-NL}.
Unlike in Part.~\ref{eq:subsec_simplified_toner_tu}, we find that the bands' speeds $c_1$ and $c_2$ for the two species are given by
\begin{align}
  \label{eq:diff_speed}
  c_1&= c + \beta_a \delta c\;, & c_2&= c - \beta_a \delta c\;.
 \end{align}
From \eref{eq:diff_speed}, we deduce that two bands fronts located at the exact same spot have different speeds.
It implies that, in the presence of nonlinear nonreciprocal terms, the bands of species $1$ and $2$ can not be located at the same spot in steady-state.
We thus expect to observe a finite spatial delay between two bands for
\eref{eq:dyn_sys_NR-NL}.
However, \eref{eq:diff_speed} also implies that we will not be able to analytically quantify the steady-state band properties of \eref{eq:toner_tu_coupled_1_NL} because of this very spatial delay.
Let us now repeat the numerical analysis performed in Part.~\ref{eq:subsec_simplified_toner_tu} for the phenomenological nonlinear Toner-Tu model.
\par
\begin{figure*}
  \includegraphics{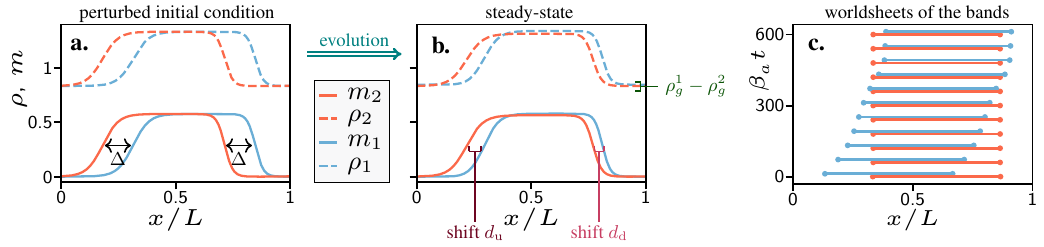}
  \vspace{-0.4cm}
  \caption{
  \textbf{Effect of nonreciprocity on flocking bands in Toner-Tu model \eref{eq:toner_tu_coupled_1_NL}.}
  Nonreciprocity in \eref{eq:toner_tu_coupled_1_NL} induces a spatial synchronization of flocking bands which is characterized by two shifts $d_{\rm{u}}$ and $d_{\rm{d}}$.
  Species $1$ is colored in blue while species $2$ is colored in red.
  Magnetization fields are represented in plain lines while density fields are represented in dashed lines.
  {\bf a.} Initial profile used in the simulations.
  It is obtained through numerical integration of \eref{eq:toner_tu_coupled_1_NL} with $\beta_a=0$.
  Note that the spatial delay $\Delta$ between the two bands will not affect the steady-state observed.
  {\bf b.} Stationary profile observed in the simulations of \eref{eq:toner_tu_coupled_1_NL} with nonreciprocity switched on ($\beta_a=0.001$).
  We define $d_{\rm{u}}$ and $d_{\rm{d}}$ as the spatial shifts between the upward and downward fronts of the two species.
  Note that the travelling bands end up overlapping and that the gas densities $\rho^1_g$ and $\rho^2_g$ have been modified due to nonreciprocity.
  {\bf c.} Time-evolution of species 1’s band length (blue) and species 2’s band length (red) in the reference frame of species 1. Starting from an initial condition where the bands are spatially shifted, they reach a configuration with fixed overlaps $d_{\rm{u}}$ and $d_{\rm{d}}$ in the steady state.
  {\bf Parameters:} $\beta_0=0$, $D=v=\varphi_g=\lambda=\xi=a_4=1$, $\rho^1_0=\rho^2_0=1.1$, $\alpha=0.2$, $dx=0.5$, $L=100$, $dt=0.002$
  }
  \label{fig:definition_NL_simplified}
\end{figure*}
\begin{figure*}
  \includegraphics{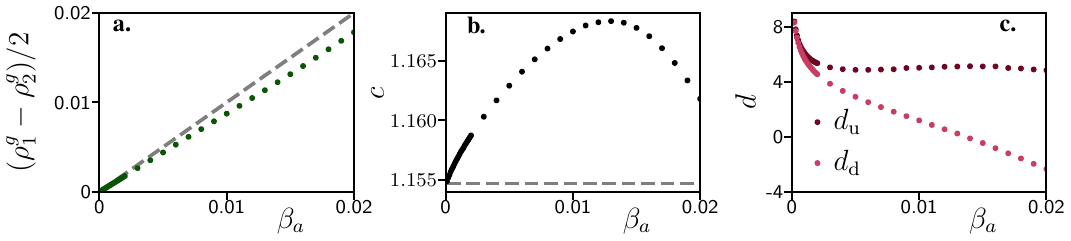}
  \vspace{-0.8cm}
  \caption{
  \textbf{Evolution of flocking bands’ characteristics with nonreciprocity in the Toner-Tu model \eref{eq:toner_tu_coupled_1_NL}.}
  Three bands’ characteristics are plotted as a function of the nonreciprocal strength $\beta_a$ for hydrodynamics \eref{eq:toner_tu_coupled_1_NL}: the difference between gas densities, the speed and the spatial shifts.
  Dots corresponds to numerical simulations of \eref{eq:toner_tu_coupled_1_NL}.
  {\bf a.} Difference between the gas densities of species $1$ and $2$. The dashed line indicates our prediction in the linear non reciprocal case $\alpha=0$.
  {\bf b.} Speed of the stationary state made of two overlapping bands as described in \Fref{fig:definition_NL_simplified} upon varying $\beta_a$.
  The dashed line corresponds to the speed of a flocking band in the single species case ($\beta_a=0$).
  {\bf c.} Steady-state spatial delays $d_{\rm{u}}$ and $d_{\rm{d}}$ as defined in \Fref{fig:definition_NL_simplified}b.
  {\bf Parameters:} $\beta_0=0$, $D=v=\varphi_g=\lambda=\xi=a_4=1$, $\rho^1_0=\rho^2_0=1.1$, $\alpha=0.2$, $dx=0.5$, $L=100$, $dt=0.002$.
  }
  \label{fig:nonlinear_bands}
\end{figure*}

We consider periodic boundary conditions and we numerically integrate \eref{eq:toner_tu_coupled_1_NL} for parameters taken in the band regime.
We start from an initial condition exhibiting a travelling band of length $L^b$ for both species.
These two travelling bands are described by equations \eref{eq:wave_form_single}-\eref{eq:wave_param_single} and we take them to be spatially separated by a distance $\Delta$ at time $t=0$.
In \Fref{fig:definition_NL_simplified}a-b, we report a typical initial condition together with its stationary solution.
Once again, independently of $\Delta$, we find that in the steady-state the travelling bands acquire an overlap with fixed spatial delays $d_{\rm{u}}$ and $d_{\rm{d}}$ between the upward and downward fronts respectively.
In \Fref{fig:definition_NL_simplified}c, we report such a convergence toward $d_{\rm{u}}$ and $d_{\rm{d}}$ in steady state for a given initial condition $\Delta$.
In \Fref{fig:nonlinear_bands}c, we report the values of $d_{\rm{u}}$ and $d_{\rm{d}}$ as a function of $\beta_a$.
Similarly to the NRASM, we find that both delays reach a finite value $d^{\star}$ in the limit $\beta_a \to 0$.
In \Fref{fig:nonlinear_bands}b, we also report the speed of the overlapping stationary flocking bands upon varying $\beta_a$.
Similarly to the NRASM, it is a nonmonotic function which first increases with nonreciprocity before decaying beyond a certain threshold.
We have thus shown that the key features of the flocking bands observed in Part.~\ref{sec:bands_NRASM} for the NRASM are captured by \eref{eq:dyn_sys_NR-NL} through a nonlinear nonreciprocal coupling.
It implies that nonlinear nonreciprocal couplings are needed to fully account for the band phenomenology observed in the NRASM.
As shown in Part.~\ref{eq:subsec_simplified_toner_tu} upon studying \eref{eq:toner_tu_coupled_1}, taking into account only linear nonreciprocal couplings $\pm\beta m_{1,2}$ will fail to reproduce both the nonmonotonic behavior of the speed and the finite spatial delay $d^\star$ at small $\beta_a$.

\begin{figure*}
  \includegraphics{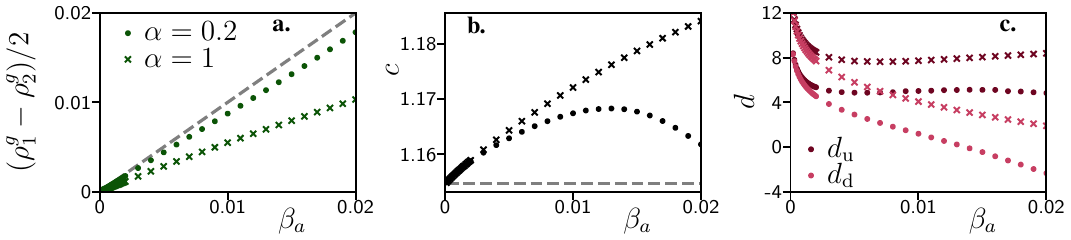}
  \vspace{-0.8cm}
  \caption{\textbf{Controlling flocks through non-reciprocity.} Three flocks' charateristics are represented as a function of nonreciprocity $\beta_a$ for different values of the nonlinear nonreciprocal term $\alpha$ in \eref{eq:toner_tu_coupled_1_NL}: the difference between gas densities, the speed and the spatial shifts.
  Dots and crosses are obtained through numerical simulations of \eref{eq:toner_tu_coupled_1_NL}. Dots correspond to $\alpha=0.2$ while crosses correspond to $\alpha=1$.
  {\bf a.} Difference between the gas densities of species $1$ and $2$ upon varying $\beta_a$. The dashed line corresponds to the linear nonreciprocal case when $\alpha=0$.
  {\bf b.} Speed of the travelling bands upon varying $\beta_a$. The dashed line correspond to the non-interacting single species case.
  {\bf c.} Steady-state spatial delays $d_{\rm{u}}$ and $d_{\rm{d}}$ upon varying $\beta_a$.
  }
  \label{fig:control_flock}
\end{figure*}

\section{Generalization to 2D flocks with continuous degree of freedom}
\label{sec:flying_XY}
In this section, we explore the generality of our results to more realistic flocking models. 
To this aim, we consider two species of motile agents evolving off-lattice and endowed with continuous orientational degrees of freedom.
These agents further experience nonreciprocal aligning interactions between their motility's directions. 
The position and orientation of the $i$-th agent ($i \in [1,..,N]$) belonging to species $\alpha$ ($\alpha \in [1,2]$) are respectively denoted $\br_i^{\alpha}$ and $\theta_i^{\alpha}$.
They evolve according to 
\begin{subequations}    
\label{eq:kuramoto_evol}
\begin{align}
    \dot{\br}_i^\alpha =& v \bu(\theta_i^\alpha) \\
    \dot{\theta}_i^\alpha =& -\sum_{(j,\delta)\in\cN^\alpha_i}J_{\alpha\delta}\sin(\theta_j^{\delta}-\theta_i^\alpha) + \sqrt{T}\eta_i^{\alpha}\;,
\end{align}
\end{subequations}
where $\cN_i^\alpha$ is the set of agents contained within radius $r_0$ of the $i$-th agent of species $\alpha$, namely \red{$\cN^\alpha_i=\{(j,\delta) \text{ st } 0<|\br_j^\delta-\br_i^\alpha|<r_0, j\in [1,..,N],\delta \in [1,2]\}$ }. 
Note that in \eqref{eq:kuramoto_evol} $v$ controls the speed of the agents and $T$ the strength of the fluctuations: $\eta_i^\alpha$ is a Gaussian white noise such that $\langle \eta_i^{\alpha}(t)\eta_j^{\delta}(s)\rangle=\delta_{ij}\delta_{\alpha\delta}\delta(t-s)$. 
Finally, $J$ is a matrix describing the interactions between species and reading
\begin{align}
    J_{\alpha\delta}=\beta \delta_{\alpha\delta} + \beta_a \epsilon_{\alpha\delta}\;,
\end{align}
where $\epsilon$ is the Levi-Civita tensor and $\beta$ ($\beta_a$) controls the intra-species (inter-species) alignment.
Note that because the Levi-Civita tensor is purely anti-symmetric the inter-species alignment is also purely nonreciprocal.
\red{We first stress that, even when $\beta_a=0$, dynamics \eref{eq:kuramoto_evol} does not feature flocking bands but rather flocking clusters due its peculiar Momentum-Conserving structure (see \cite{Chepizhko2021}).
As such, we did not search \eref{eq:kuramoto_evol} for a phenomenon similar to the synchronization of flocking bands observed in the NRASM.}

\begin{figure}
    \centering
\includegraphics{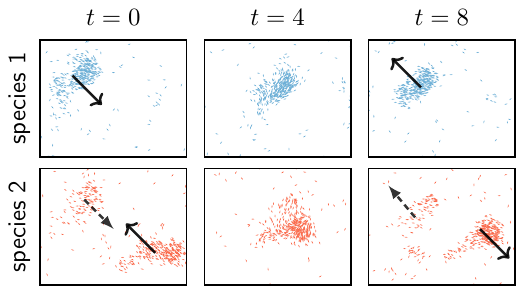}
    \caption{A collision of two clusters made of different species observed in numerical simulations of \eqref{eq:kuramoto_evol}. 
    Before the collision, at $t=0$, the two clusters are running into each others (left column) while after the collision they are running away from each others (right column). Note the presence of trailing agents belonging to species 2 inside the cluster of species 1 while the reverse is not true: no agents belonging to species 1 are trailing the cluster of species 2. This is also a feature that was observed in the NRASM (see the Rest \& Chase kymograph \Fref{fig:phase_diagram}d).
    Dashed arrows indicated the directions of motion of these trailing agents while plain arrows indicate the clusters' directions of motion.
    Parameters: $L_x=L_y=40$, $T=1$, $\rho=0.5$, $v=1$, $\beta=0.75$, $\beta_a=0.25$, $r_0=1$, $dt=0.002$.
    }
    \label{fig:kuramoto_agents}
\end{figure}

Instead, we focus on the phenomenology unveiled in the Rest \& Chase phase for the NRASM and show that it survives in dynamics \eqref{eq:kuramoto_evol} despite the embedding in dimension 2 and the continuous degree of freedom $\theta_i^{\alpha}$ driving the agents.
Performing numerical simulations of \eqref{eq:kuramoto_evol} (see details in appendix \ref{app:flying_kuramoto}), we indeed observe a phase where clusters of particles made of the same species self-propel and collide. 
Each collision between clusters of different species leads to a behavior similar to the collisions observed in the Rest \& Chase phase of the NRASM where the two clusters exchange their direction of motion (ie their degree of freedom).
We show an example of such collision in \Fref{fig:kuramoto_agents} and multiple ones in Movie 7.
In addition, the agent-based evolution \eqref{eq:kuramoto_evol} exhibits yet another feature of the NRASM which we now detail.
Fig~\ref{fig:phase_diagram}d shows that, within the Rest \& Chase phase, the clusters of species 1 are always accompanied by a small density of species 2 while the reverse is not true.
This is because species 2 aligns with species 1, enabling it to accompany cluster of species 1 as long as the density of trailing spins remains low compared to the cluster.
The reverse is not possible as species 1 anti-aligns with species 2, forbidding it to trail the clusters of species 2.
As exemplified in \Fref{fig:kuramoto_agents}, we observe a similar behavior around the clusters of our agent-based evolution \eqref{eq:kuramoto_evol}: cluster of species 1 are accompanied by a small number of trailing agents belonging to species 2 while clusters of species 2 remain untrailed.

\section{Discussion and conclusion}
In this paper, we presented a microscopic spatially-extended model of flocking with non-reciprocal interactions that can be exactly coarse-grained.
First, this model allowed us to evaluate the stability of the mean-field oscillating solution which generically emerges from nonreciprocal dynamics \cite{fruchart2021non,you2020nonreciprocity,Saha2020,Brauns2023}.
We find that it typically becomes unstable in the hydrodynamic limit when advective active terms are present: it does not survive the coupling to the conserved density field and we instead observe a new dynamical phase that we dub Chase \& Rest.
Interestingly, this phenomenology remains unchanged after the addition of a supplementary reciprocal coupling $\beta_0\neq 0$.
In addition, we highlighted the importance of nonreciprocal nonlinearities, crucial to account for this Chase \& Rest dynamics.
We also showed that nonreciprocal nonlinearities are responsible for two key features: the emergence of a spatial shift between flocks and the nonmonoticity of the flocking speed upon increasing the nonreciprocal strength.
Interestingly, the latter implies that adding nonreciprocal interactions can enhance transport with respect to the reciprocal case.
These two key features could be used to control flocks. Figure \ref{fig:control_flock} shows that an external operator able to tune the parameter $\alpha$ determining the amount of nonlinear nonreciprocity 
can control the gas densities, the spatial shifts $d_{\rm{u}}$ and $d_{\rm{d}}$ between the bands of the two species (see \Fref{fig:nonlinear_bands}c), and the flocking speed $c$. As the speed is not monotonic in $\beta_a$, the operator can in principle both accelerate and deccelerate the flocks.
As an example, this external operator could use $\alpha$ in order to impose a given spatial delay between the flocks.
This delay could later be used to separate the two species: a well-positioned trap could open upon being reached by the first species while closing right before the coming of the second species. Repeating the process could even allow for a complete separation of the species.
\par
Finally, as detailed in \Fref{fig:schematic_experiment}, experiments described by \eref{eq:hydro_NR_and_R} could be performed using self-propelled entities (bacteria, colloids, rollers...) confined into specific geometrical shapes such as racetracks, rings or annulus that ensure the restriction to a quasi-1D dynamics~\cite{geyer2019freezing,jorge2023active,bricard2013emergence}.
In these settings, one can inject two types of active entities with different properties such as surface coating or particle size that will generate nonreciprocal inter-species interactions~\cite{Liebchen2021,Maity2023}.
Our main predictions relevant for such experimental realizations are (i) the phase diagram \Fref{fig:phase_diagram} and especially the emergence of the Chase \& Rest phase in the presence of nonreciprocity and (ii) the inter-species magnetization time-correlation \eref{eq:odd_time_correlation} which can be used to characterize the presence of microscopic nonreciprocity even in phases where nonreciprocal interactions do not affect, a priori, the thermodynamic steady state.

\medskip
\medskip
\acknowledgments
D.M. acknowledges support from the Kadanoff Center for Theoretical Physics.
D.M. and V.V. also acknowledge support from the France-Chicago Center through the grant FACCTS.
Y.A., D.S. and M.F. acknowledges support from a MRSEC-funded Kadanoff–Rice fellowship and the University of Chicago Materials Research Science and Engineering Center, which is funded by the National Science Foundation under award no. DMR-2011854. 
Y.A. acknowledges support from the Zuckerman STEM Leadership Program.
V.V. acknowledges support from the Simons Foundation, the Complex Dynamics and Systems Program of the Army Research Office under grant W911NF-19-1-0268, the National Science Foundation under grant DMR-2118415 and the University of Chicago Materials Research Science and Engineering Center, which is funded by the National Science Foundation under award no. DMR-2011854. M.F. acknowledges support from the Simons Foundation. All the authors acknowledges the support of the Research Computing Center which provided the computing resources for this work.

\bibliographystyle{apsrev4-2}

\onecolumngrid
\appendix

\section{Derivation of the fluctuating hydrodynamics}
\label{app:solidification_coarse_graining}
In this appendix, we derive the fluctuating hydrodynamics \eref{eq:hydro_NR_and_R} of main text.
We start by deriving the thermodynamic average evolution \eref{eq:hydro_NR_and_R_non_fluct} in Sec.~\ref{sec_single_species} before determining the fluctuating terms arising for finite systems in Sec.~\ref{sec_single_species}. 
In both sections, we detail the computations of our method in the case of a unique species of spin with nonreciprocity switched off, \textit{i.e.} $\beta_a=\beta_0=0$.
The results can then be straightforwardly extended to include a nonreciprocal interaction with a second population of spins, \textit{i.e.} with $\beta_a\neq0$ and $\beta_0\neq 0$.

\subsection{Thermodynamic average evolution}
\label{sec_single_species}
In this section, we derive the average thermodynamic evolution \eref{eq:hydro_NR_and_R_non_fluct} of the main text.
We first illustrate our method on the case of a single species without non-reciprocity (\textit{i.e.} $\beta_0=\beta_a=0$) and detail the derivation of the corresponding thermodynamic evolution which is given by \eref{eq:hydro_non_interacting} in the main text.
Instead of using the usual Doi-Peliti techniques, we will resort to a fully classical method developed in \cite{andreanov2006field,lefevre2007dynamics,kourbane-houssene_exact_2018}.
We consider a one-dimensional lattice with $L$ different sites and a discretized time $t_j$ with $j\in\{1,..,N\}$.
In a time $dt=t_{j+1}-t_{j}$, a unique spin makes one of the three moves described in \hyperlink{rule:diffusion}{RI}, \hyperlink{rule:hopping}{RII} and \hyperlink{rule:flip}{RIII} with $\beta_0=\beta_a=0$.
We define the density $\rho_i$ and magnetization $m_i$ at site $i\in\{1,..,L\}$ as
\begin{align}
  \rho_i =& \eta_i^{+}+\eta_i^{-}, & m_i =& \eta_i^{+}-\eta_i^{-}
\end{align}
where $\eta_i^{\pm}$ are respectively the total number of $+$ and $-$ spins at site $i$.
A trajectory of the spins is thus completely determined by the set $\{\eta\}$ containing all the $\eta^{\pm}_i(t_j)$'s
\begin{align}
  \{\eta\} = \left\{ \eta^{\pm}_i(t_j)\quad\text{for}\quad i\in\{1,..,L\},\; j\in\{1,..,N\}\right\} \ .
\end{align}
More particularly, at time $t_j$, a configuration of the spins is entirely described by the set $\{\eta^j\}$ defined as
\begin{align}
  \{\eta^j\} = \left\{ \eta^{\pm}_i(t_j)\quad\text{for}\quad i\in\{1,..,L\}\right\} \ .
\end{align}
Let us define $J^{\pm}_{i}(t_{j})$ as the variation of the number of $\pm$ spins at site $i$ between time $t_j$ and $t_{j+1}$.
For a fixed trajectory of the spins, we have $J^{\pm}_{i}(t_{j})=\eta^{\pm}_i(t_{j+1})-\eta^{\pm}_i(t_j)$.
Note that because a unique spin moves during $dt$, each $J^{\pm}_i(t_j)$ takes values in $\{-1,0,1\}$ and only two of them are nonzero at the same time $t_j$.
\par
For example, when a spin $+$ at site $i$ hops to the right at time $t_j$, we have $J_{i}^{+}(t_j)=-1$ and $J_{i+1}^{+}(t_j)=1$ while all other $J_{k}^{\pm}(t_j)$ are zero.
Let us finally introduce the set $\{J\}$ containing all the $J_{i}^{\pm}(t_j)$'s
\begin{align}
  \{J\} = \left\{ J^{\pm}_i(t_j)\quad\text{for}\quad i\in\{1,..,L\},\; j\in\{1,..,N\}\right\} \ ,
\end{align}
and the set $\{J^j\}$ containing the $J_{i}^{\pm}(t_j)$'s at a given time
\begin{align}
  \{J^j\} = \left\{ J^{\pm}_i(t_j)\quad\text{for}\quad i\in\{1,..,L\}\right\} \ .
\end{align}
We start by establishing a path integral formulation for the probability $P[\{\eta\}]$ to observe a given trajectory $\{\eta\}$ of the spins.
Using a standard path integral formalism for on-lattice particle models \cite{lefevre2007dynamics,andreanov2006field}, we obtain
\begin{align}
  \label{eq:path_integral_delta}
  P[\{\eta\}]=\big{\langle}\prod_{i=1}^{L}\prod_{j=1}^{N}\delta\left(\eta^{+}_{i}(t_{j+1})-\eta^{+}_{i}(t_{j})-J^{+}_{i}(t_{j})\right)\delta\left(\eta^{-}_{i}(t_{j+1})-\eta^{-}_{i}(t_{j})-J^{-}_{i}(t_{j})\right)\big{\rangle}_{\{J\}}\;,
\end{align}
where $\langle\cdot\rangle_{\{J\}}$ indicates averaging over all the configurations in $\{J\}$.
Note that in \eref{eq:path_integral_delta}, the $\eta^{+}_{i}(t_{j})$'s and $\eta^{-}_{i}(t_{j})$'s correspond to the fixed trajectory $\{\eta\}$ while the $J_i^{\pm}(t_j)$'s are stochastic variables over which we average.
Using the integral expression of the Dirac function $\delta(s)=\int \exp(is\hat{s})d\hat{s}/(2\pi)$ into \eref{eq:path_integral_delta}, we introduce the fields $\hat{\eta}^{+}_{i}(t_{j})$ and $\hat{\eta}^{-}_{i}(t_{j})$ which are conjugated to $\eta^{+}_{i}(t_{j})$ and $\eta^{-}_{i}(t_{j})$, respectively.
We obtain
\begin{align}
  \nonumber
P[\{\eta\}]= \int \prod_{j=1}^{N}\bigg{[}&\prod_{i=1}^{L}\left[ d\hat{\eta}^{+}_{i}(t_{j})d\hat{\eta}^{-}_{i}(t_{j}) e^{\hat{\eta}^{+}_{i}(t_{j})\left[\eta^{+}_{i}(t_{j+1})-\eta^{+}_{i}(t_{j})\right]}e^{\hat{\eta}^{-}_{i}(t_{j})\left[\eta^{-}_{i}(t_{j+1})-\eta^{-}_{i}(t_{j})\right]}\right]\\
\label{eq:proba_traj_1}
&\big{\langle}\prod_{i=1}^{L}\left[e^{-\hat{\eta}^{+}_{i}(t_{j})J^{+}_{i}(t_{j})-\hat{\eta}^{-}_{i}(t_{j})J^{-}_{i}(t_{j})}\right]\big{\rangle}_{\{J^j\}}\bigg{]}\ ,
\end{align}
where $\langle\cdot\rangle_{\{J^j\}}$ is the average over all configurations $\cC\in\{J^j\}$.
\par
Denoting $\displaystyle f(\cC)=\prod_{i=1}^{L}e^{-\hat{\eta}^{+}_{i}(t_{j})J^{+}_{i}(t_{j})-\hat{\eta}^{-}_{i}(t_{j})J^{-}_{i}(t_{j})}$, we thus have
\begin{align}
  \label{eq:average_f}
  \langle \prod_{i=1}^{L}\left[e^{-\hat{\eta}^{+}_{i}(t_{j})J^{+}_{i}(t_{j})-\hat{\eta}^{-}_{i}(t_{j})J^{-}_{i}(t_{j})}\right]\rangle_{\{J^j\}} = \sum_{\cC \in \{J^j\}} f(\cC)P\left(\cC|\{\eta^j\}\right)\ ,
\end{align}
with $P\left(\cC|\{\eta^j\}\right)$ the probability to observe the set of configuration $\cC$ given the positions of the spins $\{\eta^j\}$ at the previous time $t_j$.
We now separate the possible configurations $\cC$ in $\{J^j\}$ according to the microscopic move they relate to.
We define the subset $\cN_{d}^j$, $\cN_{h}^j$, and $\cN_{s}^j$ of $\{J^j\}$ as
\begin{align}
  \cN_{d}^j =& \left\{ J^{\pm}_i(t_j)\ \text{for}\  i\in\{1,..,L\}\ \text{generated by a diffusive move \hyperlink{rule:diffusion}{\text{RI}}} \right\} \\
  \cN_{h}^j =& \left\{ J^{\pm}_i(t_j)\ \text{for}\  i\in\{1,..,L\}\ \text{generated by a hopping move \hyperlink{rule:hopping}{\text{RII}}} \right\} \\
  \cN_{s}^j =& \left\{ J^{\pm}_i(t_j)\ \text{for}\  i\in\{1,..,L\}\ \text{generated by a flipping move \hyperlink{rule:flip1}{\text{RIII}}} \right\}\ .
\end{align}
We further define $\cC_0$ as the configuration where all the $J_i^{\pm}(t_j)$ for $i\in\{1,..,L\}$ vanish: it corresponds to the case when no move is performed at time $t_j$.
We can now dispatch the sum over the configurations in \eref{eq:average_f} on the subsets $\cN_{d}^j$, $\cN_{h}^j$, $\cN_{s}^j$ and obtain
\begin{align}
  \nonumber
  \langle f \rangle_{\{J^j\}} =& f(\cC_0)P\left(\cC_0|\{\eta^j\}\right) + \sum_{\cC \in \cN_{d}^j}f(\cC)P\left(\cC|\{\eta^j\}\right) + \sum_{\cC \in \cN_{h}^j}f(\cC)P\left(\cC|\{\eta^j\}\right)\\
  \label{eq:f_av_1}
  & + \sum_{\cC \in \cN_{s}^j}f(\cC)P\left(\cC|\{\eta^j\}\right)\;.
\end{align}
Because $\cC_0$ is the zero move configuration, we have $f(\cC_0)=1$ and
\begin{align}
  \label{eq:T0}
  P\left(\cC_{0}|\{\eta^j\}\right) = 1 - \sum_{\cC \in \cN_{d}^j}P\left(\cC|\{\eta^j\}\right) - \sum_{\cC \in \cN_{h}^j}P\left(\cC|\{\eta^j\}\right) - \sum_{\cC \in \cN_{s}^j}P\left(\cC|\{\eta^j\}\right)\ .
\end{align}
Injecting \eref{eq:T0} into \eref{eq:f_av_1}, we get
\begin{align}
  \label{eq:f_av_2}
  \langle f \rangle_{\{J^j\}} = 1 + T_d + T_h + T_s\;,
\end{align}
with $T_d$, $T_h$ and $T_s$ given by
\begin{align}
  \label{eq:def_TD_TH}
  T_d =& \sum_{\cC \in \cN_{d}^j}\left(f(\cC)-1\right)P\left(\cC|\{\eta^j\}\right), & T_h =& \sum_{\cC \in \cN_{h}^j}\left(f(\cC)-1\right)P\left(\cC|\{\eta^j\}\right)\\
  \label{eq:def_TS}
  T_s =& \sum_{\cC \in \cN_{s}^j}\left(f(\cC)-1\right)P\left(\cC|\{\eta^j\}\right)\ . & &
\end{align}
We note that $T_d$, $T_h$ and $T_s$ are proportional to $dt$ through the probability $P\left(\cC|\{\eta^j\}\right)$ that a move occurred.
To order $dt$, we can thus reexponentiate \eref{eq:f_av_2} and obtain
\begin{align}
  \label{eq:f_av_exp}
  \langle f \rangle_{\{J^j\}} = \exp(T_d + T_h + T_s)+ \cO(dt^2)\,.
\end{align}
Hereafter, terms of order $\cO(dt^2)$ will be omitted for clarity.
We now determine the terms $T_d$, $T_h$ and $T_s$ through a detailed analysis of the subsets $\cN^j_d$, $\cN^j_h$ and $\cN^j_s$ respectively. \\
Let us start with $\cN^j_d$: it contains four typical configurations
\begin{itemize}
  \item $\cC_{d}^{\,i,\,1}$, when a $+$ spin diffuses from site $i$ to site $i+1$. In this case, $J_i^{+}(t_j)=-1$ and $J_{i+1}^{+}(t_j)=1$ while the remaining $J_k^{\pm}(t_j)$ are zero.
  We thus obtain $f(\cC_{d}^{\,i,\,1})=e^{\hat{\eta}^{+}_{i}(t_{j})-\hat{\eta}^{+}_{i+1}(t_{j})}$.
  The microscopic rules further give $P\left(\cC_{d}^{\,i,\,1}|\{\eta^j\}\right)= D a^{-2}\eta_i^{+}(t_j) dt$.
  \item $\cC_{d}^{\,i,\,2}$, when a $+$ spin diffuses from site $i+1$ to site $i$. In this case, $J_i^{+}(t_j)=1$ and $J_{i+1}^{+}(t_j)=-1$ while the remaining $J_k^{\pm}(t_j)$ are zero.
  We thus obtain $f(\cC_{d}^{\,i,\,2})=e^{-\hat{\eta}^{+}_{i}(t_{j})+\hat{\eta}^{+}_{i+1}(t_{j})}$.
  The microscopic rules further give $P\left(\cC_{d}^{\,i,\,2}|\{\eta^j\}\right)= Da^{-2} \eta_{i+1}^{+}(t_j) dt$.
  \item $\cC_{d}^{\,i,\,3}$, when a $-$ spin diffuses from site $i$ to site $i+1$. In this case, $J_i^{-}(t_j)=-1$ and $J_{i+1}^{-}(t_j)=1$ while the remaining $J_k^{\pm}(t_j)$ are zero.
  We thus obtain $f(\cC_{d}^{\,i,\,3})=e^{\hat{\eta}^{-}_{i}(t_{j})-\hat{\eta}^{-}_{i+1}(t_{j})}$.
  The microscopic rules further give $P\left(\cC_{d}^{\,i,\,3}|\{\eta^j\}\right)= D a^{-2}\eta_i^{-}(t_j) dt$.
  \item $\cC_{d}^{\,i,\,4}$, when a $-$ spin diffuses from site $i+1$ to site $i$. In this case, $J_i^{-}(t_j)=1$ and $J_{i+1}^{-}(t_j)=-1$ while the remaining $J_k^{\pm}(t_j)$ are zero.
  We thus obtain $f(\cC_{d}^{\,i,\,4})=e^{-\hat{\eta}^{-}_{i}(t_{j})+\hat{\eta}^{-}_{i+1}(t_{j})}$.
  The microscopic rules further give $P\left(\cC_{d}^{\,i,\,4}|\{\eta^j\}\right)= D a^{-2}\eta_{i+1}^{-}(t_j) dt$.
\end{itemize}
Using translational invariance, $T_d$ in \eref{eq:def_TD_TH} can be expressed in terms of these typical configurations $\cC_{d}^{\,i,\,k}$ for $k\in\{1,2,3,4\}$ as
\begin{align}
  \label{eq:TD_typ}
  T_d =& \sum_i\sum_{k=1}^{4}\left(f(\cC_{d}^{\,i,\,k})-1\right)P\left(\cC_{d}^{\,i,\,k}|\{\eta^j\}\right)
\end{align}
In a similar way, we now describe the single typical configuration in $\cN_h^i$
\begin{itemize}
  \item $\cC_{h}^{\,i}$, when a $+$ spin at site $i$ hops to site $i+1$. In this case, $J_i^{+}(t_j)=-1$ and $J_{i+1}^{+}(t_j)=1$ while the remaining $J_k^{\pm}(t_j)$ are zero.
  We thus obtain $f(\cC_{h}^{\,i,\,1})=e^{\hat{\eta}^{+}_{i}(t_{j})-\hat{\eta}^{+}_{i+1}(t_{j})}$.
  The microscopic rules gives $P\left(\cC_{h}^{\,i,\,1}|\{\eta^j\}\right)= \eta_{i}^{+}(t_j) v\,a^{-1}\,dt$.
\end{itemize}
Using translation invariance, $T_h$ in \eref{eq:def_TD_TH} can be expressed in terms of this typical configurations $\cC_{h}^{\,i}$ as
\begin{align}
  \label{eq:TH_typ}
  T_h =& \sum_i\left(f(\cC_{h}^{\,i})-1\right)P\left(\cC_{h}^{\,i}|\{\eta^j\}\right)
\end{align}
Finally, there are two typical configurations in $\cN_s^i$
\begin{itemize}
  \item $\cC_{s}^{\,i,\,1}$, when a $+$ spin at site $i$ flips into a $-$ spin. In this case, $J_i^{+}(t_j)=-1$ and $J_{i}^{-}(t_j)=1$ while the remaining $J_k^{\pm}(t_j)$ are zero.
  We thus obtain $f(\cC_{h}^{\,i,\,1})=e^{\hat{\eta}^{+}_{i}(t_{j})-\hat{\eta}^{-}_{i}(t_{j})}$.
  The microscopic rules further give
  \begin{align}
    \nonumber
    P\left(\cC_{s}^{\,i,\,1}|\{\eta^j\}\right)=& \eta_{i}^{+}(t_j)\gamma\,dt e^{-\beta\left[\eta_{i}^{+}(t_j)-\eta_{i}^{-}(t_j)\right]}
  \end{align}
  \item $\cC_{s}^{\,i,\,2}$, when a $-$ spin at site $i$ flips into a $+$ spin. In this case, $J_i^{+}(t_j)=1$ and $J_{i}^{-}(t_j)=-1$ while the remaining $J_k^{\pm}(t_j)$ are zero.
  We thus obtain $f(\cC_{h}^{\,i,\,2})=e^{-\hat{\eta}^{+}_{i}(t_{j})+\hat{\eta}^{-}_{i}(t_{j})}$.
  The microscopic rules further give
  \begin{align}
    \nonumber
    P\left(\cC_{s}^{\,i,\,2}|\{\eta^j\}\right)=& \eta_{i}^{-}(t_j)\gamma\,dt e^{\beta\left[\eta_{i}^{+}(t_j)-\eta_{i}^{-}(t_j)\right]}
  \end{align}
\end{itemize}
Using translation invariance, $T_s$ in \eref{eq:def_TS} can be expressed in terms of these typical configurations $\cC_{s}^{\,i,\,1}$ and $\cC_{s}^{\,i,\,2}$ as
\begin{align}
  \label{eq:TS_typ}
  T_s =& \sum_i\sum_{k=1}^{2}\left(f(\cC_{s}^{\,i,\,k})-1\right)P\left(\cC_{s}^{\,i,\,k}|\{\eta^j\}\right)\ .
\end{align}
Injecting \eref{eq:TD_typ},\eref{eq:TH_typ},\eref{eq:TS_typ} into the expression \eref{eq:f_av_exp} to get $\langle f\rangle_{\{J^{j}\}}$, we can then evaluate $P[\{\eta\}]$ in \eref{eq:proba_traj_1} as
\begin{equation}
  P[\{\eta\}]=\int \prod_{i=1}^{L}\prod_{j=1}^{N} d\hat{\eta}^{+}_{i}(t_{j})d\hat{\eta}^{-}_{i}(t_{j})\,e^{\cS}\ ,
\end{equation}
where the action $\cS$ reads
\begin{align}
  \nonumber
  \cS =& \sum_{i,j}\bigg{[}\hat{\eta}^{+}_{i}(t_j)\left(\eta^{+}_{i}(t_{j+1})-\eta^{+}_{i}(t_j)\right)+\hat{\eta}^{-}_{i}(t_j)\left(\eta^{-}_{i}(t_{j+1})-\eta^{-}_{i}(t_j)\right)\\
  \nonumber
  & +\frac{D dt}{a^2}\big{[}\eta^{+}_{i}(t_j)(e^{\hat{\eta}^{+}_{i}(t_j)-\hat{\eta}^{+}_{i+1}(t_j)}-1) + \eta^{+}_{i+1}(t_j)(e^{-\hat{\eta}^{+}_{i}(t_j)+\hat{\eta}^{+}_{i+1}(t_j)}-1) \\
  \nonumber
  &+\eta^{-}_{i}(t_j)(e^{\hat{\eta}^{-}_{i}(t_j)-\hat{\eta}^{-}_{i+1}(t_j)}-1) +\eta^{-}_{i+1}(t_j)(e^{-\hat{\eta}^{-}_{i}(t_j)+\hat{\eta}^{-}_{i+1}(t_j)}-1)\big{]}\\
  \nonumber
  &+\frac{vdt}{a}\eta^{+}_{i}(t_j) (e^{\hat{\eta}^{+}_{i}(t_j)-\hat{\eta}^{+}_{i+1}(t_j)}-1) +\gamma dt \big{[}f_s^+ (e^{\hat{\eta}^{+}_{i}(t_j)-\hat{\eta}^{-}_{i}(t_j)}-1) + f_s^-(e^{-\hat{\eta}^{+}_{i}(t_j)+\hat{\eta}^{-}_{i}(t_j)}-1)\big{]}\bigg{]}\ ,
\end{align}
with $f_s^+$ and $f_s^-$ given by
\begin{align}
  f_s^+ =& \eta^{+}_{i}(t_j) e^{-\beta(\eta^{+}_{i}(t_j)-\eta^{-}_{i}(t_j))} &
  f_s^- =&  f_s^+ \left(\eta^{+}_{i}(t_j)\rightarrow \eta^{-}_{i}(t_j),\,\eta^{-}_{i}(t_j)\rightarrow \eta^{+}_{i}(t_j)\right)\;.
\end{align}
At this point, while we would like to perform a Taylor expansion of the action at order $o(a)$ and $o(dt)$.
However, we can not do it because $\eta_i^{\pm}(t_j)$ is an integer and therefore the expressions $\eta^{\pm}_{i+1}=\eta^{\pm}_i + \partial_x \eta_i^{\pm} a$ or $\eta^{\pm}_i(t_{j+1})=\eta^{\pm}_i (t_j) + \partial_t \eta^{\pm}_i dt$ do not make sense.
We have first to smooth the $\eta_i^{\pm}(t_j)$'s into real variables.
To this aim, we will leverage the different scalings of the microscopic moves with $a$ as well as a specific property of purely diffusive particles.
First, we remark that in the $a\to 0$ limit particles will mostly diffuse on the lattice since diffusive moves, occurring at rate $\sim 1/a^2$, are dominant with respect to self-propulsion and alignment moves.
Second, we recall that, for purely diffusing particles, the stochastic variables $\eta^+_i(t_j)$ and $\eta_i^-(t_j)$ follow an independent Poisson law on each site whose average we parametrize by $\rho^+_i(t_j)$ and $\rho^-_i(t_j)$ ($(i,j)\in \{ 1,..,L\} \times \{ 1,..,N\}$).
Therefore, in the limit $a \to 0$, we can smooth the $\eta^+_i(t_j)$ and $\eta_i^-(t_j)$ by taking the mean value of $\cS$ with respect to this Poisson law.
Doing so, we obtain
\begin{align}
  \nonumber
  \langle \cS \rangle =& \sum_{i,j}\bigg{[}\hat{\eta}^{+}_{i}(t_j)\left(\langle\eta^{+}_{i}(t_{j+1})-\eta^{+}_{i}(t_j)\rangle\right)+\hat{\eta}^{-}_{i}(t_j)\left(\langle\eta^{-}_{i}(t_{j+1})-\eta^{-}_{i}(t_j)\rangle\right)\\
  \nonumber
  & +\frac{D dt}{a^2}\big{[}\langle\eta^{+}_{i}(t_j)\rangle(e^{\hat{\eta}^{+}_{i}(t_j)-\hat{\eta}^{+}_{i+1}(t_j)}-1) + \langle\eta^{+}_{i+1}(t_j)\rangle(e^{-\hat{\eta}^{+}_{i}(t_j)+\hat{\eta}^{+}_{i+1}(t_j)}-1) \\
  \nonumber
  &+\langle\eta^{-}_{i}(t_j)\rangle(e^{\hat{\eta}^{-}_{i}(t_j)-\hat{\eta}^{-}_{i+1}(t_j)}-1) +\langle\eta^{-}_{i+1}(t_j)\rangle(e^{-\hat{\eta}^{-}_{i}(t_j)+\hat{\eta}^{-}_{i+1}(t_j)}-1)\big{]}\\
  &+\frac{vdt}{a}\langle \eta^{+}_{i}(t_j) \rangle(e^{\hat{\eta}^{+}_{i}(t_j)-\hat{\eta}^{+}_{i+1}(t_j)}-1)
  \label{eq:action_average}
  +\gamma dt \big{[}\langle f_s^+ \rangle(e^{\hat{\eta}^{+}_{i}(t_j)-\hat{\eta}^{-}_{i}(t_j)}-1) +\langle f_s^- \rangle (e^{-\hat{\eta}^{+}_{i}(t_j)+\hat{\eta}^{-}_{i}(t_j)}-1)\big{]}\bigg{]}\;,
\end{align}
where $\langle \cdot \rangle$ implies averaging with the factorized independent Poisson law parametrized by $\rho^+_i(t_j)$ and $\rho^-_i(t_j)$.
Let us now evaluate \eref{eq:action_average}.
For terms linear in $\eta$, the average is given by the corresponding Poissonian parameter and we obtain
\begin{equation}
  \label{eq:linear_averages}
	\langle \eta^\pm_i(t_j)\rangle = \rho^{\pm}_i(t_j)\,, \quad \langle \eta^\pm_i(t_{j+1})\rangle = \rho^{\pm}_i(t_{j+1})\,, \quad \langle \eta^\pm_{i+1}(t_{j})\rangle = \rho^{\pm}_{i+1}(t_{j})\, .
\end{equation}
For $\langle f_s^+\rangle$ we compute that
\begin{align}
	\langle f_s^+ \rangle=& \bigg{\langle} \eta^+_i(t_j)e^{-\beta (\eta_i^+ (t_j)-\eta_i^- (t_j))}\bigg{\rangle}
	= \sum_{n,k=0}^{\infty}\frac{n\left[\rho^{+}_i(\tau)\right]^n\left[\rho^{-}_i(\tau)\right]^k}{n!k!} e^{-\rho^{+}_i(\tau)-\rho^{-}_i(\tau)-\beta(n-k)}
  \label{eq:fs+_final}
  = \,\cU(\rho_i^+ (t_j),\rho_i^- (t_j))\;,
\end{align}
where the function $\cU(x,y)$ is defined by
\begin{align}
  \nonumber
  \cU(x,y)=& x e^{-\beta + e^{-\beta}x+e^{\beta}y-x-y} \,.
\end{align}
The symmetry $+\leftrightarrow -$ then yields $\langle f_s^-\rangle$ as
\begin{align}
  \label{eq:fs-_final}
  \langle f_s^- \rangle=&\,\cU(\rho_i^- (t_j),\rho_i^+ (t_j))
\end{align}
Plugging the averages computed in \eref{eq:linear_averages} \eref{eq:fs+_final} \eref{eq:fs-_final} into expression \eref{eq:action_average} for $\langle \cS \rangle$, we get an action depending on the smooth, real variables $\rho_i^{\pm}(t_j)$
\begin{align}
  \nonumber
  \langle \cS \rangle =& \sum_{i,j}\bigg{[}\hat{\eta}^{+}_{i}(t_j)\left(\rho^{+}_{i}(t_{j+1})-\rho^{+}_{i}(t_j)\right)+\hat{\eta}^{-}_{i}(t_j)\left(\rho^{-}_{i}(t_{j+1})-\rho^{-}_{i}(t_j)\right)\\
  \nonumber
  & +\frac{D dt}{a^2}\big{[}\rho^{+}_{i}(t_j)(e^{\hat{\eta}^{+}_{i}(t_j)-\hat{\eta}^{+}_{i+1}(t_j)}-1) + \rho^{+}_{i+1}(t_j)(e^{-\hat{\eta}^{+}_{i}(t_j)+\hat{\eta}^{+}_{i+1}(t_j)}-1) \\
  \nonumber
  &+\rho^{-}_{i}(t_j)(e^{\hat{\eta}^{-}_{i}(t_j)-\hat{\eta}^{-}_{i+1}(t_j)}-1) +\rho^{-}_{i+1}(t_j)(e^{-\hat{\eta}^{-}_{i}(t_j)+\hat{\eta}^{-}_{i+1}(t_j)}-1)\big{]}
  +\frac{vdt}{a}\rho_i^{+}(t_j)(e^{\hat{\eta}^{+}_{i}(t_j)-\hat{\eta}^{+}_{i+1}(t_j)}-1) \\
  \label{eq:action_smoothed}
  & +\gamma dt \big{[}\cU \left(\rho_i^{+}(t_j),\rho_i^{-}(t_j)\right)(e^{\hat{\eta}^{+}_{i}(t_j)-\hat{\eta}^{-}_{i}(t_j)}-1) + \cU \left(\rho_i^{-}(t_j),\rho_i^{+}(t_j)\right) (e^{-\hat{\eta}^{+}_{i}(t_j)+\hat{\eta}^{-}_{i}(t_j)}-1)\big{]}\bigg{]}\ .
\end{align}
We can now take the limit of continuous time using $\rho^{\pm}_i(t_{j+1})-\rho^{\pm}_i(t_{j}) = \dot{\rho}^{\pm}_i(t) dt$ in the above expression.
Dropping the time dependence from now on, we assume that the quantities $\rho^{\pm}_i$, $\hat{\eta}_i^{\pm}$ are taken at time $t$
\begin{align}
  \nonumber
  \langle \cS \rangle =& \int dt\sum_{i}\bigg{[}\hat{\eta}^{+}_{i}\dot{\rho}^{+}_i+\hat{\eta}^{-}_{i}\dot{\rho}^{-}_i
  +\frac{D}{a^2}\big{[}\rho^{+}_{i}(e^{\hat{\eta}^{+}_{i}-\hat{\eta}^{+}_{i+1}}-1) + \rho^{+}_{i+1}(e^{-\hat{\eta}^{+}_{i}+\hat{\eta}^{+}_{i+1}}-1)
  +\rho^{-}_{i}(e^{\hat{\eta}^{-}_{i}-\hat{\eta}^{-}_{i+1}}-1) +\rho^{-}_{i+1}(e^{-\hat{\eta}^{-}_{i}+\hat{\eta}^{-}_{i+1}}-1)\big{]}\\
  &+\frac{v}{a}\rho_i^{+}(e^{\hat{\eta}^{+}_{i}-\hat{\eta}^{+}_{i+1}}-1)
  \label{eq:action_CT}
  +\gamma \big{[}\cU \left(\rho_i^{+},\rho_i^{-}\right)(e^{\hat{\eta}^{+}_{i}-\hat{\eta}^{-}_{i}}-1) + \cU \left(\rho_i^{-},\rho_i^{+}\right) (e^{-\hat{\eta}^{+}_{i}+\hat{\eta}^{-}_{i}}-1)\big{]}\bigg{]}\ .
\end{align}
We can now make the following change of variables
\begin{align}
  \rho_{i}=&\rho^{+}_{i}+\rho^{-}_{i} & m_{i}=&\rho^{+}_{i}-\rho^{-}_{i} & \hat{\rho}_{i}=&\frac{\hat{\eta}^{+}_{i}+\hat{\eta}^{-}_{i}}{2} & \hat{m}_{i}=&\frac{\hat{\eta}^{+}_{i}-\hat{\eta}^{-}_{i}}{2}\;.
\end{align}
In these new set of variables the actions $\cS$ reads
\begin{align}
  \nonumber
 \langle \cS \rangle &= \int_{0}^{T}dt\sum_{i}\bigg{[}\hat{\rho}_{i}\dot{\rho}_{i}+\hat{m}_{i}\dot{m}_{i}+\frac{D}{a^2}\rho_{i+1}(e^{-\hat{\rho}_{i}-\hat{m}_{i}+\hat{\rho}_{i+1}+\hat{m}_{i+1}}+e^{-\hat{\rho}_{i}+\hat{m}_{i}+\hat{\rho}_{i+1}-\hat{m}_{i+1}}-2) \\
 \nonumber
 & +\frac{D}{a^2}\rho_{i}(e^{\hat{\rho}_{i}+\hat{m}_{i}-\hat{\rho}_{i+1}-\hat{m}_{i+1}}+e^{\hat{\rho}_{i}-\hat{m}_{i}-\hat{\rho}_{i+1}+\hat{m}_{i+1}}-2) +\frac{D}{a^2}m_{i}(e^{\hat{\rho}_{i}+\hat{m}_{i}-\hat{\rho}_{i+1}-\hat{m}_{i+1}}-e^{\hat{\rho}_{i}-\hat{m}_{i}-\hat{\rho}_{i+1}+\hat{m}_{i+1}}) \\
 \nonumber
 & +\frac{D}{a^2}m_{i+1}(e^{-\hat{\rho}_{i}-\hat{m}_{i}+\hat{\rho}_{i+1}+\hat{m}_{i+1}}-e^{-\hat{\rho}_{i}+\hat{m}_{i}+\hat{\rho}_{i+1}-\hat{m}_{i+1}}) + \frac{v}{a} \frac{\rho_i+m_i}{2}(e^{\hat{\rho}_{i}+\hat{m}_{i}-\hat{\rho}_{i+1}-\hat{m}_{i+1}}-1) \\
 \label{eq:cS_discrete_space}
 &+ \gamma\;\mathcal{U}\left(\frac{\rho_i+m_i}{2},\frac{\rho_i-m_i}{2}\right)(e^{\hat{m}_i}-1) + \gamma\;\mathcal{U}\left(\frac{\rho_i-m_i}{2},\frac{\rho_i+m_i}{2}\right)(e^{-\hat{m}_i}-1)\;.
 \end{align}
 Before taking the limit of continuous space in the action $\langle \cS\rangle$, we need to perform a Taylor expansion of the fields using
 \begin{align}
   \nonumber
   \rho_{i+i}=&\rho_i + a \partial_x\rho_i + \frac{a^2}{2} \partial_{xx}\rho_i + o(a^2) &
   m_{i+i}=& m_i + a \partial_x m_i + \frac{a^2}{2} \partial_{xx} m_i + o(a^2)\\
   \nonumber
   \hat{\rho}_{i+i} =& \hat{\rho}_i + a \partial_x\hat{\rho}_i + \frac{a^2}{2} \partial_{xx}\hat{\rho}_i + o(a^2) &
   \hat{m}_{i+i}=& \hat{m}_i + a \partial_x \hat{m}_i + \frac{a^2}{2} \partial_{xx} \hat{m}_i + o(a^2)\;.
 \end{align}
 Plugging the above expansions into \eref{eq:cS_discrete_space},  we are now ready to take the limit $a\to 0$. We obtain
 \begin{align}
 \label{eq:integrated_action_single_species}
   \langle \cS \rangle &=  \frac{1}{a}\int_{0}^{L}\int_{0}^{T}  S[\rho,m,\hat{\rho},\hat{m}] dt dx + o(a^{-1})\ ,
 \end{align}
 where $S[\rho,m,\hat{\rho},\hat{m}]$ is given by
 \begin{align}
   \nonumber
   S =&  \hat{\rho}\dot{\rho}+\hat{m}\dot{m}+D\nabla\hat{\rho}\nabla\rho+D\nabla\hat{m}\nabla m +\frac{D}{2}\rho\left[(\nabla\hat{\rho}+\nabla\hat{m})^{2}+(\nabla\hat{\rho}-\nabla\hat{m})^{2}\right]+\frac{D}{2}m\left[(\nabla\hat{\rho}+\nabla\hat{m})^{2}-(\nabla\hat{\rho}-\nabla\hat{m})^{2}\right] \\
   \label{eq:final_S}
   &+ \gamma\;\mathcal{U}\left(\frac{\rho+m}{2},\frac{\rho-m}{2}\right)(e^{\hat{m}}-1)-v\nabla\hat{m}\frac{\rho+m}{2} + \gamma\;\mathcal{U}\left(\frac{\rho-m}{2},\frac{\rho+m}{2}\right)(e^{-\hat{m}}-1) -v\nabla\hat{\rho}\frac{\rho+m}{2}\;.
\end{align}
We are now ready to deduce the hydrodynamics of the microscopic model.
Because we took the regime $a\rightarrow 0$, we just have to do a saddle-point of the above integrand $S$ to get the evolution equations for $\rho$ and $m$.
The hydrodynamics is obtained from minimizing \eref{eq:final_S} with respect to $\hat{m}$ and $\hat{\rho}$
 \begin{align}
   \label{eq:minimization_action}
   \frac{\delta S}{\delta \hat{\rho}}&= 0 & \frac{\delta S}{\delta \hat{m}}&= 0 & \frac{\delta S}{\delta \rho}&= 0 & \frac{\delta S}{\delta m}&= 0 \ .
 \end{align}
 Conditions $\delta S/\delta \rho = \delta S/\delta m = 0$ are met for auxiliary homogeneous fields $\hat{\rho}=\hat{m}=0$ while, for such fields, $\delta S/\delta \hat{\rho} =0$ and $ \delta S/\delta \hat{m} = 0$ yields the sought after hydrodynamics for $\rho$ and $m$
 \begin{subequations}
 \label{eq:hydro_rho_exact_single_species}
 \begin{align}
   \label{eq:hydro_rho_exact_app}
	\partial_t \rho &= D \nabla^2\rho - \frac{v}{2}\nabla(m+\rho) \\
  \label{eq:hydro_m_exact_app}
 	\partial_t m &= D \nabla^2m - \frac{v}{2}\nabla(m+\rho) - 2\gamma m e^{-\beta + (\cosh(\beta)-1)\rho}\left(\cosh\left[\sinh(\beta)m\right]m-\sinh\left[\sinh(\beta)m\right]\rho\right)\;
 \end{align}
 \end{subequations}
 which corresponds to the coarse-grained evolution of a unique species of spins aligning with their own kin.

 The method that allowed us to obtain \eref{eq:hydro_rho_exact_app}-\eref{eq:hydro_m_exact_app} can be straightforwardly applied to the case of two species of spins with nonreciprocal alignment considered in \hyperlink{rule:diffusion}{RI}, \hyperlink{rule:hopping}{RII}, \hyperlink{rule:flip1}{RIII}, \hyperlink{rule:flip2}{RIV}.
 The computations are more involved but the only difference with the single species case is the expression of $f_s^{+}$ and $f_s^{-}$ which stem from the alignment dynamics.
 Let us take the example of species $1$ with rule \hyperlink{rule:flip1}{RIII} to compute $f_s^{+,1}$.
 When $\beta_a$ and $\beta_0$ are nonzero, $f_s^{+,1}$ is given by
 \begin{align}
   \nonumber
 	\langle f_s^{+,1} \rangle=& \:\bigg{\langle} \eta^{+,1}_i(t_j)e^{-\beta (\eta_i^{+,1} (t_j)-\eta_i^{-,1} (t_j))+(\beta_a -\beta_0)(\eta_i^{+,2} (t_j)-\eta_i^{-,2} (t_j))}\bigg{\rangle} \\
   \nonumber
 	=& \!\!\!\sum_{n,k,l,m=0}^{\infty}\!\!\!\frac{n\left[\rho^{+,1}_i(t_j)\right]^n\left[\rho^{-,1}_i(t_j)\right]^k\left[\rho^{+,2}_i(t_j)\right]^l\left[\rho^{-,2}_i(t_j)\right]^m}{n!\;k!\;l!\;m!} e^{-\rho^{+,1}_i(t_j)-\rho^{-,1}_i(t_j)-\rho^{+,2}_i(t_j)-\rho^{-,2}_i(t_j)-\beta(n-k)+(\beta_a-\beta_0)(l-m)}\\
   \label{eq:fs+1}
   =& \,\cU_{\beta_a,\beta_0}\left(\rho_i^{+,1} (t_j),\rho_i^{-,1} (t_j),\rho_i^{+,2} (t_j),\rho_i^{-,2} (t_j)\right)\;,
 \end{align}
 where $\cU_{\beta_a,\beta_0}(x,y,z,w)$ is given by
\begin{align}
  \cU_{\beta_a,\beta_0}(x,y,z,w)=& x e^{-\beta + e^{-\beta}x+e^{\beta}y + e^{\beta_a-\beta_0}z+ e^{\beta_0-\beta_a}w-x-y-w-z} \,.
\end{align}
By the symmetries $+\leftrightarrow -$ and $\beta_a\leftrightarrow -\beta_a$, we obtain the remaining $\langle f_s^{-,1}\rangle$, $\langle f_s^{-,2}\rangle$ and $\langle f_s^{+,2}\rangle$ as
\begin{align}
  \label{eq:fs-1}
\langle f_s^{-,1} \rangle =& \,\cU_{\beta_a,\beta_0}\left(\rho_i^{-,1} (t_j),\rho_i^{+,1} (t_j),\rho_i^{+,2} (t_j),\rho_i^{-,2} (t_j)\right)\;\\
\label{eq:fs+2}
\langle f_s^{+,2} \rangle =& \,\cU_{-\beta_a,\beta_0}\left(\rho_i^{+,2} (t_j),\rho_i^{-,2} (t_j),\rho_i^{+,1} (t_j),\rho_i^{-,1} (t_j)\right)\;\\
\label{eq:fs-2}
\langle f_s^{-,2} \rangle =& \,\cU_{-\beta_a,\beta_0}\left(\rho_i^{-,2} (t_j),\rho_i^{+,2} (t_j),\rho_i^{+,1} (t_j),\rho_i^{-,1} (t_j)\right)\;
\end{align}
Plugging expressions \eref{eq:fs+1}, \eref{eq:fs-1}, \eref{eq:fs+2}, \eref{eq:fs-2} into the single species expression \eref{eq:action_average} and taking the continuum limit, we obtain the actions $S^{1}$ and $S^{2}$ corresponding to species $1$ and $2$ respectively as
\begin{align}
  \nonumber
  S^{1} =&  \hat{\rho}_1\dot{\rho}_1+\hat{m}_1\dot{m}_1+D\nabla\hat{\rho}_1\nabla\rho_1+D\nabla\hat{m}_1\nabla m_1 +\frac{D}{2}\rho_1\left[(\nabla\hat{\rho}_1+\nabla\hat{m}_1)^{2}+(\nabla\hat{\rho}_1-\nabla\hat{m}_1)^{2}\right]\\
  \nonumber
  &+\frac{D}{2}m_1\left[(\nabla\hat{\rho}_1+\nabla\hat{m}_1))^{2}-(\nabla\hat{\rho}_1-\nabla\hat{m}_1)^{2}\right]+ \gamma\;\mathcal{U}_{\beta_a,\beta_0}\left(\frac{\rho_1+m_1}{2},\frac{\rho_1-m_1}{2},\frac{\rho_2+m_2}{2},\frac{\rho_2-m_2}{2}\right)(e^{\hat{m}_1}-1)\\
  \label{eq:action_species_1}
  &+ \gamma\;\mathcal{U}_{\beta_a,\beta_0}\left(\frac{\rho_1-m_1}{2},\frac{\rho_1+m_1}{2},\frac{\rho_2+m_2}{2},\frac{\rho_2-m_2}{2}\right)(e^{-\hat{m}_1}-1) -v\nabla\hat{\rho}_1\frac{\rho_1+m_1}{2}-v\nabla\hat{m}_1\frac{\rho_1+m_1}{2}\\
  \nonumber
  S^{2} =&  \hat{\rho}_2\dot{\rho}_2+\hat{m}_2\dot{m}_2+D\nabla\hat{\rho}_2\nabla\rho_2+D\nabla\hat{m}_2\nabla m_2 +\frac{D}{2}\rho_2\left[(\nabla\hat{\rho}_2+\nabla\hat{m}_2)^{2}+(\nabla\hat{\rho}_2-\nabla\hat{m}_2)^{2}\right]\\
  \nonumber
  &+\frac{D}{2}m_2\left[(\nabla\hat{\rho}_2+\nabla\hat{m}_2))^{2}-(\nabla\hat{\rho}_2-\nabla\hat{m}_2)^{2}\right]+ \gamma\;\mathcal{U}_{-\beta_a,\beta_0}\left(\frac{\rho_2+m_2}{2},\frac{\rho_2-m_2}{2},\frac{\rho_1+m_1}{2},\frac{\rho_1-m_1}{2}\right)(e^{\hat{m}_2}-1)\\
  \label{eq:action_species_2}
  &+ \gamma\;\mathcal{U}_{-\beta_a,\beta_0}\left(\frac{\rho_2-m_2}{2},\frac{\rho_2+m_2}{2},\frac{\rho_1+m_1}{2},\frac{\rho_1-m_1}{2}\right)(e^{-\hat{m}_2}-1)-v\nabla\hat{m}_2\frac{\rho_2+m_2}{2}-v\nabla\hat{\rho}_2\frac{\rho_2+m_2}{2}\,.
\end{align}
Minimizing the total action $S = S_1 + S_2$ with respect to the auxiliary fields as detailed in \eref{eq:minimization_action} then yields the sought after deterministic hydrodynamics \eref{eq:hydro_NR_and_R_non_fluct} of main text, which describes the agent-based dynamics \hyperlink{rule:diffusion}{RI}, \hyperlink{rule:hopping}{RII}, \hyperlink{rule:flip1}{RIII}, \hyperlink{rule:flip2}{RIV} in the thermodynamic limit.

\subsection{Fluctuations around the thermodynamic limit}
In this section, we detail the computation of the fluctuating hydrodynamics \eref{eq:hydro_NR_and_R} of the main text.
We first describe how to obtain a fluctuating hydrodynamics from the action $S$ given in \eref{eq:final_S} for a single species of spin.
Despite involving lengthy computations, the generalization to two species with nonreciprocal interactions is straightforward.
In order to obtain the thermodynamic evolution \eref{eq:hydro_rho_exact_single_species}, we minimized the action $S$ and found a minimum when $\hat{\rho}=\hat{m}=0$.
To find the fluctuating hydrodynamics, we will therefore Taylor expand the action $S$ at second order in the auxiliary fields around this minimum as
\begin{align}
\label{eq:action_second_order}
S=&
\begin{pmatrix}
    S^{\rho}_0 & S^{m}_0 \\
\end{pmatrix}
\begin{pmatrix}
    \hat{\rho} \\
    \hat{m} \\
\end{pmatrix}+
\frac{1}{2}
\begin{pmatrix}
    \partial_x\hat{\rho} & \partial_x\hat{m} & \hat{m} \\
\end{pmatrix}
\bar{M}
\begin{pmatrix}
    \partial_x\hat{\rho} \\ 
    \partial_x\hat{m} \\
    \hat{m} \\
\end{pmatrix}\;,
\end{align}
where $S^{\rho}_0$,  $S^{m}_0$ are the deviations from the coarse-grained hydrodynamics \eref{eq:hydro_rho_exact_app} and \eref{eq:hydro_m_exact_app} respectively, 
\begin{align}
S^{\rho}_0 =& \partial_t \rho - D \nabla^2\rho + \frac{v}{2}\nabla(m+\rho) \\
S^{m}_0 =& \partial_t m - D \nabla^2m + \frac{v}{2}\nabla(m+\rho) + 2\gamma m e^{-\beta + (\cosh(\beta)-1)\rho}\left(\cosh\left[\sinh(\beta)m\right]m-\sinh\left[\sinh(\beta)m\right]\rho\right)\;,
\end{align}
while the matrix $\bar{M}$ is given by 
\begin{align}
\bar{M}=\begin{pmatrix}
    2 D \rho & 2 D m & 0 \\
    2 D m & 2 D \rho & 0 \\
    0 & 0 & \gamma_{\text{r}}\; S^m_f \\
\end{pmatrix}\;,
\end{align}
with $\gamma_{\text{r}}=2\gamma  e^{-\beta-\rho+\rho\cosh(\beta)}$ as in the main text and $S^m_f$ reading 
\begin{align}
S^m_f= \frac{\rho  \cosh (m \sinh (\beta ))-m \sinh (m \sinh (\beta ))}{2}\;.
\end{align}
Inserting \eref{eq:action_second_order} into \eref{eq:integrated_action_single_species},  we obtain the probability of the fields' trajectory as 
\begin{align}
\mathcal{P}(\{\rho,m\})=\int \tilde{\mathcal{D}}\exp\bigg(\int\bigg[\frac{1}{a}\begin{pmatrix}
    S^{\rho}_0 & S^{m}_0 \\
\end{pmatrix}
\begin{pmatrix}
    \hat{\rho} \\
    \hat{m} \\
\end{pmatrix}+
\frac{1}{2a}
\begin{pmatrix}
    \partial_x\hat{\rho} & \partial_x\hat{m} & \hat{m} \\
\end{pmatrix}
\bar{M}
\begin{pmatrix}
    \partial_x\hat{\rho} \\ 
    \partial_x\hat{m} \\
    \hat{m} \\
\end{pmatrix}\bigg] dxdt\bigg)\;,
\end{align}
where $\tilde{\mathcal{D}}$ is an abbreviation for a path integration over all fields present inside the action, \textit{i.e.} here $\tilde{\mathcal{D}}=\mathcal{D}[\rho,m,\hat{\rho},\hat{m}]$.
We now perform a Hubbard-Stratonovich transformation and decouple the quadratic terms of the action by introducing three gaussian noise fields $\eta_1$, $\eta_2$ and $\eta_3$.
We obtain
\begin{align}
\label{eq:action_fluctuante_1}
\mathcal{P}(\{\rho,m\})=\int \tilde{\mathcal{D}}\exp\bigg(\int\bigg[\frac{1}{a}\begin{pmatrix}
    S^{\rho}_0 & S^{m}_0 \\
\end{pmatrix}
\begin{pmatrix}
    \hat{\rho} \\
    \hat{m} \\
\end{pmatrix}
+\frac{1}{\sqrt{a}}\begin{pmatrix}
    \eta_1 & \eta_2 & \eta_3 \\
\end{pmatrix}
\begin{pmatrix}
    \partial_x\hat{\rho} \\
    \partial_x\hat{m} \\
    \hat{m} \\
\end{pmatrix}
-\frac{1}{2}
\begin{pmatrix}
    \eta_1 & \eta_2  & \eta_3  \\
\end{pmatrix}
\bar{M}^{-1}
\begin{pmatrix}
    \eta_1 \\ 
    \eta_2 \\
    \eta_3 \\
\end{pmatrix}\bigg]dxdt\bigg)\;.
\end{align}
Integrating by parts the auxiliary fields and integrating over $\hat{\rho}$ and $\hat{m}$ in \eref{eq:action_fluctuante_1} then yields
\begin{align}
\label{eq:fluctuating_action_3}
\mathcal{P}(\{\rho,m\})=\int \tilde{\mathcal{D}}\;\delta(S_0^{\rho}+\sqrt{a}\;\partial_x\eta_1)\delta(S_0^{m}+\sqrt{a}\;\partial_x\eta_2+\eta_3)\exp\bigg(\int-\frac{1}{2}
\begin{pmatrix}
    \eta_1 & \eta_2  & \eta_3  \\
\end{pmatrix}
\bar{M}^{-1}
\begin{pmatrix}
    \eta_1 \\ 
    \eta_2 \\
    \eta_3 \\
\end{pmatrix}dxdt\bigg)\;.
\end{align}
From \eref{eq:fluctuating_action_3}, we deduce the following fluctuating hydrodynamics for the magnetization and density fields
\begin{align}
\partial_t \rho =& D \nabla^2\rho - \frac{v}{2}\nabla(m+\rho) + \sqrt{a}\;\partial_x\eta_1 \\
\partial_t m =&  D \nabla^2m - \frac{v}{2}\nabla(m+\rho) - 2\gamma m e^{-\beta + (\cosh(\beta)-1)\rho}\left(\cosh\left[\sinh(\beta)m\right]m-\sinh\left[\sinh(\beta)m\right]\rho\right) + \sqrt{a}\; \partial_x\eta_2 +\sqrt{a}\;\eta_3\;,
\end{align}
where the correlation of the Gaussian noise fields $\eta_{n}$ ($n\in\{1,2,3\}$) is given by $\langle\eta_{n}\eta_{n^{\prime}}\rangle=\bar{M}_{n,n^{\prime}}$. 
A similar derivation can be performed on the actions $S^{1}$ and $S^{2}$ given by \eref{eq:action_species_1}-\eref{eq:action_species_2}. 
It yields the fluctuating hydrodynamics in the case of two species of spins interacting nonreciprocally as 
\begin{subequations}
\label{eq:fluct_hydro_app_NR}
\begin{align}
\partial_t \rho_1 =& D\partial_{xx}\rho_1 - \tfrac{v}{2} \partial_x (m_1+\rho_1) + \sqrt{a}\;\partial_x \eta^1_1 \\
  \partial_t m_1 =& D \partial_{xx} m_1 - \tfrac{1}{2} \partial_x (m_1+\rho_1) - l(\rho_1,\rho_2,m_1,m_2,\beta_a) + \sqrt{a}\;\partial_x \eta^1_2 + \sqrt{a\:\bar{l}(\rho_1,\rho_2,m_2,\beta_a)}\; \eta^1_3 \\
\partial_t \rho_2 =& D\partial_{xx}\rho_2 - \tfrac{v}{2} \partial_x (m_2+\rho_2) + \sqrt{a}\;\partial_x \eta^2_1 \\
  \partial_t m_2 =& D \partial_{xx} m_2 - \tfrac{v}{2} \partial_x (m_2+\rho_2) - l(\rho_2,\rho_1,m_2,m_1,-\beta_a) + \sqrt{a}\;\partial_x \eta^2_2 + \sqrt{a\:\bar{l}(\rho_2,\rho_1,m_1,-\beta_a)}\; \eta^2_3\; ,
\end{align}
\end{subequations}
where $l$ and $\bar{l}$ are given by \eref{eq:landau_NR_and_R} and \eref{eq:landau_NR_and_R_bar_l} respectively. 
In \eref{eq:fluct_hydro_app_NR}, the $\eta^{\alpha}_{n}$ ($\alpha\in\{1,2\}$ and $n\in\{1,2,3 \}$) are Gaussian white noises such that $\langle\eta^{\alpha}_{n}\eta^{\alpha^{\prime}}_{n'}\rangle=\delta_{\alpha,\alpha^{\prime}}M^{\alpha}_{n,n^{\prime}}$, with the species-dependent matrix $M^{\alpha}$ being given by \eref{eq:matrix_noise_beta0}.
 \section{Mean-field evolution in absence of self-propulsion ($v=0$)}
\label{app:hopfSNIC}

\begin{figure}
    \centering
    \def\setdimen#1#2{\newdimen#1\pgfmathsetlength{#1}{#2}}

\includegraphics{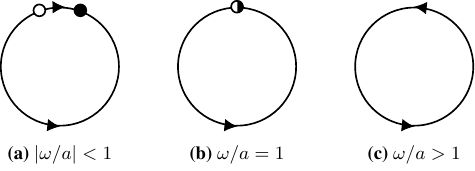}
    \caption{\textbf{Saddle-node on invariant circle.} In the Saddle-node on invariant circle (SNIC) bifurcation, a saddle-node bifurcation (panel b) merges two heteroclinic orbits (panel a) into a limit cycle (panel c).}
    \label{fig:snic}
\end{figure}

\begin{figure*}
        \centering
        \includegraphics{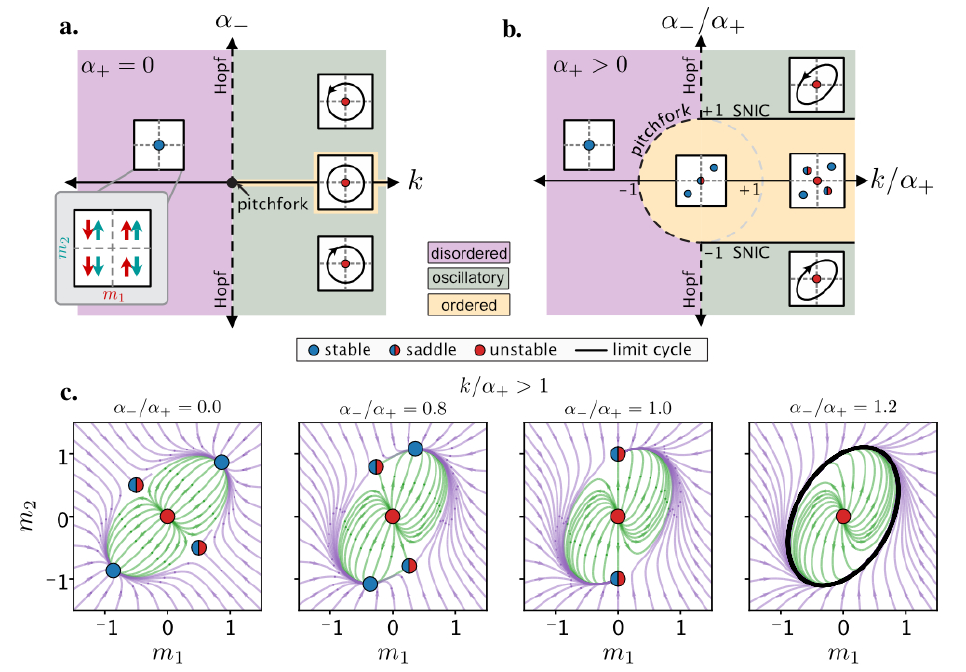}
        \caption{\textbf{Phase diagram for prototypical Hopf-SNIC system.} \textbf{a.} Phase diagram when $\alpha_+ = 0$ of Eq.~\ref{eq:ising_nonreciprocal_o2_main}. For $k < 0$, the only solution is the trivial fixed point, a.k.a. the disordered solution with $(m_1, m_2) = (0, 0)$ (purple). Crossing the $k = 0$ line, the system undergoes a pitchfork bifurcation if $\alpha_- = 0$, with steady states along a circle (orange). If the $k=0$ line is crossed with $\alpha_- \neq 0$, one instead undergoes a Hopf bifurcation to an oscillatory state (green). Schematics of the steady-state behavior in the different regimes are shown in small insets on an $(m_1, m_2)$ phase plane, illustrated by the gray inset. \textbf{b.} The phase diagram with $\alpha_+ \neq 0$ shows qualitative changes. While the three phases (disordered, oscillatory, and ordered) still exist, the range of parameters where an ordered solution exists is expanded. In addition, one can undergo a SNIC bifurcation by moving through the line $\alpha_- / \alpha_+ = \pm 1$ with $k/\alpha_+ > 0$. This bifurcation is entirely absent with $\alpha_+ = 0$. Within the circle defined by $k^2 + \alpha_-^2 = \alpha_+^2$, the trivial fixed point is a saddle-node. Outside of this area, the trivial fixed point is unstable ($k>0$) or stable ($k<0$). \textbf{c.} A series of phase portraits showing the flows (lines and arrows) and fixed points (from Eq.~\eqref{eq:ising_nonreciprocal_o2_fixedpts} in the Appendix) as $\alpha_- / \alpha_+$ increases at fixed $k / \alpha_+$. Here, $\alpha_+ = 0.5$ and $k / \alpha_+ = 2$. We see two pairs of fixed points, each containing one saddle and one stable node, approach each other as $\alpha_- / \alpha_+$ increases. When $\alpha_- / \alpha_+ = 1$, each pair of fixed points meet and annihilate, resulting in a limit cycle (thick black line), marking the SNIC bifurcation. In all subfigures, a stable fixed point is illustrated with a blue point, an unstable fixed point is illustrated with a red point, and a saddle is illustrated with a half-blue-half-red point. Limit cycles are illustrated with thick black lines.}
        \label{fig:o2phaseDiagram}
    \end{figure*}

\subsection{Details on SNIC bifurcations}
\label{subsec:SNIC_bifurcation}
A generic feature of SNIC bifurcations is the divergence of the oscillation period as the transition is approached. 
This is markedly different from a Hopf bifurcation, where the period of the oscillations remain constant close to the transition. Futhermore, SNIC bifurcations cannot be detected via a linear stability analysis at a single point in phase space. To see that, let us consider the simplest instantiation of a SNIC bifurcation: the dynamical system~\cite{Strogatz2015}
\begin{equation}
    \label{SNIC_toy}
    \dot{\theta} = f(\theta) \equiv \omega - a \sin \theta
\end{equation}
in which $\theta$ is the angle describing the position on the circle, and $\omega$ and $a$ are parameters. 
Fixed points $\theta^*$ of the dynamical system \eref{SNIC_toy} satisfy $f(\theta^*) = 0$. 
As shown in Fig.~\ref{fig:snic}, the dynamical system \eqref{SNIC_toy} has two fixed points (a saddle point and a node) when $|\omega|<|a|$. When $|\omega|>|a|$, it has a limit cycle and no fixed point. The transition between these regimes occurs through the collision of the saddle point with the node: this is the SNIC bifurcation. The period of the oscillations can be computed as~\cite{Strogatz2015}
\begin{equation}
    T = \int_{0}^{2\pi} \frac{\dd \theta}{f(\theta)}
    = \frac{2\pi}{\sqrt{\omega^2 - a^2}}
\end{equation}
which diverges as $T \sim |\omega - a|^{-1/2}$ upon approaching the bifurcation.
In the mean-field of the NRASM \eref{eq:system_zero_d}, note that the limit cycle emerges as four saddle-node collisions occur at the same time (one in each quadrant of the $m_1$-$m_2$ plane): this is because the underlying symmetry of the dynamical system is four-fold.
In the next appendix, we present an analytically tractable model inspired by, but distinct from, the NRASM that exhibits both SNIC and Hopf bifurcations, thereby illustrating their generic features.

\subsection{A simplified model for the mean-field NRASM}
\label{subsec:simple_model_mf_nrasm}

In order to illustrate the interplay between Hopf and SNIC bifurcations, we now introduce a minimal analytically solvable model capturing some essential features of Eq.~\eqref{eq:system_zero_d}. We consider two non-conserved variables $m_1, m_2$ whose time evolution obeys 
\begin{equation}
\label{eq:ising_nonreciprocal_o2_main}
    \dot{m}_i = A_{ij} m_j - |\mathbf{m}|^2 m_i; \quad \mathbb{A} = \begin{pmatrix}
        k & \alpha_+ - \alpha_- \\
        \alpha_+ + \alpha_- & k
    \end{pmatrix}.
\end{equation}
where $i = \lbrace 1, 2 \rbrace$, $|\mathbf{m}|^2 = m_1^2 + m_2^2$, and we use Einstein notation to imply summation over repeated indices. The primary feature that makes this model analytically tractable is its invariance under the vector representation of $O(2)$ rotations of the magnetization vector $(m_1, m_2)$ when $\alpha_+ = \alpha_- = 0$, and under $SO(2)$ rotations when only $\alpha_+ = 0$ (see Ref.~\cite{Marques2013} for a detailed analysis of this case).
Finally, we note that a mean-field (zero-dimensional) model similar to \eref{eq:ising_nonreciprocal_o2_main} has been studied in Ref.~\cite{Guislain2023}

It is simplest to see that \eref{eq:ising_nonreciprocal_o2_main} contains both a Hopf and a SNIC bifurcation by converting from Cartesian coordinates, $(m_1, m_2)$, to polar coordinates, $(r, \theta)$, using $m_1 = r \cos(\theta), m_2 = r \sin(\theta)$. Doing so, Eq.~\eqref{eq:ising_nonreciprocal_o2_main} becomes
\begin{subequations}
\label{eq:hopfSNICbis}
\begin{align}
     \dot{r} &= r(k + \alpha_+ \sin(2\theta)) - r^3 \label{eq:hopfNFbis}
    \\
     \dot{\theta} &= \alpha_- + \alpha_+ \cos(2 \theta). \label{eq:snicNFbis}
\end{align}
\end{subequations}
When $\alpha_+ = 0$ and $\alpha_- \neq 0$, we recognize Eq.~\eqref{eq:hopfSNICbis} as the normal form of a Hopf bifurcation with $k$ as the bifurcation parameter. 
When $\alpha_+ \neq 0$, the Hopf bifurcation remains, but becomes azimuthally asymmetric. In addition, Eq.~\eqref{eq:snicNFbis} is similar to the normal form a SNIC bifurcation, up to a rotation $\theta \to \theta + \pi / 4$ of the coordinate system, which leads to $\dot{\theta} = \alpha_- - \alpha_+ \sin(2 \theta)$ (compare to Eq.~\eqref{SNIC_toy}).

The factor of 2 in the trigonometric functions of Eq.~\eqref{eq:hopfSNICbis} reflects an additional $\mathbb{Z}_2$ symmetry compared to the normal form of a SNIC bifurcation \eqref{SNIC_toy}: two saddle-node collisions take place at the same time. This originates from the $\mathbb{Z}_2$ reflection symmetry of our variables. In Eq.~\eqref{eq:system_zero_d} of the main text, a similar situation arises but with four saddle-node collisions instead of two.

We now study the stability of the fixed points associated to \eqref{eq:ising_nonreciprocal_o2_main}.
Going back to Cartesian coordinates again and writing $\dot{m}_i= f_i(\mathbf{m})$, the Jacobian matrix $\mathbb{J}$ with elements $J_{ij} = (\partial f_i / \partial m_j)$ evaluated around a point $\mathbf{m}$ is given by
\begin{equation}
    \label{eq:ising_nonreciprocal_o2_jacbian}
    \mathbb{J} = \begin{pmatrix}
    k - 3 m_1^2 - m_2^2 & -2 m_1 m_2 + \alpha_+ - \alpha_- \\
    -2 m_1 m_2 + \alpha_+ + \alpha_- & k - 3 m_2^2 - m_1^2
    \end{pmatrix}\;.
\end{equation}
The corresponding eigenvalues read 
\begin{align}
    \lambda_\pm(\mathbf{m}) &= \dfrac{\mathrm{Tr}(\mathbb{J})}{2} \pm \sqrt{ \left( \dfrac{\mathrm{Tr}(\mathbb{J})}{2} \right)^2 - \mathrm{det}(\mathbb{J})}= k - 2(m_1^2 + m_2^2) \pm \sqrt{(m_1^2 + m_2^2)^2 - \alpha_-^2 + \alpha_+^2 - 4 \alpha_+ m_1 m_2}.
\end{align}

First, we will study the eigenvalues evaluated at the disordered solution $\mathbf{m}^*_0 = (0, 0)$, which are given by
\begin{equation}
    \lambda_\pm(\mathbf{m}_0^*) \equiv \lambda_\pm^0 =  k \pm \sqrt{\alpha_+^2 - \alpha_-^2}.
\end{equation}

When $\alpha_+ = 0$, we see that the eigenvalues become complex when $\alpha_- \neq 0$, and they gain a positive real part when $k > 0$. 
This is characteristic of a Hopf bifurcation. If $\alpha_- = 0$, the system instead undergoes a so-called circle-pitchfork bifurcation~\cite{Kness1992,Marques2013,Crawford1991} as shown in \Fref{fig:o2phaseDiagram}a.

With $\alpha_+ > 0$, the eigenvalues are purely real when $|\alpha_-| < |\alpha_+|$. In this regime, the disordered state loses its stability when the eigenvalues vanish, $\lambda_\pm^0 = 0$, giving the condition
\begin{align}
    k \pm \sqrt{\alpha_+^2 - \alpha_-^2} = 0
    \Rightarrow \dfrac{k^2}{\alpha_+^2} + \dfrac{\alpha_-^2}{\alpha_+^2} = 1.
\end{align}
This is the equation for the unit circle in the $(k/\alpha_+, \alpha_- / \alpha_+)$ plane. Outside this circle with $k < 0$, the disordered state is stable. Passing into the circle, the system undergoes a pitchfork bifurcation and finds one of two ordered solutions with $|\mathbf{m}^*| > 0$. Note that this is not the same as the circle-pitchfork bifurcation mentioned above because the nonzero $\alpha_+$ breaks the rotational symmetry of Eq.~\eqref{eq:ising_nonreciprocal_o2_main}. Inside the circle, $\lambda_+^0 > 0$ and $\lambda_-^0 < 0$, making the disordered state a saddle. Outside the circle with $k > 0$, the disordered state is fully unstable, and two other saddle nodes appear (\Fref{fig:o2phaseDiagram}b., purple and orange regions). We will give explicit expressions for these solutions below.

The system transitions to an oscillatory state when $|\alpha_-| > |\alpha_+|$ with $\mathrm{Re}(\lambda_\pm^0) > 0$. This can occur either directly from the disordered state, in which case the system undergoes a Hopf Bifurcation (\Fref{fig:o2phaseDiagram}b. moving from the purple region to the green region), or from the ordered state, in which case the system undergoes a SNIC bifurcation (\Fref{fig:o2phaseDiagram}b. moving from the orange region to the green region). The SNIC bifurcation occurs when the two saddle nodes each annihilate with a stable fixed point (\Fref{fig:o2phaseDiagram}c.). We stress that the SNIC bifurcation cannot be determined locally from the eigenvalues $\lambda_\pm$. We require $\mathrm{Re}(\lambda_\pm^0)  > 0$ only to ensure that the disordered state is not a stable solution. The SNIC bifurcation is a saddle-node bifurcation that occurs on a closed path $\cC$ called an \textit{invariant circle} because $\cC$ is left invariant by the dynamics. While the saddle-node bifurcation can be determined from a linear stability analysis, the fact that it occurs on $\cC$ cannot.

Using the Cartesian representation of our equations of motion, Eq.~\ref{eq:ising_nonreciprocal_o2_main}, we can explicitly find the positions of the non-trivial solutions when $|\alpha_-| < |\alpha_+|$. The SNIC bifurcation occurs when these solutions become complex, and thereby unphysical. We first give the answer:
\begin{equation}
\label{eq:ising_nonreciprocal_o2_fixedpts}
    \mathbf{m}^*_{\pm, \mp} = \mp \left( \dfrac{k \pm \sqrt{\alpha_+^2 - \alpha_-^2}}{2} \right)^{1/2}
    \begin{pmatrix}
        \left( 1 - \dfrac{\alpha_-}{\alpha_+}\right)^{1/2} \\
        \pm \left( 1 + \dfrac{\alpha_-}{\alpha_+}\right)^{1/2}
    \end{pmatrix}
\end{equation}
In the above, the notation $\mathbf{m}^*_{\pm, \mp}$ indicates that the two $\pm$ signs in Eq.~\ref{eq:ising_nonreciprocal_o2_fixedpts} must be assigned together, but separately from the $\mp$ sign, giving the promised 4 solutions \footnote{These solutions are $\mathbf{m}^*_{+, +}$, $\mathbf{m}^*_{+, -}$, $\mathbf{m}^*_{-, +}$, and $\mathbf{m}^*_{-, -}$}. We recognize the prefactor as $\lambda_\pm^0$, which emphasizes the fact that we require $\lambda_\pm^0 \in \mathbb{R}_+$ in order to find non-trivial solutions.

Let us now derive \eref{eq:ising_nonreciprocal_o2_fixedpts}. 
Our approach will be to assume the ansatz $(m_1, m_2) = m ( 1 , c )$ before solving for $c$, and $m$. Written explicitly, Eq.~\eref{eq:ising_nonreciprocal_o2_main} is
\begin{equation*}
    \begin{aligned}
        \partial_t m_1 &= (k - (m_1^2 + m_2^2))m_1 + (\alpha_+ - \alpha_-) m_2 = f_1 \\
        \partial_t m_2 &= (k - (m_1^2 + m_2^2))m_2 + (\alpha_+ + \alpha_-) m_1 = f_2\;,
    \end{aligned}
\end{equation*}
where, in steady state, we have $f_1 = f_2 = 0$.

Taking $m_1 f_1 + m_2 f_2 = 0$ yields
\begin{align}
\label{eq:interm_danny_ansatz_1}
    0 &= (k - (m_1^2 + m_2^2))(m_1^2 + m_2^2) + 2 \alpha_+ m_1 m_2
\end{align}
Plugging our ansatz in \eref{eq:interm_danny_ansatz_1}, we obtain $m$ as
\begin{align}
\label{eq:msquared1}
    m^2 = \dfrac{1}{1+c^2} \left( k + \dfrac{2 \alpha_+ c}{1 + c^2} \right)\;.
\end{align}
Similarly, taking $m_1 f_1 - m_2 f_2 = 0$, we also find
\begin{align}
\label{eq:msquared2}
    m^2 &= \dfrac{1}{1+c^2} \left( k - \dfrac{2 \alpha_- c}{1 - c^2} \right)\;.
\end{align}
Equating Eqs.~\eref{eq:msquared1} and \eref{eq:msquared2} then yields $c$ as
\begin{equation}
    \label{eq:solution_for_c}
    c = \pm \sqrt{\dfrac{\alpha_+ + \alpha_-}{\alpha_+ - \alpha_-}} \equiv \pm c^*.
\end{equation}
Plugging Eq.~\eref{eq:solution_for_c} into Eq.~\eref{eq:msquared1} (the same solution is found if we instead plug into Eq.~\eref{eq:msquared2}) gives
\begin{align}
        m^2 &= \dfrac{1}{1 + (c^*)^2} \left( k \pm \dfrac{2 \alpha_+ c^*}{1 + (c^*)^2} \right)
= \dfrac{1}{2} \left( 1 - \dfrac{\alpha_-}{\alpha_+} \right) \left( k \pm \sqrt{\alpha_+^2 - \alpha_-^2} \right)\;.
\end{align}
We therefore deduce the expression of $m$ as 
\begin{align}
\label{eq:final_m1_app}
    m_1 = m_{\pm, \mp} &= \mp \left( \dfrac{ k \pm \sqrt{\alpha_+^2 - \alpha_-^2}}{2} \right)^{1/2} \left( 1 - \dfrac{\alpha_-}{\alpha_+} \right)^{1/2}\;.
\end{align}
Note that we have distinguished between the sign of $c^*$ ($\pm$) and the sign of the square root of $m$ ($\mp$). This is for denoting that each signs can be chosen independently, hence giving 4 separate solutions. 
Now setting $m_2 = \pm c^* m_{\pm, \mp}$ and taking care that the sign of $c^*$ chosen is the same as the one in the term $k \pm \sqrt{\alpha_+^2 - \alpha_-^2}$, we obtain
\begin{align}
\label{eq:final_m2_app}
        m_2 
        &= \mp
        \left( \dfrac{ k \pm \sqrt{\alpha_+^2 - \alpha_-^2}}{2}\right)^{1/2}
        \left( \pm \left( \dfrac{\alpha_+ + \alpha_-}{\alpha_+ - \alpha_-} \right)^{1/2} \right)
        \left( 1 - \dfrac{\alpha_-}{\alpha_+} \right) ^{1/2} &=  \mp \left( \dfrac{ k \pm \sqrt{\alpha_+^2 - \alpha_-^2}}{2}\right)^{1/2} \left( \pm \left( 1 + \dfrac{\alpha_-}{\alpha_+} \right) ^{1/2} \right)\;.
\end{align}
Using \eref{eq:final_m2_app} and \eref{eq:final_m1_app}, we finally arrive at the solutions presented in Eq.~\eref{eq:ising_nonreciprocal_o2_fixedpts},
\begin{equation*}
    \mathbf{m}_{\pm, \mp} = \mp \left( \dfrac{k \pm \sqrt{\alpha_+^2 - \alpha_-^2}}{2} \right)^{1/2}
    \begin{pmatrix}
        \sqrt{1 - \dfrac{\alpha_-}{\alpha_+}}
        \\
        \pm \sqrt{1 + \dfrac{\alpha_-}{\alpha_+} }
    \end{pmatrix}\;.
\end{equation*}

 \section{Derivation of nonreciprocal bands}
\label{app:NR_bands}
In this appendix, we rederive system \eref{eq:dyn_sys_single} and its solution \eref{eq:wave_form_single} with parameters \eref{eq:wave_param_single}.
We then assess perturbatively how it is affected by nonreciprocity and derive \eref{eq:results_first_order} of main text.
Inserting \eref{eq:wave_form_single} into \eref{eq:dyn_sys_single}, we find
\begin{align}
  \label{eq:solvab_cond}
  0=g_0(m_l,c,\rho_g)\frac{1+\tanh(k_{\pm}\zeta)}{2} + \frac{1}{8c} g_1(k_{\pm},m_l,c) \sech(k_{\pm}\zeta)^2 + \frac{1}{8}g_2(k_{\pm},m_l)\sech(k_{\pm}\zeta)^2\tanh(k_{\pm}\zeta)\;,
\end{align}
where $g_0$, $g_1$ and $g_2$ are given by
\begin{align}
  g_0(m_l,c,\rho_g)=&\frac{dH}{dm}\bigg{|}_{m_l}=-\left(\varphi_g - \rho_g \right)m_l + \frac{v_0}{c}m_l^2 - a_4 m_l^3\\
  g_1(k_{\pm},m_l,c)=& 3 a_4 c\; m_l^3+4 c^2 k_{\pm} m_l-2 c m_l^2 k \xi -2 m_l v (2 k_{\pm} \lambda +m_l) \\
  g_2(k_{\pm},m_l)=& a_4 m_l^3-2 k \xi m_l^2- 8 m_l D k^2\;.
\end{align}
Condition \eref{eq:solvab_cond} is verified for every $\zeta$ whenever all the $g_i$'s vanish.
Therefore, the wave's parameters $k_{\pm}$, $m_l$, $c$ and $\rho_g$ are such that
\begin{align}
  \label{eq:system_cond_solvab}
  g_0(m_l,c,\rho_g)=&0, & g_1(k_{\pm},m_l,c)=&0, & g_2(k_{\pm},m_l)=&0\;.
\end{align}
One can then check that system \eref{eq:system_cond_solvab} indeed admits \eref{eq:wave_param_single} of main text as a solution.
We now assess perturbatively how the single species solution \eref{eq:wave_form_single} with parameters \eref{eq:wave_param_single} is modified by nonreciprocity.
To this aim, we insert \eref{eq:wave_form_NR} into \eref{eq:dyn_sys_NR-NL} and derive the new solvability condition.
For species $1$, we obtain that
\begin{align}
  \label{eq:solvab_cond_species_1}
  \nonumber
  0=&\frac{1}{2}\left[g_0(m^l_1,c_1,\rho^1_g)-\beta_a m^l_2+\beta_a\alpha m^l_2 m^l_1\right] + \frac{1}{2}\left[g_0(m^l_1,c_1,\rho^1_g)-\beta_a m^l_2+\beta_a\alpha m^l_2 m^l_1\right]\tanh(k_{\pm}^1\zeta)\\ &+ \frac{1}{8c_1}\left[ g_1(k_{\pm}^1,m_1^l,c_1) + 2 c_1\beta_a\alpha m_1^l m_2^l\right] \sech(k_{\pm}^1\zeta)^2 + \frac{1}{8}g_2(k_{\pm}^1,m^l_1,c_1)\sech(k_{\pm}^1\zeta)^2\tanh(k_{\pm}^1\zeta)\;.
\end{align}
The solvability condition \eref{eq:solvab_cond_species_1} for species $1$ is therefore verified whenever
\begin{align}
  \label{eq:solvab_system_species_1}
  g_0(m^l_1,c_1,\rho^1_g)=&\beta_a m_2^l-\beta_a \alpha m_1^lm_2^l, & g_1(k^1_{\pm},m^l_1,c_1)=& -2c_1\beta_a\alpha m_2^l m_1^l, & g_2(k^1_{\pm},m^l_1,c_1)=&0\;.
\end{align}
Assuming weak nonreciprocity ($\beta_a\simeq 0$), we can Taylor expand system \eref{eq:solvab_system_species_1} around the single species solution by using expansion \eref{eq:wave_param_perturb}.
Equating terms at first order in $\beta_a$, we find
\begin{align}
  \label{eq:solvab_system_species_1_TE}
  \begin{cases}
    \displaystyle
  \frac{\partial g_0}{ \partial m_l } \delta m_1^l + \frac{\partial g_0}{ \partial c } \delta c_1 + \frac{\partial g_0}{ \partial \rho_g } \delta \rho^1_g = m_l - \alpha m_l^2 \\[10pt]
  \displaystyle
  \frac{\partial g_1}{ \partial k_{\pm} } \delta k^1_{\pm} + \frac{\partial g_1}{ \partial m_l } \delta m_1^l + \frac{\partial g_1}{ \partial c} \delta c_1 = -2 c \alpha m_l^2 \\[10pt]
  \displaystyle
  \frac{\partial g_2}{ \partial k_{\pm} } \delta k^1_{\pm} + \frac{\partial g_2}{ \partial m_l } \delta m_1^l =0\;,
  \end{cases}
\end{align}
where all derivatives on the left hand side are taken at $m_l$, $c$, $k_{\pm}$ and $\rho_g$ as defined in \eref{eq:wave_param_single}.
Using the symmetry $\beta_a \leftrightarrow -\beta_a$, we also deduce the solvability condition equivalent to \eref{eq:solvab_system_species_1_TE} for species 2
\begin{align}
  \label{eq:solvab_system_species_2_TE}
  \begin{cases}
    \displaystyle
  \frac{\partial g_0}{ \partial m_l } \delta m_2^l + \frac{\partial g_0}{ \partial c } \delta c_2 + \frac{\partial g_0}{ \partial \rho_g } \delta \rho^2_g = -m_l +\alpha m_l^2 \\[10pt]
  \displaystyle
  \frac{\partial g_1}{ \partial k_{\pm} } \delta k^2_{\pm} + \frac{\partial g_1}{ \partial m_l } \delta m_2^l + \frac{\partial g_1}{ \partial c} \delta c_2 = 2 c \alpha m_l^2 \\[10pt]
  \displaystyle
  \frac{\partial g_2}{ \partial k_{\pm} } \delta k^2_{\pm} + \frac{\partial g_2}{ \partial m_l } \delta m_2^l = 0\;,
  \end{cases}
\end{align}
where all derivatives on the left hand side are, once again, taken at $m_l$, $c$, $k_{\pm}$ and $\rho_g$ as defined in \eref{eq:wave_param_single}.
We now make the following change of variables
\begin{align}
  U_m=&\frac{\delta m_1^l + \delta m_2^l}{2}, & U_c=&\frac{\delta c_1 + \delta c_2}{2} , & U_k=&\frac{\delta k^1_{\pm} + \delta k^2_{\pm}}{2} , & U_{\rho}=&\frac{\delta \rho^1_g + \delta \rho^2_g}{2} \\
  V_m=&\frac{\delta m_1^l - \delta m_2^l}{2}, & V_c=&\frac{\delta c_1 - \delta c_2}{2} , & V_k=&\frac{\delta k^1_{\pm} - \delta k^2_{\pm}}{2} , & V_{\rho}=&\frac{\delta \rho^1_g - \delta \rho^2_g}{2}\;,
\end{align}
and further defines the vectors $U$, $V$ and $S$ as
\begin{align}
  U=&\begin{pmatrix}
    U_c \\ U_k \\ U_m \\ U_{\rho}
  \end{pmatrix}, & V=&\begin{pmatrix}
    V_c \\ V_k \\ V_m \\ V_{\rho}
  \end{pmatrix}, & S=&\begin{pmatrix}
    m_l-\alpha m_l^2 \\ 2 c \alpha m_l \\ 0
  \end{pmatrix}\;.
\end{align}
Summing together \eref{eq:solvab_system_species_1_TE} and \eref{eq:solvab_system_species_2_TE} before dividing by $2$ then yields
\begin{align}
  \label{eq:u_system}
  J_{\pm}U=0\;,
\end{align}
while substracting them before dividing by $2$ yields
\begin{align}
  \label{eq:v_system}
  J_{\pm}V=S\;,
\end{align}
where $J$ is akin to a Jacobian matrix which is given by
{
\everymath{\displaystyle}
\begin{align}
  \label{eq:J_def}
  J_{\pm}=\begin{pmatrix}
    \frac{\partial g_0}{\partial c} & 0 & \frac{\partial g_0}{\partial m_l} & \frac{\partial g_0}{\partial \rho_g} \\
    \frac{\partial g_1}{\partial c} & \frac{\partial g_1}{\partial k_{\pm}} & \frac{\partial g_1}{\partial m_l} & 0 \\
    0 & \frac{\partial g_2}{\partial k_{\pm}} & \frac{\partial g_2}{\partial m_l} & 0 \\
  \end{pmatrix}=\begin{pmatrix}
    \frac{\partial g_0}{\partial c} & 0 & \frac{\partial g_0}{\partial m_l} & \frac{\partial g_0}{\partial \rho_g} \\
    \frac{\partial g_1}{\partial c} & 0 & \frac{\partial g_1}{\partial m_l} & 0 \\
    0 & \frac{\partial g_2}{\partial k_{\pm}} & \frac{\partial g_2}{\partial m_l} & 0 \\
  \end{pmatrix}\;.
\end{align}
}
To obtain the last equality in \eref{eq:J_def}, we have used that $\displaystyle\frac{\partial g_1}{\partial k_{\pm}}$ vanishes at $(c,k_{\pm},m_l)$.
Let us first consider the case of linear nonreciprocity when $\alpha=0$.
As we are looking for steady-state fronts, we can further set the component $V_c$ to zero.
In this regime, system \eref{eq:v_system} becomes a $3\times 3$ system which reads
{
\everymath{\displaystyle}
\begin{align}
  \label{eq:system_alpha_0}
\begin{pmatrix}
  0 & \frac{\partial g_0}{\partial m_l} & \frac{\partial g_0}{\partial \rho_g} \\[8pt]
  0 & \frac{\partial g_1}{\partial m_l} & 0 \\[8pt]
  \frac{\partial g_2}{\partial k_{\pm}} & \frac{\partial g_2}{\partial m_l} & 0
\end{pmatrix} \begin{pmatrix}
  V_k \\[20pt] V_m \\[20pt] V_{\rho}
\end{pmatrix}=\begin{pmatrix}
  m_l \\[20pt] 0 \\[20pt] 0
\end{pmatrix}
\end{align}
}
Note that, as $g_0$ does not depend on $k_{\pm}$, the solution of \eref{eq:system_alpha_0} for both the upward front $k_{+}$ and the downward front $k_{-}$ is then simply
\begin{align}
  \label{eq:sol_v_alpha_0}
  V_k&=0, & V_{m}&=0 & V_{\rho}&=\frac{m_l}{\frac{\partial g_0}{\partial \rho_g}} =-1\;.
\end{align}
We now have to look for the solution of the linear system \eref{eq:u_system}.
Considering the $\pm$ cases, there is a total of 5 different equations in \eref{eq:u_system} (remember that $g_0$ does not depend on $k_{\pm}$).
As we are considering steady-state bands, $U_c$, $U_m$ and $U_{\rho}$ have to be independent of the $\pm$ front selected.
This is not the case for $U_{k_{\pm}}$ which can take different values $U_{k_{+}}$ and $U_{k_{-}}$ depending on the front considered.
Therefore, we also have a total of 5 different variables.
Because system \eref{eq:u_system} has an equal number of equations and variables as well as no source term, its only solution is zero
\begin{align}
  \label{eq:sol_u_alpha_0}
  U_c=U_m=U_{k_{+}}=U_{k_{-}}=U_{\rho}=0\;.
\end{align}
With \eref{eq:sol_v_alpha_0}-\eref{eq:sol_u_alpha_0} we have just shown that, in the case $\alpha=0$ and at first order in $\beta_a$, the single-species flocking bands are modified according to \eref{eq:results_first_order} of main text.
\par
Let us now study the case $\alpha\neq 0$.
We first have to look for the solution of the linear system \eref{eq:v_system}.
Considering the $\pm$ cases, there is a total of 5 different equations in \eref{eq:u_system}.
As we are considering steady-state bands, $V_c$, $V_m$ and $V_{\rho}$ have to be independent of the $\pm$ front selected.
This is not the case for $V_{k_{\pm}}$ which can take different values $V_{k_{+}}$ and $V_{k_{-}}$ depending on the front considered.
We thus also have a total of 5 different variables.
System \eref{eq:v_system} has an equal number of equations and variables as well as a source term.
Its only solution turns out to be such that $V_c \neq 0$, which in particular implies \eref{eq:diff_speed} of main text.
In particular, this means that whenever $\alpha\neq 0$, the bands of the two species have different velocities when they are located at the same position.
Therefore, in the steady-state, we expect to observe a stationary spatial shift induced by nonlinearity between the two bands.
This stands in contrast with the linear case $\alpha=0$.
 \section{Spatial synchronization of bands}
\label{app:synchronization}
In this appendix we derive \eref{eq:synchronization} of main text.
We assume that we initially have two overlapping flocking bands.
At first order in $\beta_a$, their profile is given by \eref{eq:wave_form_single} with \eref{eq:wave_param_single} being perturbed by non reciprocity according to \eref{eq:results_first_order}
\begin{align}
  \label{eq:rho_1_first_order}
  \rho_1=&\varphi_g+\beta_a+\frac{v m_l}{4c}\left[1+\tanh(k_{+}(\zeta-x_1(t)))\right]\left[1+\tanh(k_{-}(\zeta-L^b-x_1(t)))\right]\\
  \label{eq:m_1_first_order}
  m_1=&\frac{m_l}{4}\left[1+\tanh(k_{+}(\zeta-x_1(t)))\right]\left[1+\tanh(k_{-}(\zeta-L^b-x_1(t)))\right]\\
  \label{eq:rho_2_first_order}
  \rho_2=&\varphi_g-\beta_a+\frac{v ml}{4c}\left[1+\tanh(k_{+}(\zeta-x_2(t)))\right]\left[1+\tanh(k_{-}(\zeta-L^b-x_2(t)))\right]\\
  \label{eq:m_2_first_order}
  m_2=&\frac{m_l}{4}\left[1+\tanh(k_{+}(\zeta-x_2(t)))\right]\left[1+\tanh(k_{-}(\zeta-L^b-x_2(t)))\right]\;,
\end{align}
where $\zeta=x-ct$, $L^b$ is the bands' size and $x_1(t)$, $x_2(t)$ are the positions of the two bands.
Without loss of generality, we will assume that $x_2(t)>x_1(t)$.
Note that when $\beta_a=0$, we know that $x_1(t)$ and $x_2(t)$ remain constant.
We will therefore assume a separation of time scales such that $\dot{x_1}(t)=\beta_a dx_1/ds$ with $s=\beta_a t$ being an adimensionalized timescale.
Inserting \eref{eq:rho_1_first_order}, \eref{eq:m_1_first_order} and \eref{eq:m_2_first_order} into the second line of system \eref{eq:toner_tu_coupled_1} and integrating between $\zeta=x_1(t)$ and $\zeta=x_1(t)+L^b$, we obtain at first order in $\beta_a$
\begin{align}
  \label{eq:collective_x1}
  -\left(\int_{x_1(t)}^{x_1(t)+L^b}\frac{dm_1(\zeta)}{d\zeta}d\zeta\right)\frac{d x_1}{ds}= \int_{x_1(t)}^{x_1(t)+L^b}m_1(\zeta)d\zeta-\int_{x_1(t)}^{x_1(t)+L^b}m_2(\zeta)d\zeta\;.
\end{align}
Let us now perform further approximations. We consider first the integral on the left hand side that we denote $\hat{\alpha}_0$
\begin{align}
  \hat{\alpha}_0=&-\int_{x_1(t)}^{x_1(t)+L^b}\frac{dm_1(\zeta)}{d\zeta}d\zeta\\
  \hat{\alpha}_0\simeq&- \int_{x_1(t)}^{x_1(t)+L^b} m_l\frac{d}{d\zeta}\frac{1+\tanh(k_{+}(\zeta-x_1))}{2} - \int_{x_1(t)}^{x_1(t)+L^b} m_l\frac{d}{d\zeta}\frac{1+\tanh(k_{-}(\zeta-x_1-L^b))}{2} \\
  \label{eq:collective_coordinate_alpha_0}
  \hat{\alpha}_0\simeq& -\frac{m_l}{2}\left(\tanh(k_{+}L^b)+\tanh(k_{-}L^b)\right)
\end{align}
We now express the last two integrals on the right hand side of \eref{eq:collective_x1} as
\begin{align}
  \label{eq:collective_coordinate_alpha_1}
  \int_{x_1(t)}^{x_1(t)+L^b}m_1(\zeta)d\zeta \simeq& L^b m_l\;,\\
  \label{eq:collective_coordinate_alpha_2}
  \int_{x_1(t)}^{x_1(t)+L^b}m_2(\zeta)d\zeta \simeq& (L^b - x_2 + x_1)m_l\;.
\end{align}
Inserting \eref{eq:collective_coordinate_alpha_0}-\eref{eq:collective_coordinate_alpha_1}-\eref{eq:collective_coordinate_alpha_2} into \eref{eq:collective_x1}, we obtain
\begin{align}
  \label{eq:collective_coordinate_x1_final}
  \hat{\alpha}_{0} \frac{d x_1}{ds} = -(x_1-x_2)m_l\;.
\end{align}
Repeating the same step for species $2$, one shows that the equivalent equation for $x_2$ reads
\begin{align}
  \label{eq:collective_coordinate_x2_final}
  \hat{\alpha}_{0} \frac{d x_2}{ds} = -(x_2-x_1)m_l\;.
\end{align}
Subtracting \eref{eq:collective_coordinate_x2_final} to \eref{eq:collective_coordinate_x1_final}, we obtain the time evolution of the delay $\Delta(s)=x_2(s)-x_1(s)$ as
\begin{align}
  \label{eq:collective_coordinate_delay_final}
  \hat{\alpha}_{0}\frac{d \Delta}{ds} = - 2m_l\Delta\;,
\end{align}
which shows that $\Delta$ relaxes exponentially to zero: the two flocking bands spatially synchronize together.
\begin{figure}
\includegraphics{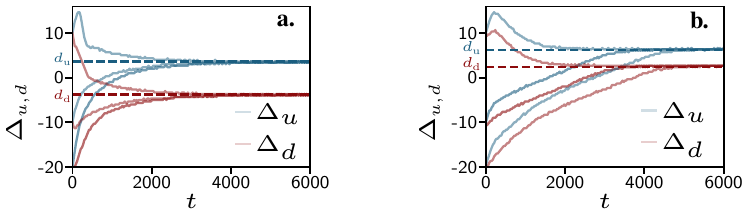}
\caption{Time-evolution of the spatial shifts $\Delta_u$ and $\Delta_d$ showing their respective convergence toward $d_{\rm{u}}$ and $d_{\rm{d}}$ for different initial values of $\Delta$. {\bf a.} Time-evolution of $\Delta_u$ and $\Delta_d$ for the linear nonreciprocal Toner-Tu model \eref{eq:toner_tu_coupled_1}. Parameters: $D=v=\phi_g=\lambda=\xi=a_4=1$, $\rho^1_0=\rho^2_0=1.1$, $dx=0.5$, $L=200$, $dt=0.002$, $\beta_a=0.02$. {\bf b.} Time-evolution of $\Delta_u$ and $\Delta_d$ for the NRASM \eref{eq:hydro_NR_and_R_non_fluct}. Parameters: $D=\gamma=1$, $v=2$, $\rho_0^1=\rho_0^2=1.2$, $L=100$, $dx=0.5$, $dt=0.002$, $\beta_a=0.02$.}
\end{figure}
 \section{Derivation of the travelling-wave ODE for the NRAIM}
\label{app:dynamical_NRAIM}
In this appendix, we derive the dynamical system \eref{eq:wave_nraim_1}-\eref{eq:wave_nraim_2} for travelling waves in the nonreciprocal active spin model.
We will focus on deriving \eref{eq:wave_nraim_1} but a similar path can be followed to obtain \eref{eq:wave_nraim_2}.
Following \cite{solon_pattern_2015}, we look for solitonic solutions of \eref{eq:hydro_NR_and_R_non_fluct} in the form $\rho_{\alpha}=\rho_{\alpha}(x-(c+v/2)t)$ and $m_{\alpha}=m_{\alpha}(x-(c+v/2)t)$ where $\alpha\in\{1,2\}$.
The first two lines of \eref{eq:hydro_NR_and_R_non_fluct} concerning the time-evolution of $\rho_1$ and $m_1$ then become
\begin{align}
  \label{eq:dyn_sys_nraim_1}
  0 =& D\rho_1^{\prime\prime}+c \rho_1^{\prime} -\frac{v}{2}m_1^{\prime} \\
  \label{eq:dyn_sys_nraim_2}
  0=& Dm_1^{\prime\prime}+c m_1^{\prime} -\frac{v}{2}\rho_1^{\prime}-2g(\rho_1,\rho_2,m_1,m_2)\;,
\end{align}
where prime indicates a derivative with respect to $\zeta=x-(c+v/2)t$.
Equation \eref{eq:dyn_sys_nraim_1} on the density is formally solved as
\begin{align}
  \rho_1=\rho^1_g+\frac{v}{2c}\sum_{k=0}^{\infty}\left(-\frac{D}{c}\right)^k m_1^{(k)}\;,
\end{align}
where $\rho_g^1$ is the density inside the disordered phase and the subscript $(k)$ indicates the $k$-th derivative with respect to $\zeta$.
As in \cite{solon_pattern_2015}, we will hereafter stop at second order in gradients and assume that
\begin{align}
  \label{eq:development_rho_1}
  \rho_1 &= \rho^1_g+\frac{v}{2c} m_1 - \frac{Dv}{2c^2} m_1^{\prime} + \frac{D^2v}{2c^3} m_1^{\prime\prime} \;, & \rho_1^{\prime}&= \frac{v}{2c} m_1^{\prime} - \frac{Dv}{2c^2} m_1^{\prime\prime}\;.
\end{align}
We now Taylor-expand \eref{eq:dyn_sys_nraim_2} at third order in $m_{\alpha}$ and first order in $\delta\rho_{\alpha}=\rho_{\alpha}-\rho_0$ ($\alpha \in \{1,2\}$).
We obtain
\begin{align}
  \nonumber
  0 =&- \frac{2}{6} \gamma e^{\rho_0 (\cosh (\beta )+\cosh (\beta_a))-\beta -2 \rho_0} \bigg{(}m_1^3\sinh^2(\beta)[3-\rho_0\sinh(\beta)]+m_1^2 m_2 \sinh(\beta ) \sinh (\beta_a)[3\rho_0\sinh(\beta)-6]
  \\
  \nonumber
  &+6 \delta\rho_1 m_1 [\cosh(\beta)-\sinh(\beta)-1+\rho_0\sinh(\beta)-\rho_0\cosh(\beta)\sinh(\beta)] -6 \delta\rho_2 m_1[\cosh(\beta_a)-1][\rho_0\sinh(\beta)-1]
  \\
  \nonumber
  &+6 m_1 [1- \rho_0 \sinh (\beta )] -3 m_1 m_2^2 \sinh^2 (\beta_a )[\rho_0\sinh(\beta)-1]+6\delta\rho_1 m_2 \sinh(\beta_a)[1-\rho_0+\rho_0\cosh(\beta)]
  \label{eq:unsimplified_dyn_sys_rho}
  \\ & +6 \delta\rho_2 m_2 \sinh(\beta_a)\rho_0[\cosh(\beta_a)-1]+ m_2^3 \rho_0 \sinh ^3(\beta_a) +6 m_2 \rho_0 \sinh (\beta_a)\bigg{)}+ Dm_1^{\prime\prime} + c m_1^{\prime}-\frac{v}{2}\delta\rho_1^{\prime}\;.
\end{align}
At first order in $\beta_a$, we can further simplify \eref{eq:unsimplified_dyn_sys_rho} into
\begin{align}
  \nonumber
  0 =&- \frac{2}{6} \gamma e^{\rho_0 (\cosh (\beta )+1)-\beta -2 \rho_0} \bigg{(}m_1^3\sinh^2(\beta)[3-\rho_0\sinh(\beta)]+m_1^2 m_2 \sinh(\beta )\beta_a[3\rho_0\sinh(\beta)-6]
  \\
  \nonumber
  &+6 \delta\rho_1 m_1 [\cosh(\beta)-\sinh(\beta)-1+\rho_0\sinh(\beta)-\rho_0\cosh(\beta)\sinh(\beta)]
  \\
  \label{eq:unsimplified_dyn_sys_rho_2}
  &+6 m_1 [1- \rho_0 \sinh (\beta )]+6\delta\rho_1 m_2 \beta_a[1-\rho_0+\rho_0\cosh(\beta)]
  +6 m_2 \rho_0 \beta_a\bigg{)}+ Dm_1^{\prime\prime} + c m_1^{\prime}-\frac{v}{2}\delta\rho_1^{\prime}\;.
\end{align}
As we are interested in the onset of travelling waves, we can further set $\beta=\argsh(\rho_0^{-1})$ in \eref{eq:unsimplified_dyn_sys_rho_2} to get
\begin{align}
  \nonumber
  0 =&- \frac{2}{6} \gamma e^{ \sqrt{1+\rho_0^2} -\argsh(\rho_0^{-1}) - \rho_0} \bigg{(}2\frac{m_1^3}{\rho_0^2}-3m_1^2 m_2 \frac{\beta_a}{\rho_0}-6 \frac{\delta\rho_1 m_1}{\rho_0}+
  6\delta\rho_1 m_2 \beta_a[1-\rho_0+\sqrt{1+\rho_0^2}]
  +6 m_2 \rho_0 \beta_a\bigg{)} \\
  \label{eq:unsimplified_dyn_sys_rho_3}
  &+ Dm_1^{\prime\prime} + c m_1^{\prime}-\frac{v}{2}\delta\rho_1^{\prime}\;.
\end{align}
Finally, because we will focus on the limit $\rho_0\to\infty$ where $\beta$ vanishes as $\beta\sim \rho_0^{-1}$, we also have to set $\beta_a=\bar{\beta}_a/\rho_0$ so that $\beta$ and $\beta_a$ have a similar scaling.
Doing so in \eref{eq:unsimplified_dyn_sys_rho_3} then yields
\begin{align}
  \nonumber
  0 =&- \frac{2}{6} \gamma e^{ \sqrt{1+\rho_0^2} -\argsh(\rho_0^{-1}) - \rho_0} \bigg{(}2\frac{m_1^3}{\rho_0^2}-3m_1^2 m_2 \frac{\bar{\beta}_a}{\rho_0^2}-6 \frac{\delta\rho_1 m_1}{\rho_0}+
  6\frac{\delta\rho_1 m_2}{\rho_0} \bar{\beta}_a[1-\rho_0+\sqrt{1+\rho_0^2}]
  +6 m_2 \bar{\beta}_a\bigg{)} \\
  \label{eq:unsimplified_dyn_sys_rho_4}
  &+ Dm_1^{\prime\prime} + c m_1^{\prime}-\frac{v}{2}\delta\rho_1^{\prime}\;.
\end{align}
Using \eref{eq:development_rho_1}, we can express $\delta\rho_1$ and $\delta\rho_1^{\prime}$ as functions of $m_1$ according to
\begin{align}
  \label{eq:development_delta_rho_1}
  \delta\rho_1 &= \rho^1_g-\rho_0 + \frac{v}{2c} m_1 - \frac{Dv}{2c^2} m_1^{\prime} + \frac{D^2v}{2c^3} m_1^{\prime\prime} \;, & \delta\rho_1^{\prime}&= \frac{v}{2c} m_1^{\prime} - \frac{Dv}{2c^2} m_1^{\prime\prime}\;.
\end{align}
Inserting \eref{eq:development_delta_rho_1} into \eref{eq:unsimplified_dyn_sys_rho_4}, we obtain
\begin{align}
  \label{eq:complicated_magn}
  \eta_1 m_1^{\prime\prime}=\frac{1}{D(1+\frac{v^2}{4c_1^2})}\Gamma_1 m_1^{\prime} + \frac{1}{D(1+\frac{v^2}{4c_1^2})}\alpha_1\;,
\end{align}
where $\eta_1$, $\Gamma_1$ and $\alpha_1$ are complicated functions of $m_1$, $m_2$, $\rho_0$ and the parameters.
Dividing \eref{eq:complicated_magn} by $\eta_1 4c_1^2/(D(4c_1^2+v^2))$, we obtain
\begin{align}
  \label{eq:complicated_magn_2}
  D\left(1+\frac{v^2}{4c_1^2}\right) m_1^{\prime\prime}=\frac{\Gamma_1}{\eta_1} m_1^{\prime} + \frac{\alpha_1}{\eta_1}\;.
\end{align}
We now Taylor expand $\Gamma_1/\eta_1$ at first order in $m_1$, $m_2$. We get
\begin{align}
  \label{eq:dev_friction}
  \frac{\Gamma_1}{\eta_1}=\Gamma_1^0 + \Gamma_1^{10}m_1 - \bar{\beta}_a\Gamma_1^{01}m_2\;.
\end{align}
Doing likewise for $\alpha_1/\eta_1$ at third order in $m_1$ and $m_2$, we get
\begin{align}
  \label{eq:dev_potential}
  \frac{\alpha_1}{\eta_1}=\alpha_1^{10}m_1- \bar{\beta}_a\alpha_1^{01}m_2 - \alpha_1^{20}m_1^2 + \alpha_1^{30}m_1^3 + \bar{\beta}_a\alpha_1^{11}m_2m_1- \bar{\beta}_a\alpha_1^{21}m_1^2m_2+\bar{\beta}_a\alpha_1^{02}m_2^2+\bar{\beta}_a\alpha_1^{03}m_2^3+\bar{\beta}_a\alpha_1^{12}m_1m_2^2\;.
\end{align}
Developing the $\alpha_1^{ij}$ in \eref{eq:dev_potential} at zero-th order in $\bar{\beta}_a$ before retaining only the first non vanishing order in $\rho_0$, we obtain
\begin{align}
  \label{eq:alphas_potential}
  \alpha_1^{10}=&2\gamma \Delta_1,\qquad  \alpha_1^{01}=2\gamma\left(\Delta_1-1\right),\qquad \alpha_1^{20}=\frac{\gamma  v \left(4 c_1^2+8 \gamma  \Delta_1 D+v^2\right)}{c_1 \rho_0 \left(4 c_1^2+v^2\right)},\qquad  \alpha_1^{11}=\gamma  \frac{v \left(4 c_1^2+8 \gamma  (2 \Delta_1-1) D+v^2\right)}{c_1 \rho_0 \left(4 c_1^2+v^2\right)}\\
  \alpha_1^{30}=&\frac{2 \gamma  \left(16 c_1^6+8 c_1^4 v^2+c_1^2 \left(24 \gamma  D v^2+v^4\right)+6 \gamma  D v^2 \left(8 \gamma  \Delta_1 D+v^2\right)\right)}{3 c_1^2 \rho_0^2 \left(4 c_1^2+v^2\right)^2},\qquad \alpha_1^{02}=0,\qquad \alpha_1^{03}=0,\qquad \alpha_1^{12}=0\\
  \alpha_1^{21}=&\gamma\frac{  \left(16 c_1^6+8 c_1^4 v^2+c_1^2 \left(32 \gamma  D v^2+v^4\right)+8 \gamma  D v^2 \left(4 \gamma  (3 \Delta_1-1) D+v^2\right)\right)}{c_1^2 \rho_0^2 \left(4 c_1^2+v^2\right)^2}\;,
\end{align}
where $\Delta_1=\lim_{\rho_0\to\infty}\frac{\rho_0-\rho_g^1}{\rho_0}$.
Doing likewise for the $\Gamma_1^{ij}$ in \eref{eq:dev_friction}, we get
\begin{align}
  \label{eq:alphas_friction}
  \Gamma_1^{10}=&8\gamma\frac{D v}{\rho_0 \left(4 c_1^2+v^2\right)},& \Gamma_1^{01}=&8 \gamma\frac{D v}{\rho_0 \left(4 c_1^2+v^2\right)},& \Gamma^{0}_1=& \frac{v^2-4 c_1^2}{4 c_1}\;.
\end{align}
We finally find that the travelling wave equation for $m_1(\zeta)$ is given by
\begin{align}
  \label{eq:dyn_sys_wave_m1}
  D\left(1+\frac{v^2}{4c_1^2}\right) m_1^{\prime\prime}=\left(\Gamma_1^0 + \Gamma_1^{10}m_1 - \bar{\beta}_a\Gamma_1^{01}m_2\right) m_1^{\prime} +\alpha_1^{10}m_1 - \bar{\beta}_a\alpha_1^{01}m_2 + \alpha_1^{20}m_1^2 + \alpha_1^{30}m_1^3 + \bar{\beta}_a\alpha_1^{11}m_2m_1- \bar{\beta}_a\alpha_1^{21}m_1^2m_2\;,
\end{align}
where the expression for the $\alpha^{ij}_1$, $\Gamma^{ij}_1$ are given in \eref{eq:alphas_potential} and \eref{eq:alphas_friction}.
We can further cast \eref{eq:dyn_sys_wave_m1} into
\begin{align}
  \label{eq:dyn_sys_wave_m1_fin}
  D\left(1+\frac{v^2}{4c_1^2}\right) m_1^{\prime\prime}&= -\left(\bar{f}_1(m_1)-\bar{\beta_a} \Gamma_1^{01} m_2\right)m^{\prime}_1 - \frac{d\bar{H}_1(m_1)}{dm_1} - \bar{\beta}_a \alpha_{1}^{01} m_2 + \bar{\beta}_a \alpha_{1}^{11} m_2 m_1 - \bar{\beta}_a \alpha_{1}^{21} m_1^2 m_2 \;,
\end{align}
where $\bar{f}_1(m_1)$ and $\bar{H}_1(m_1)$ are given by
\begin{align}
  \bar{f}_1(m_1)=&\Gamma_1^0+\Gamma_1^{10}m_1, & \bar{H}_1(m_1)=& -\frac{\alpha_1^{10}}{2}m^2_1 + \frac{\alpha_1^{20}}{3}m^3_1 - \frac{\alpha_1^{30}}{4}m^4_1\;.
\end{align}
To find the equation equivalent to \eref{eq:dyn_sys_wave_m1_fin} for $m_2(\zeta)$, one just has to use the symmetries $1\leftrightarrow 2$ (including in the expressions of the $\alpha_1^{ij}$ and $\Gamma_1^{ij}$) and $\bar{\beta}_a\leftrightarrow-\bar{\beta}_a$.
Note that \eref{eq:dyn_sys_wave_m1_fin} corresponds to equation \eref{eq:wave_nraim_1} reported in main text.
 \section{Floquet analysis}
\label{app:floquet}
In this appendix, we detail the Floquet analysis of \eref{eq:hydro_NR_and_R_non_fluct} around the limit cycle present in the homogeneous dynamical system \eref{eq:system_zero_d}.

\subsection{Linear stability analysis}

We perform a linear stability analysis of the hydrodynamic equations.
To this aim, we expand the magnetizations and the densities as
\begin{equation*}
  \rho_\alpha(t) = \rho_0^\alpha + \delta \rho_\alpha(x,t)
  \;
  \text{and}
  \;
  m_\alpha(t) = m_\alpha^0(t) + \delta m_\alpha(x,t)
\end{equation*}
in which $\alpha=1,2$ labels the species, and 
where $X_0(t)=(\rho_1^0, \rho_2^0, m_1^0, m_2^0)$ is a (possibly time-dependent) solution of \eref{eq:hydro_NR_and_R_non_fluct}.
In Fourier space, the linearized equations of motion of the deviations $\delta X(q) = (\delta m_1^q, \delta m_2^q, \delta \rho_1^q, \delta \rho_2^q)^T$ around this solution read 
\begin{equation}
  \label{eq_linearized}
  \partial_t \delta X(q) = J(X_0(t),q) \, \delta X(q)
\end{equation}
in which $J(X_0(t),q)$ is the Jacobian of the linearized equations of motion (given explicitly in Appendix \ref{app:floquet}), and $q$ is the wavevector. When the solution $X_0$ is a fixed point (i.e., does not depend on time), then the eigenvalues and eigenvectors of $J(X_0,q)$ determine its linear stability. When the solution is a limit cycle (namely, when there is some period $T$ such that $X_0(t) = X_0(t+T)$), then we need to use Floquet theory to determine the linear stability of the solution. Basically, this consists in analyzing instead the eigenvalues of the matrix $\OO_q(T)$, where $\OO_q(t)$ is the solution of Eq.~\eqref{eq_linearized} with initial condition $\OO_q(0) = \mathbb{I}_{2\times2}$ (the identity matrix). This is explained in detail in Appendix \ref{app:floquet_theory}.
This analysis shows that the two growth rates of the hydrodynamics Eq. \eref{eq:hydro_NR_and_R_non_fluct} in the non-motile case ($v=0$) are $\sigma_1^{\prime}=\sigma_1-Dq^2$ and $\sigma_2^{\prime}=\sigma_2-Dq^2$, where $\sigma_1$ and $\sigma_2$ are the growth rate that characterize the stability of the simpler non-extended dynamical system \eref{eq:system_zero_d}.

Explicitly, Eq.~\eqref{eq_linearized} of the main text reads
\begin{align}
  \label{eq:evolution_deviations}
  \partial_t \begin{pmatrix}
    \delta m_1^q \\
    \delta m_2^q \\
    \delta \rho_1^q \\
    \delta \rho_2^q \\
\end{pmatrix} = \begin{pmatrix}
  M^0(t)-Dq^2\mathbb{I}_{2\times2} & A(t)\\
  0 & -Dq^2\mathbb{I}_{2\times2} \\
\end{pmatrix}\begin{pmatrix}
  \delta  m_1^q \\
  \delta m_2^q \\
  \delta \rho_1^q \\
  \delta \rho_2^q \\
\end{pmatrix}\;,
\end{align}
where the subscript $q$ indicates the $q$-th Fourier component, $\mathbb{I}_{2\times2}$ is the two by two identity matrix, $M^0(t)$ is the stability matrix of the dynamical system
\begin{align}
  \label{eq:system_zero_d_app}
  \begin{cases}
  \dot{m}_1=&-2g(\rho_0^1,\rho_0^2,m_1,m_2) \\ \dot{m}_2=&-2g(\rho_0^2,\rho_0^1,m_2,-m_1)
  \end{cases}
\end{align}
and $A(t)$ is given by
\begin{align}
    A(t)=-2\begin{pmatrix}
  \partial g\left(\rho_0^1,\rho_0^2,m^0_1(t),m^0_2(t)\right)/\partial \rho_0^1 & \partial g\left(\rho_0^1,\rho_0^2,m^0_1(t),m^0_2(t)\right)/\partial \rho_0^2 \\
  \partial g\left(\rho_0^2,\rho_0^1,m^0_2(t),-m^0_1(t)\right)/\partial \rho_0^2 & \partial g\left(\rho_0^2,\rho_0^1,m^0_2(t),-m^0_1(t)\right)/\partial \rho_0^2 \\
    \end{pmatrix}\;.
\end{align}
The four growth rates of \eref{eq:evolution_deviations}, which we denote by $\sigma^{\prime}_1$, $\sigma^{\prime}_2$, $\sigma^{\prime}_3$ and $\sigma^{\prime}_4$, are given by the real part of the eigenvalues of the dynamical matrix in \eref{eq:evolution_deviations}.
From \eref{eq:evolution_deviations}, we remark that the density components $\delta\rho_1^q$ and $\delta\rho_2^q$ readily corresponds to two stable modes hence $\sigma^{\prime}_3=-Dq^2$ and $\sigma^{\prime}_4=-Dq^2$.
In steady-state $\delta\rho_1^q$ and $\delta\rho_2^q$ thus vanish and the remaining two growth rates are equal to the growth rates of a simpler system where the densities are kept constants:
\begin{align}
  \label{eq:evolution_deviations_simpler}
  \partial_t \begin{pmatrix}
    \delta m_1^q \\
    \delta m_2^q \\
\end{pmatrix} = \left(M^0(t)-Dq^2\mathbb{I}_{2\times2}\right)\begin{pmatrix}
  \delta  m_1^q \\
  \delta m_2^q \\
\end{pmatrix}\;.
\end{align}
The solution to \eref{eq:evolution_deviations_simpler} is given by
\begin{align}
  \begin{pmatrix}
    \delta  m_1^q(t) \\
    \delta m_2^q(t) \\
  \end{pmatrix}=\TExp[M^0(t)-Dq^2t\mathbb{I}_{2\times2}]\begin{pmatrix}
    \delta  m_1^q(0) \\
    \delta m_2^q(0) \\
  \end{pmatrix}
\end{align}
where $\TExp[.]$ stands for the time-ordered exponential.
Further using that $[M_0(t),\mathbb{I}_{2\times2}]=0$, we obtain that
\begin{align}
  \label{eq:simplified_solution_homogeneous_system}
  \begin{pmatrix}
    \delta  m_1^q(t) \\
    \delta m_2^q(t) \\
  \end{pmatrix}=e^{-Dq^2t}\mathbb{I}_{2\times2}\TExp[M^0(t)]\begin{pmatrix}
    \delta  m_1^q(0) \\
    \delta m_2^q(0) \\
  \end{pmatrix}\;.
\end{align}
Calling $\sigma_1$ and $\sigma_2$ the growth rates of the homogeneous dynamical system \eref{eq:system_zero_d} around $m_1^0(t)$ and $m_2^0(t)$, we can write $\TExp[M^0(t)]$ as
\begin{align}
  \label{eq:growth_time_exp}
  \TExp[M^0(t)]=P\begin{pmatrix}
    e^{\sigma_1 t} & 0 \\
    0 & e^{\sigma_2 t} \\
  \end{pmatrix}P^{-1}\;.
\end{align}
Inserting \eref{eq:growth_time_exp} into \eref{eq:simplified_solution_homogeneous_system}, we obtain
\begin{align}
  \label{eq:growth_final_app}
  \begin{pmatrix}
    \delta  m_1^q(t) \\
    \delta m_2^q(t) \\
  \end{pmatrix}=&P\begin{pmatrix}
    e^{\sigma_1 t-Dq^2t} & 0 \\
    0 & e^{\sigma_2 t-Dq^2t} \\
  \end{pmatrix}P^{-1}\begin{pmatrix}
    \delta  m_1^q(0) \\
    \delta m_2^q(0) \\
  \end{pmatrix}\;,
\end{align}
where $P$ is an unknown -and possibly time-dependent- change of basis matrix.

\subsection{Floquet theory}
\label{app:floquet_theory}
\begin{figure}
\includegraphics{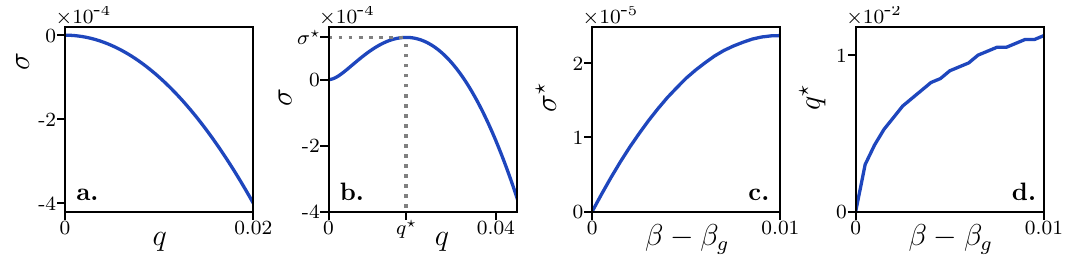}
\caption{Highest growth rate obtained from the numerical Floquet stability analysis of hydrodynamics \eref{eq:hydro_NR_and_R_non_fluct} around the limit cycle observed in the homogeneous dynamical system \eref{eq:system_zero_d}.
  {\bf a.} passive case, $v=0$. The periodic limit cycle remains stable.
  {\bf b.,c.,d.} active case, $v=1$.
  {\bf b.} the periodic limit cycle undergoes a type II instability at small wavelength.
  {\bf c.} most unstable growth rate $\sigma^{\star}$ as a function of the distance to the onset of existence $\beta_g$.
  {\bf d.} most unstable mode $q^{\star}$ as a function of the distance to the onset of existence $\beta_g$.
  Parameters: $\beta=0.9$, $\beta_a=0.4$, $D=\gamma=\rho^1_{0}=\rho^2_{0}=1$. We choose these parameters to lie inside the region $\beta_g<\beta<\beta_h$ of the phase diagram (see \Fref{fig:phase_diagram}) where none of the homogeneous constant phases are stable.}
  \label{fig:floquet_analysis}
\end{figure}

We start by recalling the Floquet method before detailing our numerical implementation.
Let us define the vectors $\bX_0(t)$ and $\delta \bX(x,t)$ as
\begin{align}
  \bX_0(t)=\begin{pmatrix}
    m_1^h(t) \\
    m_2^h(t) \\
    \rho^1_0 \\
    \rho^2_0 \\
\end{pmatrix}, \qquad \delta \bX(x,t)=\begin{pmatrix}
  \delta m_1(x,t) \\
  \delta m_2(x,t) \\
  \delta\rho^1(x,t) \\
  \delta\rho^2(x,t) \\
\end{pmatrix}\;,
\end{align}
with $m_1^h(t)$ and $m_2^h(t)$ being the $T$-periodic solution of dynamical system \eref{eq:system_zero_d}.
Linearizing \eref{eq:hydro_NR_and_R_non_fluct} around $\bX_0(t)$ in Fourier space then yields
\begin{align}
  \label{eq:linear_evolution_floquet}
  \partial_t \delta\bX_q = L_q(\bX_0(t))\delta\bX_q
\end{align}
where $\delta\bX_q$ is the $q$-th Fourier mode of $\delta \bX$ and $L_q(\bX_0(t))$ is a linear $T$-periodic operator.
Formally integrating \eref{eq:linear_evolution_floquet} between $t=0$ and $t=T$ gives
\begin{align}
  \delta\bX_q(T)=\OO_q (T)\delta \bX_q(0)
\end{align}
where $\OO_q(T)$ is a matrix whose $i$-th column corresponds to the solution of \eref{eq:linear_evolution_floquet} with an initial condition $\delta\XX^j_q(0)=\delta_{ij}$ taken at time $T$.
Using the $T$-periodicity of the operator $L_q(\bX_0(t))$, we can repeat this integration procedure to obtain the deviations at time $t=nT$,
\begin{align}
  \label{eq:floquet_total_evolution}
  \delta\bX_q(nT)=\OO_q (T)^n\delta \bX_q(0)\;,
\end{align}
with $n$ being a positive integer.
Decomposing $\OO_q (T)$ onto its eigenmodes, we can write it as
\begin{align}
  \label{eq:decomposition_O_operator}
  \OO_q (T)=P\begin{pmatrix}
    e^{\sigma_1 T} & 0 & 0 & 0 \\
    0 & e^{\sigma_2 T} & 0 & 0 \\
    0 & 0 & e^{\sigma_3 T} & 0 \\
    0 & 0 & 0 & e^{\sigma_4 T} \\
  \end{pmatrix}P^{-1}\;,
\end{align}
where $P$ is as an unspecified change of basis matrix and the $\sigma_j$ are the growth rates of $\OO_q (T)$.
Inserting \eref{eq:decomposition_O_operator} into \eref{eq:floquet_total_evolution}, we obtain the deviation at time $t=nT$ as
\begin{align}
    \label{eq:deviations_final}
  \delta\bX_q(nT)=P\begin{pmatrix}
    e^{\sigma_1 nT} & 0 & 0 & 0 \\
    0 & e^{\sigma_2 nT} & 0 & 0 \\
    0 & 0 & e^{\sigma_3 nT} & 0 \\
    0 & 0 & 0 & e^{\sigma_4 nT} \\
  \end{pmatrix}P^{-1}\delta \bX_q(0)\;.
\end{align}
We readily see on \eref{eq:deviations_final} that the stability of the deviation in the long time limit is determined by the growth rates of the operator $\OO_q (T)$.
However, in most cases, it is quite hard to determine $\OO_q (T)$ analytically as its construction involves solving a time-inhomogeneous differential equation.
Therefore, in this appendix, we will resort to a numerical determination of $\OO_q (T)$ and its corresponding growth rates.
We start by determining each column of $\OO_q (T)$ through a numerical integration of \eref{eq:linear_evolution_floquet}.
We then compute the growth rates of the obtained matrix by finding its eigenvalues.
In \Fref{fig:floquet_analysis}a. and b., we report the growth rate as a function of the wavenumber $q$ both in the passive ($v=0$) and active ($v=1$) cases.
We focused on parameters lying inside the region $\beta_g<\beta<\beta_h$ of the phase diagram where none of the homogeneous constant phases are stable.
We found that, while the oscillating solution $\bX_0(t)$ remains stable when $v=0$, it undergoes a type II instability at small $q$ when $v=1$.
In addition, in \Fref{fig:floquet_analysis}c. and d., we report the highest growth rate $\sigma^{\star}$ along with its corresponding wavenumber $q^{\star}$ as a function of the distance to the onset of flocking $\beta-\beta_g$.
We found that both $\sigma^{\star}$ and $q^{\star}$ vanish around the transition.
In particular, $q^{\star}$ seems to behave as $\sim \sqrt{\beta-\beta_g}$, which is also a signature of a type II instability.

\section{Derivation of time-correlation functions}
\label{app:fluctuating_hydro}
In this appendix, we use our fluctuating hydrodynamics \eref{eq:hydro_NR_and_R} to study the correlation functions of the magnetization fields in the disordered phase ($\beta<\beta_g$) when $v=0$.
In this regime, we know from section \ref{sec:phase_diagram} that the average thermodynamic fields are homogeneous and given by $\rho_{\alpha}=\rho_0$, $m_{\alpha}=0$.
Therefore, in the disordered phase, nonreciprocity do not affect the thermodynamic density and magnetization fields described by \eref{eq:hydro_NR_and_R_non_fluct}.
Nonetheless, finite system will exhibit fluctuations around the latter thermodynamic evolution \eref{eq:hydro_NR_and_R_non_fluct}.
We here detail how these fluctuations, which are captured by our fluctuating hydrodynamics \eref{eq:hydro_NR_and_R}, reveals the presence of nonreciprocal interactions at the microscopic level.
To this aim, we start by linearizing \eref{eq:hydro_NR_and_R} around $\rho_{\alpha}=\rho_0$, $m_{\alpha}=0$ when $v=0$.
Introducing the deviations $\delta\rho_{\alpha}=\rho_{\alpha}-\rho_0$ and $\delta m_{\alpha}=m_{\alpha}$, we find their evolution at linear order as 
\begin{subequations}
\label{eq:hydro_fluct_disorder}
\begin{align}
\partial_t\delta\rho_1=&D\partial_{xx}\delta\rho_1 + \sqrt{2Da\rho_0}\;\partial_x\eta_1^1\;, \\
\partial_t\delta m_1=&D\partial_{xx} \delta m_1 -\tilde{\gamma}(\rho_0,\rho_0) \chi\; \delta m_1 -\tilde{\gamma}(\rho_0,\rho_0)\vartheta\; \delta m_2 +\sqrt{2Da\rho_0}\;\partial_x\eta_2^1 + \sqrt{a\tilde{\gamma}(\rho_0,\rho_0)}\;\eta_3^1\;, \\
\partial_t\delta\rho_2=&D\partial_{xx}\delta\rho_2 + \sqrt{2Da\rho_0}\;\partial_x\eta_1^2\;, \\
\partial_t\delta m_2=&D\partial_{xx} \delta m_2 -\tilde{\gamma}(\rho_0,\rho_0) \chi\; \delta m_2 +\tilde{\gamma}(\rho_0,\rho_0)\vartheta\; \delta m_1 +\sqrt{2Da\rho_0}\;\partial_x\eta_2^2 + \sqrt{a\tilde{\gamma}(\rho_0,\rho_0)}\;\eta_3^2 \;,
\end{align}
\end{subequations}
where $\chi=1-\sinh(\beta)$, $\vartheta=\sinh(\beta_a)$, $\tilde{\gamma}$ is given by \eref{eq:gamma_eff_reciprocal_plus_nonreciprocal} and $\eta^{\alpha}_n$ are uncorrelated Gaussian white noises such that $\langle\eta^{\alpha}_n(x,t)\eta^{\beta}_{n^{\prime}}(x^{\prime},t^{\prime})\rangle=\delta_{\alpha,\beta}\delta_{n,n^{\prime}}\delta(x-x^{\prime})\delta(t-t^{\prime})$.
Taking the Fourier transform of \eref{eq:hydro_fluct_disorder}, we obtain the evolution of the $(q,\omega)$-th Fourier modes of the density $\delta \rho^{q,\omega}_{\alpha}$ and magnetization $\delta m^{q,\omega}_{\alpha}$ fields as
\begin{subequations}
\label{eq:hydro_fluct_disorder_fourier}
\begin{align}
i\omega \delta\rho^{q,\omega}_1=&-Dq^2\rho^{q,\omega}_1 + iq\sqrt{2Da\rho_0}\;\eta_1^1(q,\omega)\;, \\
i\omega\delta m^{q,\omega}_1=& -\left(Dq^2 +\tilde{\gamma}(\rho_0,\rho_0) \chi\;\right)m^{q,\omega}_1 -\tilde{\gamma}(\rho_0,\rho_0)\vartheta\; m^{q,\omega}_2 +iq\sqrt{2Da\rho_0}\;\eta_2^1(q,\omega) + \sqrt{a\tilde{\gamma}(\rho_0,\rho_0)}\;\eta_3^1(q,\omega)\;, \\
i\omega\delta\rho^{q,\omega}_2=&-Dq^2\delta\rho^{q,\omega}_2 + iq\sqrt{2Da\rho_0}\;\eta_1^2(q,\omega)\;, \\
i\omega\delta m^{q,\omega}_2=&-\left(Dq^2 +\tilde{\gamma}(\rho_0,\rho_0) \chi\right) m^{q,\omega}_2 +\tilde{\gamma}(\rho_0,\rho_0)\vartheta\; m^{q,\omega}_1 +iq\sqrt{2Da\rho_0}\;\eta_2^2(q,\omega) + \sqrt{a\tilde{\gamma}(\rho_0,\rho_0)}\;\eta_3^2(q,\omega) \;,
\end{align}
\end{subequations}
where $\eta^{\alpha}_n(q,\omega)$ are the Fourier transform of the gaussian white noises $\eta^{\alpha}_n(x,t)$ such that $\langle\eta^{\alpha}_n(q,\omega)\eta^{\beta}_n(q^{\prime},\omega^{\prime})\rangle=4\pi^2\delta_{\alpha,\beta}\delta_{n,n^{\prime}}\delta(q+q^{\prime})\delta(\omega+\omega^{\prime})$.
Inverting \eref{eq:hydro_fluct_disorder_fourier}, we obtain the inter(intra)-species magnetization correlations as
\begin{align}
\label{eq:cross_corr_m_fourier}
\langle \delta m^{q,\omega}_1\delta m^{q^{\prime},\omega^{\prime}}_2\rangle =&\frac{a4\pi^2\delta(q+q^{\prime})\delta(\omega+\omega^{\prime})i\omega\vartheta\left[\hat{\gamma}^2+4\hat{\gamma}Dq^2\right]}{\omega^4+2\omega^2\left[Dq^2+\hat{\gamma}(\chi-\vartheta)\right]\left[Dq^2+\hat{\gamma}(\chi+\vartheta)\right]+\left[(Dq^2+\hat{\gamma}\chi)^2+\hat{\gamma}^2\vartheta^2\right]^2}\;,\\
\label{eq:auto_corr_m_fourier}
\langle \delta m^{q,\omega}_1\delta m^{q^{\prime},\omega^{\prime}}_1\rangle =&\frac{a2\pi^2\delta(q+q^{\prime})\delta(\omega+\omega^{\prime})\left[\hat{\gamma}+4Dq^2\right]}{(Dq^2+\hat{\gamma}\chi)^2+\omega^2}\;,
\end{align}
where $\hat{\gamma}$ is a shorthand notation for $\hat{\gamma}=\tilde{\gamma}(\rho_0,\rho_0)$. 
Inverting back \eref{eq:cross_corr_m_fourier}-\eref{eq:auto_corr_m_fourier} in time, we obtain the correlation functions
\begin{align}
\label{eq:cross_corr_m_fourier_last}
\langle \delta m_1(x,0)\delta m_2(x,t)\rangle =& \sin(\vartheta\hat{\gamma}t)a\int \frac{dq}{2\pi}e^{-t(Dq^2+\hat{\gamma}\chi)}\frac{4Dq^2+\hat{\gamma}}{4Dq^2+4\hat{\gamma}\chi}\;, \\
\label{eq:auto_corr_m_fourier_last}
\langle \delta m_1(x,0)\delta m_1(x,t)\rangle =& a\int \frac{dq}{2\pi}e^{-t(Dq^2+\hat{\gamma}\chi)}\frac{4Dq^2+\hat{\gamma}}{4Dq^2+4\hat{\gamma}\chi} \;.
\end{align}
Performing the integration over $q$ then yields
\begin{align}
\langle \delta m_1(0)\delta m_2(t)\rangle =& \sin(\vartheta\hat{\gamma}t)a\mathcal{G}(t,\hat{\gamma}\chi)\;, &
\langle \delta m_1(0)\delta m_1(t)\rangle =& a\mathcal{G}(t,\hat{\gamma}\chi) \;,
\end{align}
where we have used translational invariance to omit the spatial dependency of the correlation functions \eref{eq:cross_corr_m_fourier_last}-\eref{eq:auto_corr_m_fourier_last} and where $\mathcal{G}(t,u)$ is given by 
\begin{align}
\label{eq:mathcal_g}
\mathcal{G}(t,u)=-\frac{1}{2\sqrt{D\pi}\sqrt{t}}e^{-tu}+\frac{\sqrt{u}}{8\sqrt{D}}\frac{4\chi-1}{\chi}\erfc\left(\sqrt{tu}\right)\;.
\end{align}
Looking at Eq~\eref{eq:mathcal_g}, we remark that $\mathcal{G}$ diverges as $1/\sqrt{t}$ at small time: the correlation functions $\langle \delta m_1(0)\delta m_2(t)\rangle$ and $\langle \delta m_1(0)\delta m_1(t)\rangle$ are ill-defined in this regime.
This is a known issue arising in stochastic field theories with conserved noises \cite{tauber2014critical,mccomb2003renormalization} which is solved by introducing a cutoff in the Fourier modes above $q\sim 1/a$ to take into account the finiteness of the lattice size.
Instead of following this route, we will hereafter consider only the ratio between  $\langle \delta m_1(0)\delta m_2(t)\rangle$ and $\langle \delta m_1(0)\delta m_1(t)\rangle$: this ratio remain well defined, even in the limit $t\to0$, and we can compare it to the same quantitiy measured in the agent-based simulations.
In particular we can express $\langle \delta m_1(0)\delta m_2(t)\rangle$ in terms of $\langle \delta m_1(0)\delta m_2(t)\rangle_{\beta_a=0}$ as
\begin{align}
\label{eq:cross_corr_final}
\langle \delta m_1(0)\delta m_2(t)\rangle =&\sin(\vartheta\hat{\gamma}t)\frac{\mathcal{G}(t,\hat{\gamma}\chi)}{\mathcal{G}(t,\hat{\gamma}_0\chi)}\langle \delta m_1(0)\delta m_1(t)\rangle_{\beta_a=0}\;
\end{align}
where $\langle \delta m_1(0)\delta m_1(t)\rangle_{\beta_a=0}$ is the intra-species magnetization correlation in the absence of nonreciprocity and $\hat{\gamma}_0=\tilde{\gamma}(\rho_0,\rho_0)$ with $\tilde{\gamma}$ given by \eref{eq:gamma_eff_reciprocal_plus_nonreciprocal} (note that here we consider the case $\beta_0=0$).
From Eq~\eref{eq:cross_corr_final}, we read the expression of $\mathcal{C}(t,\hat{\gamma},\hat{\gamma}_0)$ in the main text as 
\begin{align}
\label{eq:mathcal_c_final}
    \mathcal{C}(t,\hat{\gamma},\hat{\gamma}_0)=\frac{\mathcal{G}(t,\hat{\gamma}\chi)}{\mathcal{G}(t,\hat{\gamma}_0\chi)}\langle \delta m_1(0)\delta m_1(t)\rangle_{\beta_a=0}\;.
\end{align}
Rewriting \eref{eq:cross_corr_final}, we obtain a theoretical prediction for the ratio of the magnetization correlations as 
\begin{align}
\label{eq:ratio_corr_final}
\frac{\mathcal{G}(t,\hat{\gamma}_0\chi)}{\mathcal{G}(t,\hat{\gamma}\chi)}\frac{\langle \delta m_1(0)\delta m_2(t)\rangle}{\langle \delta m_1(0)\delta m_1(t)\rangle_{\beta_a=0}}=\sin(\hat{\gamma}\vartheta t)\;.
\end{align}
In \eref{eq:ratio_corr_final}, the left hand side can be evaluated from numerical simulations and compared with the theoretical prediction of the right hand side.
We perform such a comparison on \Fref{fig:time_corr} and show that \eref{eq:ratio_corr_final} indeed provides a quantitative description of the magnetization correlations in the disordered phase of the agent-based simulations.

\begin{figure}
\centering
\includegraphics{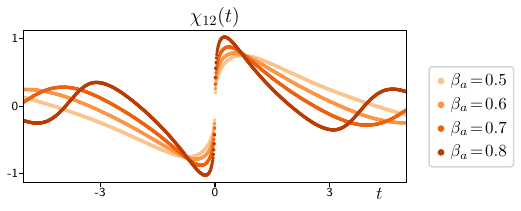}
\caption{Normalized anti-symmetric part of the magnetization time-correlation $\chi_{12}(t)$ as defined in \eref{eq:thermodynamic_bound} obtained from microscopic simulations for different values of $\beta_a$.
}
\label{fig:chi}
\end{figure}

To conclude this appendix, we now briefly discuss the universal thermodynamic bound recently introduced in Refs.~\cite{Ohga2023,Vu2023}. It establishes a relationship between the normalized time-antisymmetric part of the cross-correlations $\chi_{12}(t)$ and the maximum cycle affinity $\cF^t_{max}$ in the Markov transitions of the underlying model.
In the case of the NRASM, this relationship takes the form of the following inequality
\begin{align}
\label{eq:thermodynamic_bound}
    \bigg|\frac{\langle m_1(t)m_2(0)\rangle - \langle m_2(t)m_1(0)\rangle}{2 \sqrt{ \langle m_1(0)m_1(0)-m_1(t)m_1(0)\rangle \langle m_2(0)m_2(0)-m_2(t)m_2(0)\rangle}}\bigg|=|\chi_{12}(t)|\leq \frac{1}{2\pi}\cF^t_{\text{max}}\;.
\end{align}
As an alternative to $\cF^t_{\text{max}}$, the bound on the right hand side of \eref{eq:thermodynamic_bound} can also be expressed in terms of the rate of entropy production~\cite{Shiraishi2023,Vu2023}.
We plot the quantity $\chi_{12}(t)$ computed from agent-based simulations of \hyperlink{rule:diffusion}{RI}-\hyperlink{rule:flip2}{RIV} in \Fref{fig:chi}. 
In our case, the $t \to 0$ version of bound~\eref{eq:thermodynamic_bound} does not provide any information as $\chi_{12}(t \to 0) \to 0$. 
However, $\chi_{12}(t)$ takes finite values at finite delays, which gives a lower bound for the cycle with the highest affinity in the microscopics, and for the rate of entropy production.

\section{Numerical methods}
\subsection{Microscopic agent-based simulations}
\label{app:microscopic_simu}

In this appendix, we detail the microscopic simulations of the active lattice gas dynamics \hyperlink{rule:diffusion}{RI} to \hyperlink{rule:flip1}{RIV} which were used to obtain \Fref{fig:kymo_micro}. 
We used a random sequential update with discretized time and $N\sim 10^4$ particles. 
At each time step $dt$, we repeat N times the following chain of instructions 
\begin{itemize}
\item[LI] Draw a particle uniformly
\item[LII] Draw a float uniformly between 0 and 1
\item[LII] Update the position and spin of the particle according to the stochastic dynamic \hyperlink{rule:diffusion}{RI} to \hyperlink{rule:flip1}{RIV}
\end{itemize}
Note that, in \Fref{fig:kymo_micro}, the lattice size is of order $a\sim 0.003$ while the total length is $L = 100$: there is thus a high number of sites $N_{sites}\sim 3 \times 10^4$. 
Furthermore, we had to impose very small time step $dt \sim 1.1 \times 10^{-5}$ in order to keep the probabilty $Da^{-2} dt$ of a
diffusive move \hyperlink{rule:diffusion}{RI} of order 1.
As shown in \Fref{fig:kymo_micro}, the microscopic simulations exhibit the same phenomenology as their corresponding hydrodynamics \eref{eq:hydro_NR_and_R_non_fluct}. 
In particular, we observe all the four phases reported in the phase diagram of \Fref{fig:phase_diagram} in the main text: swap phase, bands, Chase \& Rest and inhomogeneous oscillations.

\subsection{Numerical methods for PDEs}
\label{app:numerics}
The numerical integrations of PDEs were carried out using a semi-spectral method with a semi-implicit Euler scheme. We used an anti-alias 3/2 rule and computed all the non-linearities in direct space before transforming them back in Fourier space.
The numerical integration of dynamical system \eref{eq:homogeneous_1} was performed using Mathematica \texttt{NDSolve}.

\begin{itemize}
    \item {\bf Movie 1.} Swap phase observed in the NRASM in the nonmotile case ($v=0$). The movie was obtained through direct numerical integration of \eref{eq:hydro_NR_and_R_non_fluct}. {\bf Parameters:} $L=100$, $\rho=1$, $D=1$, $v=0$, $\beta=1$, $\beta_a=0.3$, $\gamma=1$, $dx=0.2$, $dt=0.002$.
    \item {\bf Movie 2.} Band phase observed in the NRASM. The movie was obtained through direct numerical integration of \eref{eq:hydro_NR_and_R_non_fluct}.{\bf Parameters:} $L=100$, $\rho=1$, $D=1$, $v=1$, $\beta=1$, $\beta_a=0.1$, $\gamma=1$, $dx=0.05$, $dt=0.0002$.
    \item {\bf Movie 3.} Rest \& Chase phase observed in the NRASM. The movie was obtained through direct numerical integration of \eref{eq:hydro_NR_and_R_non_fluct}. {\bf Parameters:} $L=100$, $\rho=1$, $D=1$, $v=1$, $\beta=1$, $\beta_a=0.3$, $\gamma=1$, $dx=0.05$, $dt=0.0002$.
    \item {\bf Movie 4.} Inhomogeneous oscillations observed in the NRASM. The movie was obtained through direct numerical integration of \eref{eq:hydro_NR_and_R_non_fluct}. {\bf Parameters:} $L=1200$, $\rho=1$, $D=1$, $v=1$, $\beta=0.92$, $\beta_a=0.35$, $\gamma=1$, $dx=0.2$, $dt=0.01$.
    \item {\bf Movie 5.} Inhomogeneous oscillations observed in the NRASM. The movie was obtained through direct numerical integration of \eref{eq:hydro_NR_and_R_non_fluct}. {\bf Parameters:} $L=100$, $\rho=1$, $D=1$, $v=1$, $\beta=1.1$, $\beta_a=0.4$, $\gamma=1$, $dx=0.05$, $dt=0.001$.
    \item {\bf Movie 6.} Coarsening of the Rest \& Chase phase observed in the NRASM. The movie was obtained through direct numerical integration of \eref{eq:hydro_NR_and_R_non_fluct}. {\bf Parameters:} $L=150$, $\rho=1$, $D=1$, $v=1$, $\beta=1.15$, $\beta_a=0.3$, $\gamma=1$, $dx=0.05$, $dt=0.002$.
\end{itemize}

\subsection{Nonreciprocal flocking model}
\label{app:flying_kuramoto}
In order to simulate the nonreciprocal flocking model \eref{eq:kuramoto_evol}, we used a Molecular Dynamics algorithm leveraging an Ito forward stepping for the stochastic terms. 
We implemented periodic boundary conditions and partitioned space in squared chunks of size $r_0$ in order to speed up the computation of the interaction between the spins.
Finally, as an important bottleneck of MD lies in the generation of random numbers, we used IntelMKL Gaussian generators to accelerate the drawing of the $\eta_i^{\alpha}$.
\begin{itemize}
    \item {\bf Movie 7.} Rest \& Chase phase observed in the nonreciprocal flocking model obtained by numerically integrating \eref{eq:kuramoto_evol}.
    {\bf Parameters:} $L_x=L_y=40$, $T=1$, $\rho=0.5$, $v=1$, $\beta=0.75$, $\beta_a=0.25$, $r_0=1$, $dt=0.002$.
\end{itemize} 

\end{document}